\pdfoutput=1
\documentclass[12pt]{article}
\usepackage{cite}
\usepackage{graphicx}
\usepackage{latexsym}   
\usepackage{mathrsfs}  
\usepackage[scr=rsfs,cal=boondox]{mathalfa}
\usepackage[overload]{textcase}

\font\grande=cmr9.5 scaled \magstep4
\font\medio=cmr9.5 scaled \magstep2
\outer\def\beginsection#1\par{\medbreak\bigskip
      \message{#1}\leftline{\bf#1}\nobreak\medskip
\vskip-\parskip
      \noindent}

\begin{document}
\bibliographystyle{unsrt}

\titlepage

\vspace{1cm}
\begin{center}
{\grande  The invisible low-frequency gravitons and the audio band}\\
\vspace{0.5cm}
\vspace{1.5 cm}
Massimo Giovannini \footnote{e-mail: massimo.giovannini@cern.ch}\\
\vspace{0.5cm}
{{\sl   Istituto Nazionale di Fisica Nucleare, Milan-Bicocca, 20126 Milan, Italy}}\\ 
\vspace*{1cm}
\end{center}
\vskip 0.3cm
\centerline{\medio  Abstract}
\vskip 0.5cm
The low-frequency gravitons correspond to typical wavelengths that 
left the Hubble radius during the early inflationary stages of expansion and reentered after matter radiation equality. 
Consequently the temperature and the polarization anisotropies of the cosmic microwave background constrain the tensor-to-scalar-ratio in the aHz region but since the audio band and the MHz domain are sensitive to the post-inflationary expansion rate, the low-frequency determinations of the tensor-to-scalar-ratio can be combined with the high-frequency constraints. In this framework we examine the possibility that the low-frequency gravitons remain invisible in the aHz region but are still potentially detectable at much higher frequencies.  Because the  number of $e$-folds associated with the exit of the cosmic microwave background wavelengths depends both on the slow-roll parameters and on the total expansion rate after inflation, this approach leads to a set of lower bounds on the tensor-to-scalar-ratio. 
\noindent
\vspace{5mm}
\vfill
\newpage

\renewcommand{\theequation}{1.\arabic{equation}}
\setcounter{equation}{0}
\section{Introduction}
\label{sec1}
The tensor modes of the geometry affect all the angular power spectra of the Cosmic Microwave Background (CMB) anisotropies but the distinctive feature of relic gravitons in the aHz region\footnote{We shall be using the prefixes of the International System of Units so that, for instance, $1\, \mathrm{aHz} = 10^{-18}\,\,\mathrm{Hz}$, $1\, \mathrm{fHz} = 10^{-15}$ Hz and so on.} should appear from the analysis of the polarization anisotropies that have been originally discovered in the first releases of the WMAP collaboration \cite{WMAP1,WMAP2,WMAP3} and later confirmed by other ground based experiments as well as by the relatively recent observations of the Planck satellite \cite{PL1}. However the low-frequency gravitons not only affect the $E$-mode  but also the $B$-mode polarization which is instead absent in the case of curvature inhomogeneities; in spite of some claims subsequently withdrawn \cite{B1}, so far there is no evidence for the presence of a $B$-mode polarization caused by the low-frequency gravitons. There are, of course, sources of $B$-mode polarization that have been successfully detected in the CMB like the one associated with the 
gravitational lensing of the primary anisotropies \cite{SPT}. These $B$-mode signals 
are however not related with the ones generated by the relic gravitons in the aHz region.
The current analyses suggest that the temperature and the polarization anisotropies of the Cosmic Microwave Background (CMB) are caused by the curvature inhomogeneities that are Gaussian and (at least predominantly) adiabatic; these are, in a nutshell, the distinctive features of the so-called adiabatic paradigm whose formulation can be traced back to the pioneering analyses of Peebles and Yu \cite{PY}. In this framework various bounds on the aHz gravitons have been deduced both by the WMAP team \cite{WMAP1,WMAP2,WMAP3} and by the subsequent experiments (see e.g. \cite{RR1,RR2,RR3}).

In practice the bounds on the aHz gravitons take the form 
of limits on the so-called tensor-to-scalar-ratio $r_{T}$ since, in the adiabatic paradigm\footnote{Entropic (or non-adiabatic) modes are strictly absent in the minimal version of the concordance paradigm. Depending on 
the scenario the entropic modes can be up to five \cite{nad1,nad2,nad3,nad4}. In the presence 
of non-adiabatic modes a tensor-to-scalar-ratio must be introduced for each of the modes ad for their correlations. We shall stick to the adiabatic paradigm also because its non-adiabatic extensions are strongly constrained \cite{RR1,RR2,RR3}.}, the production of relic gravitons is associated with decreasing frequency spectra whose amplitudes and slopes are simultaneously fixed at a conventional pivot wavenumber $k_{p}$  by the relation $r_{T}=  {\mathcal A}_{T}/{\mathcal A}_{{\mathcal R}}$ where ${\mathcal A}_{T}$ and ${\mathcal A}_{{\mathcal R}}$ denote, respectively,  the amplitudes of the tensor and of the scalar power spectra.  The lowest range of comoving frequencies corresponds therefore to $\nu= {\mathcal O}(\nu_{p})$ where $\nu_{p} = k_{p}/(2\pi) = 3.092\,\,\mathrm{aHz}$ and $k_{p} = 0.002\, \mathrm{Mpc}^{-1}$. All the CMB experiments setting limits on $r_{T}$ are therefore constraining the aHz domain and these bounds will get progressively more stringent in the future. While the WMAP collaboration could set upper limits $r_{T} < {\mathcal O}(0.1)$, the most recent determinations suggest $r_{T} < {\mathcal O}(0.06)$ or even $r_{T} < {\mathcal O}(0.03)$ \cite{RR1,RR2,RR3}. In the concordance scenario (often dubbed $\Lambda$CDM paradigm where $\Lambda$ denotes the dark energy component and CDM stands for the cold dark matter contribution) the spectral slope $n_{T}$ and the slow-roll parameter $\epsilon$  are expressible in terms of $r_{T}$ according to the so-called consistency relations stipulating that $n_{T} \simeq - 2\epsilon \simeq - r_{T}/8$. The stochastic backgrounds of relic gravitons are however not peculiar of the $\Lambda$CDM case and have been actually suggested well before the formulation of any of the current scenarios aiming at a specific account of the early stages of the evolution of our Universe \cite{HH1,HH2,HH3}.

As already mentioned above, the frequencies ${\mathcal O}(\mathrm{few})$ aHz correspond to wavelengths that left the Hubble radius well before the onset of the radiation-dominated phase and reentered after matter-radiation equality \cite{AA4,AA5}. However, even assuming the presence of an early inflationary stage, the high-frequency range of the spectrum chiefly depends on the post-inflationary evolution which is observationally unaccessible \cite{AA7,AA8,AA9}. Nonetheless, the maximal comoving frequency of the relic graviton spectrum can be generally expressed as:
\begin{equation}
\nu_{max} = \overline{\nu}_{max}(r_{T}, \, {\mathcal A}_{{\mathcal R}})/{\mathcal D}(\delta_{i}, \, \xi_{i}),
\label{INT1}
\end{equation}
where $\overline{\nu}_{max}$ depends on $r_{T}$ and on the 
amplitude of curvature inhomogeneities ${\mathcal A}_{{\mathcal R}}$; in Eq. (\ref{INT1}) ${\mathcal D}(\delta_{i}, \, \xi_{i})$ accounts for the post-inflationary evolution and its specific form is not essential for the purposes of this introduction (see, however, the discussion  of appendix \ref{APPA}).  Since there could be various distinct epochs, the subscripts appearing in the expansion rates (i.e. $\delta_{i}$) and in their relative durations (i.e. $\xi_{i}$) count the successive expanding phases. Assuming that all the $\delta_{i}$ go to $1$ we have that, within the present notations, $\nu_{max} \to \overline{\nu}_{max} = {\mathcal O}(270) \,\, (r_{T}/0.06)^{1/4} \,\mathrm{MHz}$ and  Eq. (\ref{INT1}) eventually reproduces the maximal frequency of the spectrum obtained when the inflationary stage of expansion is suddenly replaced by a radiation-dominated phase. In this case the spectral energy density in critical units (denoted hereunder\footnote{ The spectral energy density in critical units (specifically defined later on) is denoted by $\Omega_{gw}(\nu, \tau_{0})$.  It is however customary to deal directly with $h_{0}^2 \, \Omega_{gw}(\nu, \tau_{0})$ since the latter quantity does not depend on the indetermination of the present Hubble rate. We also note that the present value of the scale factor is normalized as $a_{0} =1$ and this means that at $\tau_{0}$ the comoving and the physical frequencies coincide.} by $h_{0}^2 \, \Omega_{gw}(\nu,\tau_{0})$) is quasi-flat as a function of the comoving frequency.  If the post-inflationary expansion rate is different from radiation $h_{0}^2 \Omega_{gw}(\nu,\tau_{0})$ is generally not flat and whenever the expansion rate is faster than radiation the spectral energy density in critical units decreases. Conversely when the expansion rate is slower than radiation $h_{0}^2 \Omega_{gw}(\nu,\tau_{0})$ increases as a function $\nu$.  As we shall see the post-inflationary evolution also affect the maximal frequency of the spectrum illustrated in Eq. (\ref{INT1}):  $\nu_{max}$ is either larger or smaller than ${\mathcal O}(270)$ MHz depending on the expansion rate (see, in this respect, the appendix \ref{APPA} and the discussion therein).
 
Also the total number of inflationary $e$-folds depends on the post-inflationary evolution and if we focus on the number of $e$-folds associated with the crossing of the CMB scales $k = {\mathcal O}(k_{p})$ we have that\footnote{In the present paper $\ln{x}$ denotes 
the natural logarithm of a generic variable $x$; $\log{x}$ denotes instead the common logarithm
of the same quantity. As a consequence of Eq. (\ref{INT2}), as we shall see, it follows that $N_{k} > {\mathcal O}(60)$ when the post-inflationary evolution is slower than radiation 
while $N_{k} < {\mathcal O}(60)$ is the post-inflationary evolution is faster than radiation. }
\begin{equation}
 N_{k} = {\mathcal O}(60) + \frac{1}{4} \ln{\biggl(\frac{r_{T}}{0.06}\biggr)} + \ln{{\mathcal D}(\delta_{i}, \, \xi_{i})},
 \label{INT2}
 \end{equation}
where the consistency relations have been assumed. Since $N_{k}$ enters directly the predictions of the spectral index and of the tensor-to-scalar ratio we can conclude that the post-inflationary evolution ultimately affects all the inflationary observables including the spectral index of curvature inhomogeneities.

In short the main purpose of this analysis 
is to answer the following question: by how much can we reduce $r_{T}$ without suppressing the signal in the audio and in the MHz bands? From a phenomenological viewpoint the interplay between the aHz region and the high-frequency range rests on various of observations that are progressively limiting the degree of arbitrariness of the high-frequency shape of $h_{0}^2\, \Omega_{gw}(\nu,\tau_{0})$. Starting in 2004  the wide-band detectors obtained a series of limits on the spectral energy density of the relic gravitons \cite{www1,www2,www3,www4} for a typical pivot frequency smaller than $300$ Hz. While different spectral slopes lead to slightly modified bounds we should have that $h_{0}^2 \Omega_{gw}(\nu,\tau_{0}) < {\mathcal O}(10^{-9})$ for comoving frequencies falling in the audio band (i.e. bewteen few Hz and $10$ kHz). Moreover the pulsar timing arrays (PTA) recently reported an evidence potentially attributed to the relic gravitons: the four collaborations currently investigating the nHz band (between few nHz and $0.1\,\,\mu\mathrm{Hz}$)
report compatible determinations of $h_{0}^2\, \Omega_{gw}(\nu,\tau_{0})$ \cite{NANO,PPTA,EPTA,IPTA} implying   $10^{-9.09} \leq h_{0}^2 \, \Omega_{gw}(\nu,\tau_{0}) \leq 10^{-8.07}$. There are finally indirect bounds on the high-frequency branch of the spectrum coming from big-bang nucleosynthesis: since the additional relativistic species increase the expansion rate at the nucleosynthesis time it is possible to set a bound on the possible presence of relic gravitons and this constraint is customarily phrased in terms of a specific integral of $h_{0}^2\, \Omega_{gw}(\nu,\tau_{0})$ \cite{BBN1,BBN2,BBN3} between $10^{-2}$ nHz and $\nu_{max}$ whose precise value depends, as already stressed in Eq. (\ref{INT1}), on the details of the post-inflationary evolution. 
 
Equations (\ref{INT1}) and (\ref{INT2}) suggest that the low-frequency 
limits on $r_{T}$ are closely related with the high-frequency determinations through ${\mathcal D}(\delta_{i}, \xi_{i})$. Some time ago \cite{CC1} it has been pointed out that the low-frequency and high-frequency determinations of the relic graviton backgrounds could be eventually combined in a synergic perspective in order to improve the determinations of the cosmological parameters. In the concordance paradigm the low-frequency signal is often maximized by assuming the largest $r_{T}$ compatible with the current data. In this analysis we instead focus on the possibility that $r_{T}(k_{p}) \ll 0.06$ and this choice implies that the low-frequency gravitons might be eventually invisible in the aHz region without jeopardizing the possibility of a larger signal in the audio band or even in the GHz domain.

The layout of this paper is the following. In section \ref{sec2} we introduced the timeline of the comoving horizon and the connections between 
$r_{T}(k,\tau)$ and $\Omega_{gw}(k,\tau)$ in the different frequency domains relevant for the present analysis. In view of its relevance, the case of single-field scenarios satisfying the consistency relations is also discussed at the end of section \ref{sec2}. In section \ref{sec3} the most constraining physical situations are addressed in preparation for the bounds of section \ref{sec4} where a number of more specific examples is also presented. Section \ref{sec5} contains the concluding remarks and, to avoid digressions,  some useful results have been collected in the appendices \ref{APPA} and \ref{APPB}.

\renewcommand{\theequation}{2.\arabic{equation}}
\setcounter{equation}{0}
\section{The timeline of the comoving horizon and $r_{T}$}
\label{sec2}
\subsection{The timeline of the comoving horizon}
The expressions of the maximal frequency and of the number of $e$-folds given 
in Eqs. (\ref{INT1})-(\ref{INT2}) follow from the general form of the comoving horizon 
when the inflationary stage is supplemented by a post-inflationary 
evolution that does not necessarily coincide with the radiation-dominated 
plasma.
\begin{figure}[!ht]
\centering
\includegraphics[height=8cm]{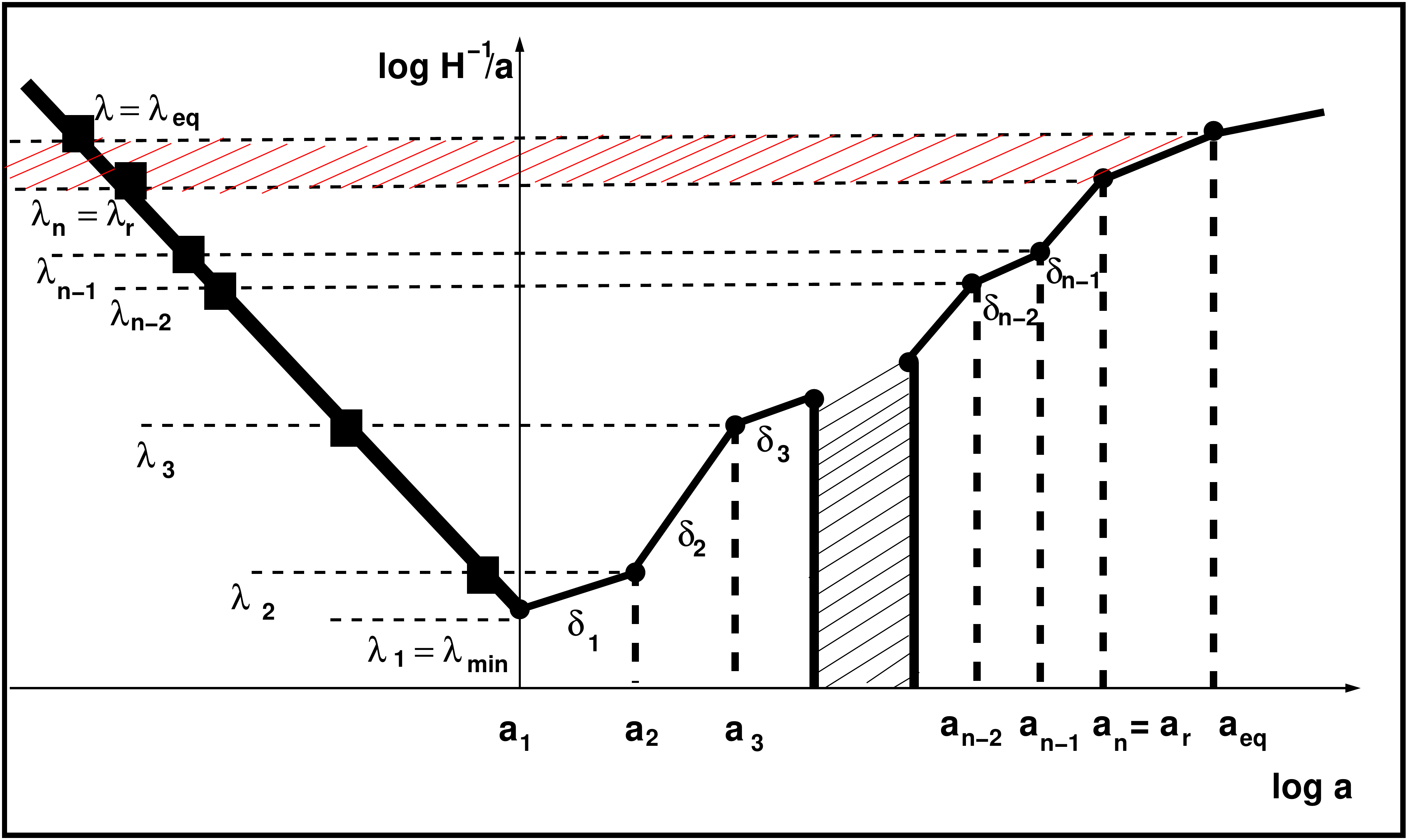}
\caption[a]{The common logarithm of the comoving horizon is illustrated as 
a function of the common logarithm of the scale factor. The left region of the cartoon corresponds to the inflationary evolution while the post-inflationary stage occurs 
for $a> a_{1}$. Between $a_{1}$ and $a_{n}$ there are $n-1$ successive sages of expansion characterized by the rates $\delta_{i}$; we consistently use the convention that $a_{n}$ coincides with $a_{r}$ (i.e. the beginning of the conventional radiation-dominated evolution). It is also understood throughout the discussion that the post-inflationary stage is bounded by the curvature scale of big-bang nucleosynthesis so that $H_{r} \geq 10^{-44} \, M_{P}$. The presence of a non-standard evolution 
between $a_{1}$ and $a_{n}$ potentially modifies the number of $e$-folds as well as the 
spectral energy density of the produced gravitons. The shaded stripe corresponds to the bunch of wavelengths exiting the horizon during inflation and reentering between radiation dominance 
and matter-radiation equality.}
\label{FIGU1}      
\end{figure}
In Fig. \ref{FIGU1} we illustrate the comoving horizon as a function of the scale 
factor (common logarithms are assumed on both axes).
The thick (diagonal) line at the left of the cartoon corresponds to the 
inflationary evolution and the {\em filled squares} define the exit of a given (comoving) wavelength while {\em the disks} in the right portion of the plot denote reentry of the 
selected scale. According 
 to Fig. \ref{FIGU1} the wavelengths smaller than $\lambda_{r}$ reenter before radiation dominance while the wavelengths $\lambda> \lambda_{r}$ (illustrated by a shaded stripe) reenter between the onset of radiation dominance and the epoch of matter--radiation equality. 

For $\lambda < \lambda_{r}$ the wavelength ${\mathcal O}(\lambda_{min})$ corresponds to comoving frequencies ${\mathcal O}(\nu_{max})$, i.e. the maximal frequency of the spectrum already mentioned in Eq. (\ref{INT1}). The scales  $\lambda_{r} < \lambda <\lambda_{eq}$ were still larger than the comoving horizon prior to matter--radiation equality and exited about $N_{k}$ $e$-folds before the end of inflation; the corresponding wavenumbers range therefore between $0.05\,\, \mathrm{Mpc}^{-1}$ and $0.002\,\, \mathrm{Mpc}^{-1}$ and the corresponding number of $e$-folds is given in Eq. (\ref{INT2}). By considering the timeline of Fig. \ref{FIGU1} we can obtain the specific form of $N_{k}$ which is in fact derived in appendix \ref{APPA} (see, in particular, Eq. (\ref{APA7})):
\begin{equation}
N_{k} = 59.4  + \frac{1}{4} \ln{\biggl(\frac{\epsilon_{k}}{0.001}\biggr)}  - \ln{\biggl(\frac{k}{0.002\,\,\mathrm{Mpc}^{-1}}\biggr)} + \frac{1}{2}\sum_{i}^{n-1} \, \biggl(\frac{\delta_{i} -1}{\delta_{i} + 1}\biggr) \, \ln{\xi_{i}} - \frac{1}{2} \ln{\biggl(\frac{H_{1}}{H_{k}}\biggr)}.
\label{NK1}
\end{equation}
Some of the numerical factors  discussed in appendix \ref{APPA} have been purposely neglected in Eq. (\ref{NK1}) and $\epsilon_{k}$ is the value of the slow-roll parameter 
when the scale ${\mathcal O}(k^{-1})$ crosses the comoving horizon. In Eq. (\ref{NK1}) 
the various $\xi_{i}$ measure the duration of each post-inflationary stage of expansion and since 
the expansion rate is always decreasing we conclude that\footnote{Note that $\xi_{n-1} = H_{n}/H_{n-1}$;
but $H_{n}= H_{r}$ since, by construction, the value of $a_{n}$ coincides with $a_{r}$, i.e. the scale 
factor at radiation dominance.}
\begin{equation}
\xi_{i} = \frac{H_{i+1}}{H_{i}} < 1, \qquad \prod_{i=1}^{n-1} \xi_{i} = \xi_{1}\,\,\xi_{2}\,\,.\,.\,.\,\,\xi_{n-2} \,\, \xi_{n-1} = \xi_{r} = H_{r}/H_{1} < 1.
\label{NK2}
\end{equation}
The fourth term at the right-hand side of Eq. (\ref{NK1}) 
is the natural logarithm of ${\mathcal D}^{-1}(\delta_{i}, \xi_{i})$ already introduced in  Eq. (\ref{APA4}) and here rewritten for immediate convenience:
\begin{equation}
{\mathcal D}(\delta_{i}, \, \xi_{i}) = \prod_{i=1}^{n-1}\, 
\xi_{i}^{- \frac{(\delta_{i} -1)}{2\, (\delta_{i} +1)}}.
\label{NK3}
\end{equation}
Concerning Eqs. (\ref{NK1})--(\ref{NK2}) and (\ref{NK3}) we remark that a reduction 
of $\epsilon_{k}$ also implies a reduction of $N_{k}$ but this effect 
is overall secondary in view of the scales of the problem: for 
a reduction of $4$ orders of magnitude of $\epsilon_{k}$ we have 
that $N_{k}$ gets reduced of $1$ order of magnitude which 
is practically immaterial for the present purposes. In case the consistency relations are enforced, a reduction 
of $\epsilon_{k}$ also implies a suppression of $r_{T}$.

If the post-inflationary plasma is only dominated by radiation then  in Eq. (\ref{NK3})
all the $\delta_{i}$ go to $1$ and the fourth term in Eq. (\ref{NK1}) disappears. Conversely 
when some of the $\delta_{i}$ are smaller than $1$ then the expansion rate gets slower than radiation
and $N_{k}$ increases. For $\delta_{i} > 1$ we may have also the opposite effect suggesting 
an overall reduction of $N_{k}$. However, as discussed in section \ref{sec3}, when $\delta_{i} >1$ the spectral energy density is generally decreasing and a large spectral energy density at high and intermediate frequencies is not expected. 
In both situations, as stressed in appendix \ref{APPB}, the Hubble 
rate at the exit of the given scale ($H_{k}$ in Eq. (\ref{NK1})) 
coincides in practice with $H_{1}$ (i.e. the expansion rate at the end of inflation) so that the last term in Eq. (\ref{NK1}) does not contribute to $N_{k}$.

The timeline of Fig. \ref{FIGU1} also determines the expression of the maximal frequency $\nu_{max}$ (see Eq. (\ref{INT1}))
where $\overline{\nu}_{max}$ now corresponds to the maximal frequency in the case 
where all the $\delta_{i} \to 1$. The value of $\overline{\nu}_{max}$ can then be estimated in the case of a post-inflationary evolution dominated by radiation and it is given by:
\begin{equation}
\overline{\nu}_{max} = 195.38 \, {\mathcal C}(g_{s}, g_{\rho}) \, \biggl(\frac{{\mathcal A}_{{\mathcal R}}}{2.41\times 10^{-9}}\biggr)^{1/4}\,\,
\biggl(\frac{\epsilon_{k}}{0.001}\biggr)^{1/4} \,\, \biggl(\frac{h_{0}^2 \, \Omega_{R\,0}}{4.15\times 10^{-5}}\biggr)^{1/4} \,\,\mathrm{MHz}.
\label{NK5}
\end{equation}
Equation (\ref{NK5}) does not assume a specific relation between $\epsilon_{k}$ and $r_{T}$ however, 
if the consistency relations are enforced, we can always trade $\epsilon_{k}$ for $r_{T}$ and the value of $\overline{\nu}_{max}$ becomes\footnote{As also discussed in appendix \ref{APPA}, the impact of ${\mathcal C}(g_{s}, g_{\rho})$ is minor;
for typical values of the late-time parameters (i.e. $g_{\rho, \, r} = g_{s,\, r} = 106.75$ 
and  $g_{\rho, \, eq} = g_{s,\, eq} = 3.94$) we have that ${\mathcal C}(g_{s}, g_{\rho}) =0.7596$ and the determination of $\overline{\nu}_{max}$ of Eq. (\ref{NK6}) moves from $\overline{\nu}_{max} = 271.93$ MHz to $206.53$ MHz. }: 
\begin{equation}
\overline{\nu}_{max} = 271.93 \, {\mathcal C}(g_{s}, g_{\rho}) \, \biggl(\frac{{\mathcal A}_{{\mathcal R}}}{2.41\times 10^{-9}}\biggr)^{1/4}\,\,
\biggl(\frac{r_{T}}{0.06}\biggr)^{1/4} \,\, \biggl(\frac{h_{0}^2 \, \Omega_{R\,0}}{4.15\times 10^{-5}}\biggr)^{1/4} \,\,\mathrm{MHz}.
\label{NK6}
\end{equation}

\subsection{The spectral energy density and the tensor-to-scalar-ratio}
For typical wavelengths larger 
than the Hubble radius (see Fig. \ref{FIGU1}) the tensor 
to scalar ratio is practically constant while the spectral 
energy density is suppressed. In the opposite limit 
the two quantities are both time dependent. Moreover, 
while the actual definition of $r_{T}$ is largely unambiguous 
(except for the conventional choice of the pivot scale) the 
different prescriptions of the energy density do not generally agree 
for typical wavelengths larger than the Hubble radius. However, 
as recently pointed out  \cite{MGEN},
a consistent definition of the energy momentum pseudo-tensor 
follows from the variation of the second-order action with respect 
to the background metric. This approach, originally suggested 
in \cite{FPP}, must be combined with an appropriate averaging scheme \cite{MGEN}
(see also \cite{H1,H2}). In the case of the relic gravitons 
the averaging of the different combinations of the field operators follows from the quantum mechanical expectation 
values  \cite{DF,CC1,MGEN} so that 
the spectral energy density in critical units is eventually given by:
\begin{equation}
\Omega_{gw}(k,\tau) = \frac{1}{24 H^2 \, a^2} \biggl[ Q_{T}(k,\tau) + k^2 \, P_{T}(k,\tau) \biggr],
\label{TS1}
\end{equation}
and it depends both on the wavenumber and on the conformal time coordinate $\tau$;
$a(\tau)$ denotes the scale factor of a conformally flat background geometry and $H$ is the standard Hubble rate which is also related to its  conformal time counterpart as
${\mathcal H} = a\, H$. We recall, in this respect, that ${\mathcal H} = a^{\prime}/a$ where the prime denotes a derivation with respect to $\tau$. The two tensor power spectra appearing in Eq. (\ref{TS1}) are:
\begin{equation}
P_{T}(k,\tau) = \frac{ 4 \ell_{P}^2}{ \pi^2} \, k^3\, \bigl|\, F_{k}(\tau)\bigr|^2,\qquad\qquad Q_{T}(k,\tau) = \frac{ 4 \ell_{P}^2}{ \pi^2} \, k^3\, \bigl|\, G_{k}(\tau)\bigr|^2.
\label{TS2}
\end{equation}
In Eq. (\ref{TS2}) the mode functions $F_{k}(\tau)$ and $G_{k}(\tau)$  obey
\begin{equation}
G_{k} = F_{k}^{\prime}, \qquad G_{k}^{\prime} = -2 \,{\mathcal H} \,G_{k} - k^2 \, F_{k}, 
\label{TS3}
\end{equation}
The amplitude of the tensor power spectra is related to the 
one of the curvature inhomogeneities via the tensor-to-scalar-ratio which is, in general, scale-dependent 
and time-dependent:
\begin{equation}
r_{T}(k, \tau) = P_{T}(k,\,\tau)/P_{{\mathcal R}}(k,\tau), \qquad P_{{\mathcal R}}(k,\tau) = \frac{k^3}{2\pi^2} |\overline{F}_{k}(\tau)|^2.
\label{TS4}
\end{equation}
From Eqs. (\ref{TS2}) and (\ref{TS4}) we therefore have that $r_{T}(k,\tau)$ can also be written as
\begin{equation}
r_{T}(k, \tau) = 8 \ell_{P}^2 \bigl| F_{k}(\tau) \bigr|^2/\bigl| \overline{F}_{k}(\tau) \bigr|^2. 
\label{TS6}
\end{equation}
In Eqs. (\ref{TS4})--(\ref{TS6}) the mode functions $\overline{F}_{k}$ and $\overline{G}_{k}$ are associated with the curvature inhomogeneities and 
their evolution is formally similar to the one of Eq. (\ref{TS3}) 
\begin{equation}
\overline{G}_{k} = \overline{F}_{k}^{\prime}, \qquad \overline{G}_{k}^{\prime} = -2 \,{\mathcal F} \,\overline{G}_{k} - 
k^2 \, \overline{F}_{k}, 
\label{TS5}
\end{equation}
where now ${\mathcal F} = z^{\prime}/z$. The explicit form of $z(\tau)$ depends on the matter 
content when the corresponding wavelengths crossed the comoving horizon. In the case of single-field models $z= z_{\varphi} = a \, \varphi^{\prime}/{\mathcal H}$ while for an irrotational relativistic fluid  the term $k^2 \overline{F}_{k}$ in Eq. (\ref{TS5}) gets modified as  $c_{st}^2 \, k^2 \overline{F}_{k} $; furthermore $z = z_{t} = a^2 \sqrt{\rho_{t} + p_{t}}/({\mathcal H}\, c_{st})$ \cite{lukash} where $p_{t}$ and $\rho_{t}$ are, respectively, the total pressure and the total energy density of the relativistic fluid.

\subsubsection{Wavelengths larger than the Hubble radius}
Since the initial conditions of the Einstein-Boltzmann hierarchy (required for the calculations 
of the temperature and polarization anisotropies) are customarily set 
well before matter-radiation equality, Eqs. (\ref{TS4})--(\ref{TS6}) must be computed when the 
relevant wavelengths are larger than the Hubble radius\footnote{ Indeed, before 
matter-radiation equality,  $r_{T}(k,\tau)$ is used to set the initial conditions 
of the Einstein-Boltzmann hierarchy in CMB applications when 
the relevant wavelengths are still larger than the comoving horizon.}
i.e. for $\tau_{ex} \leq \tau < \tau_{re}$ and $k \ll a\, H$. 
In this regime  Eqs. (\ref{TS3}) and (\ref{TS5}) are independently solved and the result is:
\begin{equation}
F_{k}(\tau) = \frac{e^{- i k\tau_{ex}}}{ a_{ex}\sqrt{2 \, k}} \,{\mathcal Q}_{k}(\tau_{ex}, \tau), 
\qquad\qquad \overline{F}_{k}(\tau) = \frac{e^{- i k\tau_{ex}}}{ z_{ex} \sqrt{2 \, k}} \, \overline{{\mathcal Q}}_{k}(\tau_{ex}, \tau),
\label{TS7}
\end{equation}
where the subscripts imply that the various quantities are evaluated 
when $\tau \to \tau_{ex}$ ( i.e. when the relevant wavelengths cross the 
comoving horizon fro $a < a_{1}$ in Fig. \ref{FIGU1}). The functions 
${\mathcal Q}_{k}(\tau_{ex}, \tau)$ and $\overline{{\mathcal Q}}_{k}(\tau_{ex}, \tau)$ 
are defined as:
\begin{eqnarray}
{\mathcal Q}_{k}(\tau_{ex}, \tau) &=& 1 - ( i k + {\mathcal H}_{ex}) \, a_{ex}^2\,\int_{\tau_{ex}}^{\tau} 
\frac{\tau^{\prime}}{a^2(\tau^{\prime})},
\nonumber\\
\overline{{\mathcal Q}}_{k}(\tau_{ex}, \tau) &=& 1 - ( i k + {\mathcal F}_{ex}) \, z_{ex}^2\,\int_{\tau_{ex}}^{\tau} 
\frac{\tau^{\prime}}{z^2(\tau^{\prime})}.
\label{TS8}
\end{eqnarray}
The results of Eqs. (\ref{TS7})--(\ref{TS8}) follow by imposing the appropriate (quantum mechanical) initial conditions for $\tau< \tau_{ex}$ and by then solving Eqs. (\ref{TS3}) and (\ref{TS5}) across $\tau_{ex}$. If we then insert Eqs. (\ref{TS7})--(\ref{TS8}) into Eq. (\ref{TS6}) we obtain 
the explicit form of $r_{T}(k,\tau)$ valid for $k < a\, H$ and $\tau_{ex} \leq \tau < \tau_{re}$
\begin{equation}
r_{T}(k,\tau) = 8 \, \ell_{P}^2 \biggl(\frac{z_{ex}}{a_{ex}}\biggr)^2 \, \frac{| {\mathcal Q}_{k}(\tau_{ex}, \tau)|^2}{|\overline{{\mathcal Q}}_{k}(\tau_{ex}, \tau)|^2}.
\label{TS9}
\end{equation}
In the case of single field inflationary models 
and for the timeline of the comoving horizon illustrated in Fig. \ref{FIGU1} we obtain\footnote{In Eq. (\ref{TS10}) the crossing condition $\tau_{ex} \simeq 1/k$ has been used. We also recall that, in the case of single-field case, from the Friedmann equations $2 \,\dot{H} = - \ell_{P}^2 \dot{\varphi}^2$; to get Eq. (\ref{TS10}) we must also use the identity $(\varphi^{\prime}/{\mathcal H}) = (\dot{\varphi}/H)$.}:
\begin{equation}
r_{T}(k,\tau) = 8 \, \ell_{P}^2 \biggl(\frac{\dot{\varphi}^2}{H^2}\biggr)_{ex} \simeq 16 \, \epsilon_{k}, \qquad 
\qquad \epsilon_{k} = - \biggl(\frac{\dot{H}}{H^2}\biggr)_{ex},
\label{TS10}
\end{equation}
since ${\mathcal Q}_{k}(\tau_{ex}, \tau) \simeq \overline{{\mathcal Q}}_{k}(\tau_{ex}, \tau) \to 1$ 
for the timeline of Fig. \ref{FIGU1}.  The result of Eq. (\ref{TS10}) follows if the initial conditions of the scalar and tensor mode functions are set when the background is already inflating; this means, in practice, that the total number of $e$-folds is larger than $N_{k}$ (see also Eq. (\ref{NK1})). In case the total number of $e$-folds is close to $N_{k}$ the field operators 
corresponding to the scalar and tensor modes are not necessarily in the vacuum \cite{fluid1,fluid2} and 
 the  protoinflationary transition can break the consistency relations. The tensor-to-scalar-ratio of Eq. (\ref{TS10}) will then include a dependence upon the sound speed of the pre-inflationary phonons; instead of $r_{T}(k,\tau) \simeq 16 \, \epsilon_{k}$ we will have $r_{T}(k,\tau) \simeq 16 \, \epsilon_{k}\, c_{st}$. 

According to Eq. (\ref{TS10}) the tensor-to-scalar-ratio is approximately 
constant for wavelengths larger than the Hubble radius; in the same limit the spectral energy density of Eq. (\ref{TS1}) is instead suppressed and 
from Eq. (\ref{TS7}) we can show that:
\begin{equation} 
\Omega_{gw}(k,\tau) = \frac{k^4}{12\, \pi^2\, H^2\, \overline{M}_{P}^2\, a^2 \, a_{ex}^2}\biggl[ \bigl| {\mathcal Q}_{k}(\tau_{ex},\tau)\bigr|^2 + \frac{\bigl| {\mathcal Q}_{k}^{\prime}(\tau_{ex},\tau)\bigr|^2}{k^2}\biggr].
\label{TS11}
\end{equation}
Since the integral appearing in $Q_{k}(\tau_{ex}, \tau)$ can be evaluated by parts 
\begin{equation}
\int_{\tau_{ex}}^{\tau} \, \frac{d\tau^{\prime}}{a^2(\tau^{\prime})} = \frac{1}{3 - \overline{\epsilon}}\biggl(\frac{1}{a^3\, H} - \frac{1}{a_{ex}^3\, H_{ex}}\biggr), \qquad \qquad \overline{\epsilon} = \int_{\tau_{ex}}^{\tau} \, d\tau^{\prime}\,\frac{\epsilon(\tau^{\prime})}{a^2(\tau^{\prime})}/\int_{\tau_{ex}}^{\tau} \frac{d\tau^{\prime}}{a^2(\tau^{\prime})},
\label{TS12}
\end{equation}
inserting Eq. (\ref{TS12}) into Eq. (\ref{TS11}) (and bearing in mind that $H_{ex}\, a_{ex} \simeq k$)
the spectral energy density in critical units becomes:
\begin{eqnarray}
\Omega_{gw}(k,\tau) &=& \frac{2 \, |k\,\tau|^2}{3 \pi}\, 
\biggl(\frac{H_{ex}}{M_{P}}\biggr)^2 \,\, 
{\mathcal M}_{k}(\tau,\overline{\epsilon})
\nonumber\\
{\mathcal M}_{k}(\tau, \overline{\epsilon}) &=& 1 + 2\biggl(\frac{a_{ex}}{a}\biggr)^2 +  \frac{2}{(3 - \overline{\epsilon})^2}\biggl[ \biggl(\frac{a_{ex}}{a}\biggr)^2 k\tau -1\biggr]^2  - \frac{2}{(3 - \overline{\epsilon})}\biggl[ \biggl(\frac{a_{ex}}{a}\biggr)^2 k\tau -1\biggr],
\label{TS13}
\end{eqnarray}
implying, as expected, that $\Omega_{gw}(k,\tau)$ is suppressed in the 
two concurrent limits $\tau_{ex} \leq \tau < \tau_{re}$ and $k\tau\ll 1$.
We finally stress hat Eq. (\ref{TS12}) holds for $\overline{\epsilon} \neq 3$ since for $\overline{\epsilon} \to 3$ there is a logarithmic enhancement that is irrelevant in this case but that must be taken into account when the wavelengths are shorter than the Hubble radius. Only if $a^2(\tau) \simeq 1/{\mathcal H}$ the contribution of the integrand of Eq. (\ref{TS12}) is relevant; it corresponds to an extended stiff phase and, in this case, the spectral energy density and the other observables inherit a logarithmic correction.

\subsubsection{Wavelengths shorter than the Hubble radius}
When the wavelengths are shorter than the Hubble radius 
$F_{k}(\tau)$ and $G_{k}(\tau)$ exhibits standing oscillations for $\tau\geq \tau_{re}$
\begin{equation}
F_{k}(\tau) = \frac{e^{- i k \, \tau_{ex}}}{a \sqrt{2 \, k}} \, {\mathcal Q}_{k}(\tau_{ex}, \tau_{re}) \, \biggl(\frac{a_{re}}{a_{ex}}\biggr) \biggl\{ \frac{{\mathcal H}_{re}}{k} \sin{(k \Delta\tau)}
+ \cos(k\Delta\tau) \biggr\},\qquad\qquad \Delta \tau =( \tau- \tau_{re}), 
\label{TS14}
\end{equation}
where ${\mathcal Q}_{k}(\tau_{ex}, \tau_{re})$ has been already defined in Eq. (\ref{TS8}) and it is 
now evaluated for $\tau \to \tau_{re}$. Equation (\ref{TS14}) holds when all the corresponding wavelengths are shorter than the Hubble radius (i.e. for $k \tau \gg 1$) and in the same approximation $G_{k}(\tau)$ becomes:
\begin{equation}
G_{k}(\tau) = \frac{e^{- i k \, \tau_{ex}}}{a}\,\, \sqrt{\frac{k}{2}}\,\, {\mathcal Q}_{k}(\tau_{ex}, \tau_{re}) \, \biggl(\frac{a_{re}}{a_{ex}}\biggr) \biggl\{ \frac{{\mathcal H}_{re}}{k} \cos{(k\Delta\tau)}
- \sin(k\Delta\tau) \biggr\}.
\label{TS15}
\end{equation}
Equations (\ref{TS14})--(\ref{TS15}) assume and expanding background (i.e.  $a_{re} \gg a_{ex}$) but they are otherwise general since the  rates at $\tau_{ex}$ and $\tau_{re}$ have not been specified. After inserting Eqs. (\ref{TS14})--(\ref{TS15}) into Eq.  (\ref{TS1}) we obtain:
\begin{equation}
\Omega_{gw}(k,\tau) = \frac{2 \, k^4}{3\, \pi \,a^4\, H^2 \,M_{P}^2} \bigl| {\mathcal Q}(\tau_{ex}, \tau_{re})\bigr|^2 \biggl(\frac{a_{re}}{a_{ex}}\biggr)^2 \biggl( 1 + \frac{{\mathcal H}_{re}^2}{k^2}\biggr) \biggl[ 1 + {\mathcal O}\biggl(\frac{{\mathcal H}}{k}\biggr)\biggr].
\label{TS16}
\end{equation}
Equations (\ref{TS14})--(\ref{TS15}) and  (\ref{TS16}) hold up to corrections 
${\mathcal O}({\mathcal H}/k)$ that are small for wavelengths shorter than the Hubble radius. As long as the relevant wavelengths appearing in Fig. \ref{FIGU1} do their first crossing during the inflationary stage where $k \tau_{ex} = {\mathcal O}(1)$. It can happen however that 
$k \tau_{re} \ll 1$ if the second crossing takes place when 
$\epsilon(a) \to {\mathcal O}(2)$, i.e. close to a radiation-dominated 
stage of expansion. To understand this relevant limit we recall that Eq. (\ref{TS3}) 
can also be written in a decoupled form:
 \begin{equation}
 f_{k}^{\prime\prime} + \biggl[ k^2 - \frac{a^{\prime\prime}}{a} \biggr] f_{k} =0, \qquad \qquad g_{k} = f_{k}^{\prime} - {\mathcal H} f_{k},
 \label{TS17}
 \end{equation}
where $f_{k}(\tau) =a(\tau) F_{k}(\tau)$ and $g_{k}(\tau) = a(\tau) G_{k}(\tau)$. In the language of Eq. (\ref{TS17}) the solutions given in Eqs. (\ref{TS7}) and (\ref{TS14})--(\ref{TS15}) hold, respectively, for $k^2 \ll  |\, a^{\prime\prime}/a|$ and for $k^2 \gg  |\, a^{\prime\prime}/a|$. The turning points where the analytical behaviour of the solution changes are defined by $k^2 \simeq 
|\,a^{\prime\prime}/a|$ that can also be rewritten as: 
\begin{equation}
k^2 = a^2 \, H^2 \biggl[ 2 - \epsilon(a)\biggr], 
\label{TS18}
\end{equation}
where, as before, $\epsilon(a) = - \dot{H}/H^2$ is the slow-roll parameter\footnote{While during inflation $\epsilon\ll 1$,  in the post-inflationary phase 
the background decelerates (but still expands) and $\epsilon(a) = {\mathcal O}(1)$.}. 
When $\epsilon \neq 2$ both turning points are regular and this means that the two 
solutions of Eq. (\ref{TS18}) are in fact $k \tau_{ex} = {\mathcal O}(1)$ and $k \tau_{re} = {\mathcal O}(1)$. For instance when a given wavelength crosses the Hubble radius during inflation we have that $\epsilon \ll 1$ and $k \simeq a_{ex} \, H_{ex}$ 
that also means, by definition, $k \tau_{ex} \simeq 1$. Similarly 
if the given wavelength reenters in a decelerated stage of expansion different from radiation we also have that $k \simeq a_{re} \, H_{re}$. However, if the reentry occurs in the radiation stage (or close to it) we have that $\epsilon_{re}\to 2$ and the condition (\ref{TS18}) implies that $k \tau_{re} \ll 1$. In Eq. (\ref{TS16}) the two situations are distinct since ${\mathcal H}_{re}^2/k^2 = {\mathcal O}(1)$ when $\epsilon_{re} \neq 2$ while ${\mathcal H}_{re}^2/k^2 \gg 1$ for $\epsilon_{re} \to 2$. We shall get back to this point in section \ref{sec3} where Eq. (\ref{TS16}) will be used to deduce the analytic form of the spectral slopes in the different frequency domains. 
For short wavelengths the mode function for the curvature inhomogeneities  is given by
\begin{equation}
\overline{F}_{k}(\tau) = \frac{e^{- i k \, \tau_{ex}}}{z \sqrt{2 \, k}} \, \overline{{\mathcal Q}}_{k}(\tau_{ex}, \tau_{re}) \, \biggl(\frac{z_{re}}{z_{ex}}\biggr) \biggl\{ \frac{{\mathcal F}_{re}}{k} \sin{[k \Delta\tau]} + \cos[k\Delta\tau] \biggr\},
\label{TS19}
\end{equation}
and if we now insert Eqs. (\ref{TS14}) and (\ref{TS19}) into Eq. (\ref{TS6}) we obtain 
the wanted form of $r_{T}(k, \tau)$ valid for $ \tau \geq \tau_{re}$ in the short-wavelength limit (i.e. for $k \tau> 1$):
\begin{eqnarray}
r_{T}(k,\tau) &=& 8 \ell_{P}^2\, \biggl[\frac{z(\tau)}{a(\tau)}\biggr]^2 \, \biggl(\frac{a_{re}}{a_{ex}}\biggr)^2 \biggl(\frac{z_{ex}}{z_{re}}\biggr)^2 {\mathcal G}^2(k \Delta\tau),
\nonumber\\
{\mathcal G}(k \Delta\tau) &=& \frac{{\mathcal F}_{re}\sin{(k \Delta\tau)} + k\,\cos(k\Delta\tau)}{{\mathcal H}_{re}\sin{(k \Delta\tau)} + k\cos(k\Delta\tau)}.
\label{TS20}
\end{eqnarray}
In the limit $\epsilon_{re} \to 2$ we also have ${\mathcal H}_{re}/k \simeq {\mathcal F}_{re}/k \gg 1$ and, in this case, ${\mathcal G}(k \Delta \tau) \to 1$. Conversely, when $\epsilon_{re} \neq 2$ we have instead that 
${\mathcal H}_{re}/k \simeq {\mathcal F}_{re}/k = {\mathcal O}(1)$; also in this situation 
${\mathcal G}(k\Delta\tau) = {\mathcal O}(1)$. We can therefore deduce from Eq. (\ref{TS20}) that:
\begin{equation}
r_{T}(k, \tau) = 16 \, \epsilon_{k} \, \frac{\epsilon(\tau)}{\epsilon_{re}}, \qquad \tau \geq \tau_{re}, \qquad k\tau > 1.
\label{TS21}
\end{equation}
It seems that, in practice, $\epsilon(\tau)$ is piecewise constant after inflation and it is of the order of $\epsilon_{re}$ so that $r_{T}(k,\tau) \to 16 \epsilon_{k}$ even for short wavelengths. The constancy 
of $\epsilon(\tau)$ is more or less obvious when the background expands as simple power-law but if the reentry of the wavelength takes place when the inflaton potential is still dominant (and oscillating) $\epsilon(\tau)$ is still approximately constant. For this purpose we can first write $\epsilon(\tau)$ in terms of the inflaton potential $V(\varphi)$, i.e. 
\begin{equation}
\epsilon(\tau) = - \dot{H}/H^2 = 3 \dot{\varphi}^2/(\dot{\varphi}^2 + 2 V).
\label{TS22}
\end{equation}
As suggested long ago the coherent oscillations of the inflaton imply the approximate 
constancy of the corresponding energy density\footnote{Indeed we have, in general, 
that $\dot{\rho}_{\varphi} + 3 H \dot{\varphi}^2 =0$ where $\rho_{\varphi} = \dot{\varphi}^2/2 + V$ 
is the energy density of the inflaton. However since $\dot{\rho}_{\varphi} \simeq 0$ and $3 H\dot{\varphi}^2 \ll \dot{\rho}_{\varphi}$
we can also write that $\dot{\varphi}^2 = 2 (V_{max} - V)$ where $V_{max} = V(\varphi_{max})$.} \cite{osc};
for immediate convenience  the inflaton potential around its minimum can be parametrized as: 
\begin{equation}
V(\varphi) = V_{0} (\varphi/\overline{M}_{P})^{2 q},\qquad \rightarrow \qquad \dot{\varphi} = \sqrt{2 V_{max}}\sqrt{1 - x^{2 q}},
\label{TS23}
\end{equation}
where $x = \varphi/\varphi_{max}$. If the numerator and the denominator 
of Eq. (\ref{TS22}) are averaged over one period of oscillations (say between $\varphi=0$ and $\varphi= \varphi_{max}$)  $\epsilon(\tau)$ becomes
\begin{equation}
\epsilon(\tau) = \frac{3 \int_{0}^{1} \sqrt{ 1 - x^{2q}} dx}{ \int_{0}^{1} dx/\sqrt{ 1 - x^{2q}}} = \frac{3 \,q }{q +1}.
\label{TS24}
\end{equation}
Thus, from Eqs. (\ref{TS21}) and  Eq. (\ref{TS24}), $\epsilon(\tau)/\epsilon_{re} \to 1$ when the reentry occurs during a phase driven by the coherent inflaton oscillations. With the same technique the average expansion rate during a phase 
of coherent oscillations follws from Eq. (\ref{TS24}); in particular  
\begin{equation}
{\mathcal H}^{\prime} = \frac{1- 2 q}{q+1} {\mathcal H}^2 \qquad \Rightarrow \qquad a(\tau) = (\tau/\tau_{1})^{\delta}, \qquad \mathrm{where}\qquad \delta = (q+1)/(2 q -1).
\label{TS25}
\end{equation}
 This result implies 
that if the wavelengths exit during a stage dominated by the coherent oscillations of the inflaton we can expect that the slope of the spectral energy density can be determined according to Eq. (\ref{TS25}).  The evolution of the comoving horizon 
in Fig. \ref{FIGU1} assumes a sequence of different expanding stages characterized 
by the constancy of the expansion rate. A fully equivalent strategy is to consider 
the continuous variation of $\delta$ implying 
\begin{equation}
\frac{1}{\delta(\tau)} = - 1 - \frac{1}{2} \frac{\partial \ln{\rho_{t}}}{\partial\ln{a}} = -1 + \epsilon(\tau),
\label{TS26}
\end{equation}
where $\rho_{t}(a)$ denotes the total energy density governing the 
post-inflationary evolution prior to radiation.  In the case of inflaton-dominated 
oscillations $\rho_{t}(a) = \rho_{\varphi}$ and 
\begin{equation}
 \delta(a)  = 1/[\epsilon(a)  -1]= (q+1)/( 2 q -1).
\label{TS26a}
\end{equation}
By going back to Fig. \ref{FIGU1} we therefore have that when the given wavelength crosses the Hubble radius prior to radiation dominance the value of $\delta$ is scale-dependent $\delta_{k}= \delta(\tau_{re}) = \delta(1/k)$. This relation follows by recalling that 
during the post-inflationary stage illustrated in the cartoon of Fig. \ref{FIGU1}
$\delta(a) \neq 1$ which also implies $\epsilon(a)\neq 2$ in Eq. (\ref{TS18}).

\subsection{Consistency relations and some examples}
In single-field scenarios the spectral index $n_{s}(k)$, the tensor-to-scalar-ratio $r_{T}(k)$ and the tensor spectral index $n^{low}_{T}(k)$ obey the consistency relations\footnote{ 
In Eq. (\ref{PP2}) $n_{T}^{low}(k)$ denotes the low-frequency spectral index associated with the wavelengths reentering during the radiation stage. To avoid confusions we anticipate that in section \ref{sec3} one (or more) high-frequency spectral indices will also be introduced. The high-frequency spectral indices involve the wavelengths that reentered the effective horizon prior to radiation dominance, i.e. for $a< a_{r}$ in Fig. \ref{FIGU1}.} 
\begin{equation}
n_{s}(k) = 1 - 6 \epsilon_{k} + 2 \overline{\eta}_{k}, \qquad r_{T}(k) = 16 \, \epsilon_{k},\qquad n^{low}_{T}(k) = - 2 \, \epsilon_{k},
\label{PP2}
\end{equation}
where $\epsilon_{k}= \epsilon(1/k)$ and $\overline{\eta}_{k} = \overline{\eta}(1/k)$ denote the slow-roll parameters 
evaluated when the bunch of wavelengths corresponding to the CMB scales exited the comoving horizon approximately $N_{k}$ $e$-folds prior to the end of inflation. In more general terms it is well known that the slow-roll parameters are all time-dependent (or field-dependent) and they are defined, within the notations employed here, as:
\begin{equation}
\epsilon = - \frac{\dot{H}}{H^2} = \frac{\overline{M}_{P}^2}{2} \biggl(\frac{V_{\,\,, \varphi}}{V} \biggr)^2, 
\qquad 
\eta = \frac{\ddot{\varphi}}{H \, \dot{\varphi}} = \epsilon - \overline{\eta}, \qquad \qquad \overline{\eta} = \overline{M}_{P}^2 \biggl(\frac{V_{\,,\varphi\varphi}}{V}\biggr).
\label{PP5}
\end{equation}
According to the current limits, the tensor-to-scalar-ratio and the scalar spectral index are determined as \cite{RR1,RR2,RR3}
\begin{equation}
r_{T}(k,\tau_{ex}) < \overline{r}_{T}, \qquad\qquad n_{s}(k,\tau_{ex}) = \overline{n}_{s},
\label{PP1}
\end{equation}
where $\overline{r}_{T}$ ranges between ${\mathcal O}(0.06)$ and ${\mathcal O}(0.03)$ while 
$0.96448 < \overline{n}_{s} < 0.96532$ with a central value corresponding to $0.9649$. Once more, by definition, in Eq. (\ref{PP1}) $r_{T}(k,\tau_{ex}) = r_{T}(k,1/k) = r_{T}(k)$ and similarly  $n_{s}(k,\tau_{ex}) = n_{s}(k,1/k) = n_{s}(k)$.
For the monomial potentials $\epsilon_{k}$ and $\overline{\eta}_{k}$ are of the same order 
and these scenarios are practically excluded by current data. We shall rather consider potentials with typical form given by:
\begin{equation}
V(\Phi) =  M^4 \,\, v(\Phi),\qquad\qquad \Phi= \varphi/\overline{M}_{P},
\label{PAR1}
\end{equation}
where $M$ denotes the energy scale of the potential. Inflation occurs for $\Phi\gg 1$ (and in this limit we therefore have $v(\Phi) \to 1$) while it ends for $\Phi\ll 1$. Using the parametrization of Eq. (\ref{PAR1}) the expressions of Eq. (\ref{PP5}) 
get even simpler; for instance $\epsilon(\Phi) =(1/2) (v_{,\,\Phi}/v)^2$ and so on and so forth (see also the subsection \ref{APPB2} and the discussions therein). We can then write $v(\Phi)$ as the ratio of two functions 
scaling (approximately but not exactly) with the  same power for $\Phi \gg 1$. An example along this direction is:
\begin{equation}
v(\Phi) = \frac{\beta^{p} \Phi^{2 q}}{[ 1 + \beta^2\Phi^{\frac{4 q}{p}}]^{\frac{p}{2}}},\qquad 4 q > p,\qquad \beta>0.
\label{PP3}
\end{equation}
In Eq. (\ref{PP3}) $\beta$, $p$ and $q$ are the parameters of the potential and, for technical reasons, we consider the case $4\, q>p$. 
A $q$-dependent  oscillating stage takes place for $\Phi \ll 1$ where the potential can be written as $v(\Phi) = \beta^{p} \Phi^{2 \, q}$.
By following the same strategy different concrete examples can be concocted like, for instance,
\begin{equation}
v(\Phi) = \frac{\bigl(e^{\gamma \Phi} -1)^{2 q}}{\bigl(e^{\frac{4 \gamma q}{p} \,\Phi} +1 \bigr)^{\frac{p}{2}}},\qquad 4 q > p, \qquad \beta >0.
\label{PP4}
\end{equation}
\begin{figure}[!ht]
\centering
\includegraphics[height=8cm]{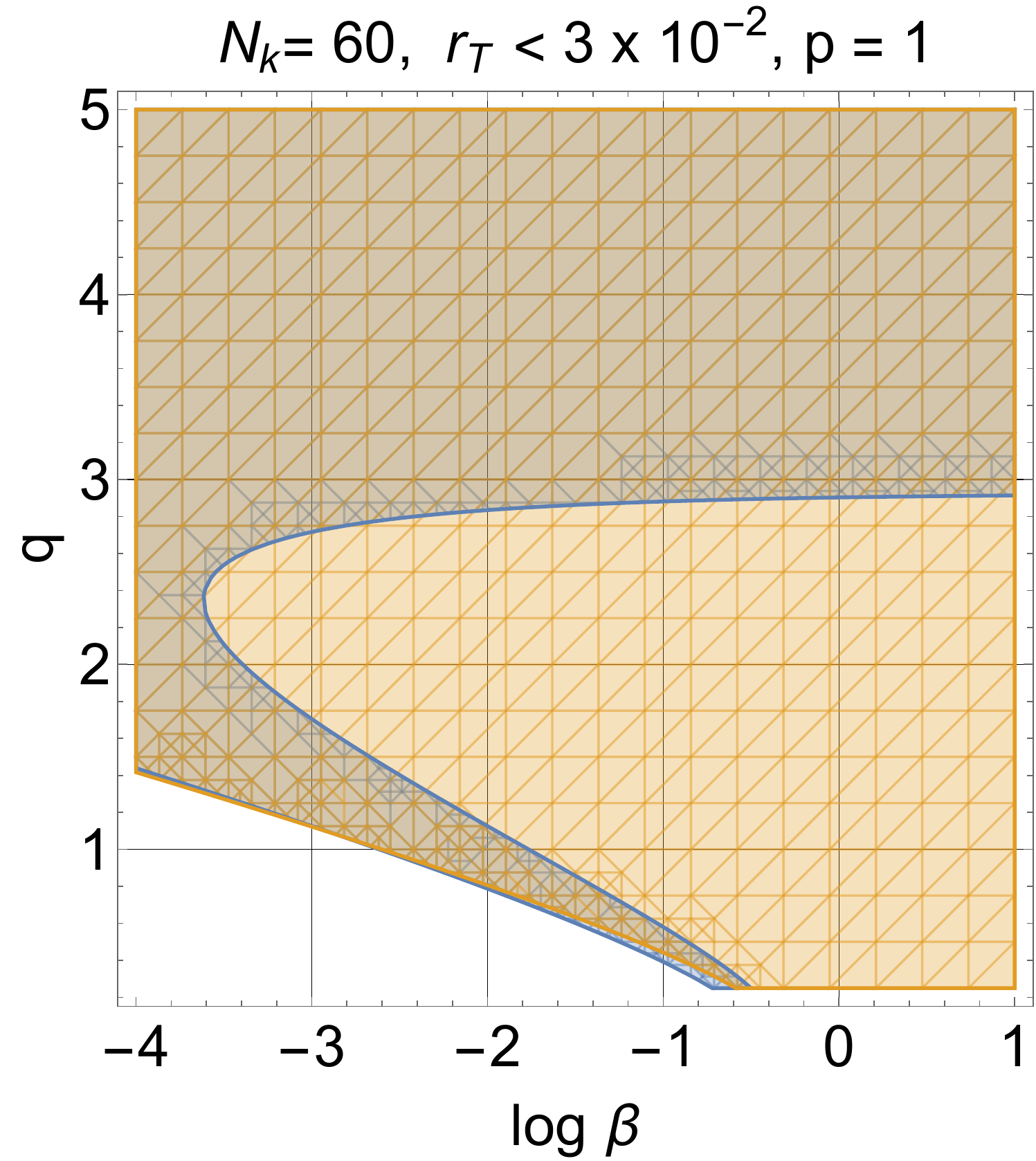}
\includegraphics[height=8cm]{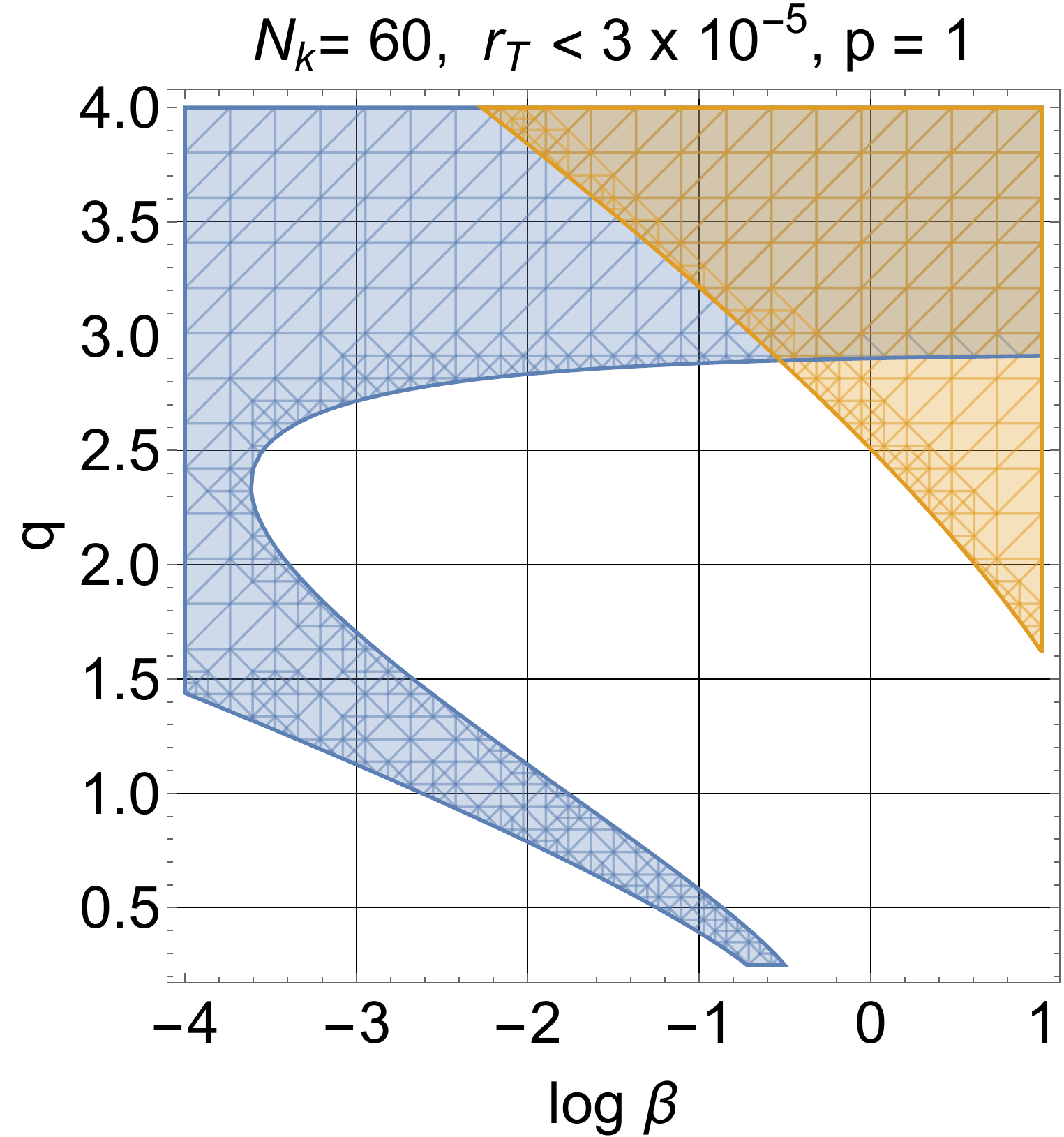}
\caption[a]{If $N_{k}$ is fixed, Eqs. (\ref{PPS1})--(\ref{PPS2}) together with the constraints 
of Eq. (\ref{PP1}) define the allowed region of the parameter 
space which is illustrated here in the plane $(\log{\beta}, \, q)$.
In the plot at the left we consider the overlap between the regions defined by $0.96448 < \overline{n}_{s} < 0.96532$ and by $r_{T} < 3 \times 10^{-2}$. In the plot at the right the limit on $r_{T}(k)$ is reduced from $r_{T} < 3\times 10^{-2}$ to
$r_{T} < 3 \times 10^{-5}$. As the limit on $r_{T}(k)$ becomes more stringent the overlap shrinks and it is localized in the region of large $q$.}
\label{FIGU2}      
\end{figure} 
While the examples along the lines of Eqs. (\ref{PP3})--(\ref{PP4}) can be multiplied, for the present purposes, different functional forms of the potential do not radically modify the scaling of the slow-roll parameters and of the tensor-to-scalar ratio. To investigate this point it is interesting to compute $r_{T}(k)$ and $n_{s}(k)$ directly in terms of $N_{k}$ and of the other parameters of the potential. 
For this purpose Eq. (\ref{PP2}) must be evaluated when the wavelengths compatible with 
the pivot scale cross the Hubble radius during inflation (see 
Fig. \ref{FIGU1}). Thanks to the results of appendix \ref{APPB} (see in particular Eqs. (\ref{PP12})--(\ref{PP13})) $n_{s}(k)$ becomes:
\begin{equation}
n_{s}(k) = n_{s}(N_{k}) =1 - \frac{12 q^2 \, \beta^{ - 2/(1+ 2q/p)}}{[ 4\, q\, ( p + 2 q)\, N_{k}/p]^{(p+ 4 q)/(p+ 2 q)}}- \frac{p +4 q}{(p + 2\, q) \, N_{k}}.
\label{PPS1}
\end{equation}
By always referring to  Eqs. (\ref{PP12})--(\ref{PP13}) the expressions of $r_{T}(k)$ and $n_{T}^{low}(k)$ are given by:
\begin{eqnarray}
r_{T}(k) &\equiv& r_{T}(N_{k}) = \frac{32 \, q^2 \,  \beta^{ - 2/(1+ 2q/p)}}{[ 4\, q\, ( p + 2 q)\, N_{k}/p]^{(p+ 4 q)/(p+ 2 q)}}, 
\nonumber\\
n^{low}_{T}(k) &\equiv& n^{low}_{T}(N_{k})
= - \frac{4 \, q^2 \, \beta^{ - 2/(1+ 2q/p)}}{[ 4\, q\, ( p + 2 q)\, N_{k}/p]^{(p+ 4 q)/(p+ 2 q)}}.
\label{PPS2}
\end{eqnarray}
The results of Eqs. (\ref{PPS1})--(\ref{PPS2}) are illustrated in Figs. \ref{FIGU2}, \ref{FIGU3} and \ref{FIGU4} 
for different values of $q$, $p$ and $\beta$.  In Fig. \ref{FIGU2} the overlap between the two regions defines 
the portion of the parameter space where  the scalar spectral index is phenomenologically viable and the 
limit on $r_{T}(k)$ is safely enforced. By comparing the left and the right plots in Fig. \ref{FIGU2} we see 
that the overlap shrinks as soon as the bounds on $r_{T}$ become progressively more demanding. In the left plot of Fig. \ref{FIGU2} the region bounded by a straight line and a curve follows 
by requiring that $n_{s}(k)$ falls within the $1\,\sigma$ observational limits set by the Planck collaboration complemented by the lensing observations, i.e. $\overline{n}_{s} = 0.9649\pm 0.0042$. The addition of the baryon acoustic 
oscillations would imply a slightly different figure (i.e. $\overline{n}_{s} = 0.9665\pm0.0038$)
which is however not essential for the illustrative purposes of this discussion. The 
second region bounded by a straight line in the left plot of Fig. \ref{FIGU2} 
corresponds to the requirement $r_{T}<0.03$. In the right plot the limit on $r_{T}$ 
(i.e. $r_{T} < 3 \times 10^{-5}$) defines the approximate triangular shape in the upper corner 
while the other shaded region coincides with the one of the left plot.

In Fig. \ref{FIGU2} we fixed $p \to 1$ and $N_{k} = 60$ and this is consistent with 
the determinations of $N_{k}$ discussed in appendix \ref{APPA} in the absence of any post-inflationary 
stage of expansion deviating from the dominance of radiation. We shall get back to this choice 
at the end of section \ref{sec4} and question its validity in the case of a long stage of post-inflationary expansion 
slower than radiation.
\begin{figure}[!ht]
\centering
\includegraphics[height=8cm]{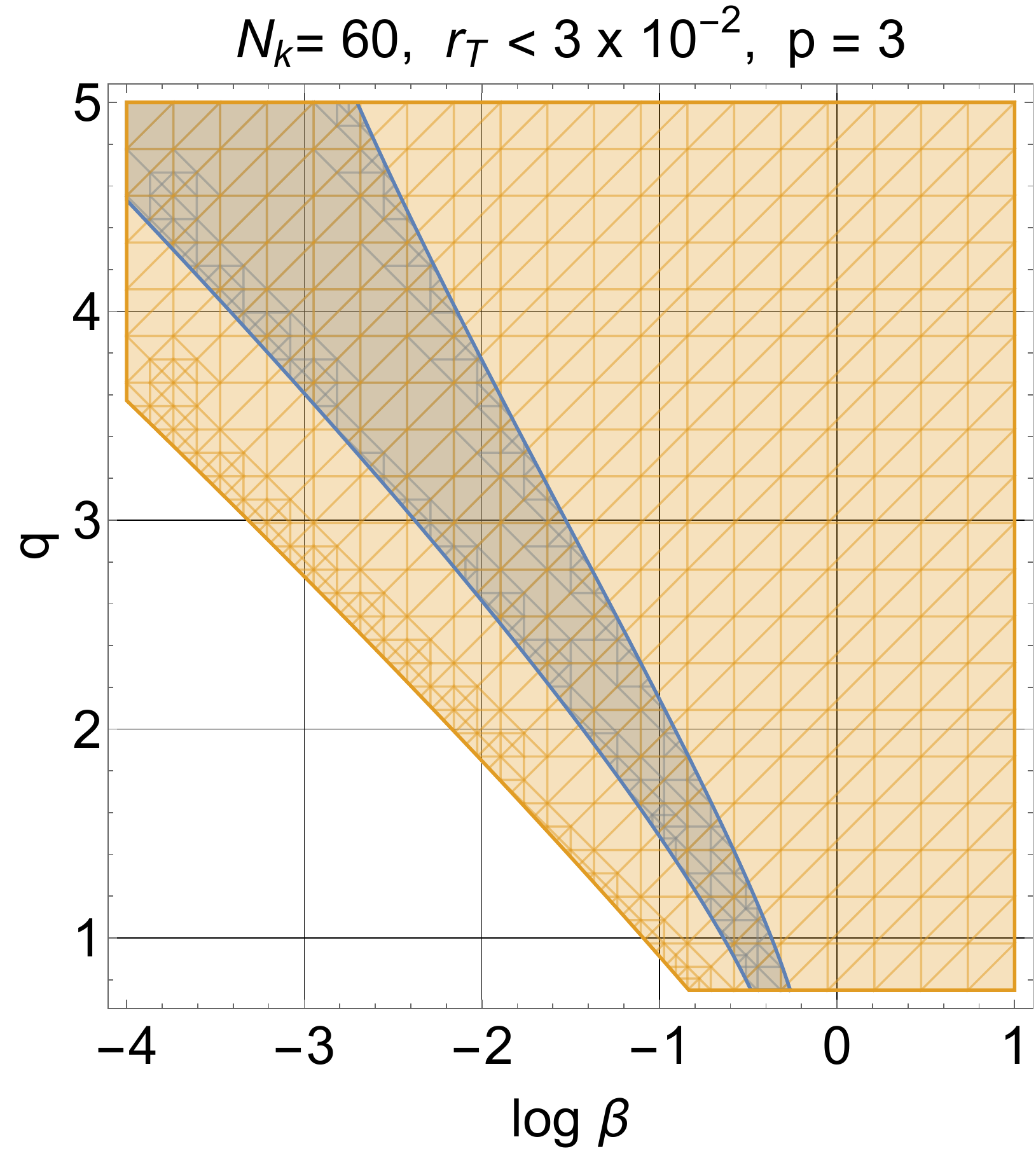}
\includegraphics[height=8cm]{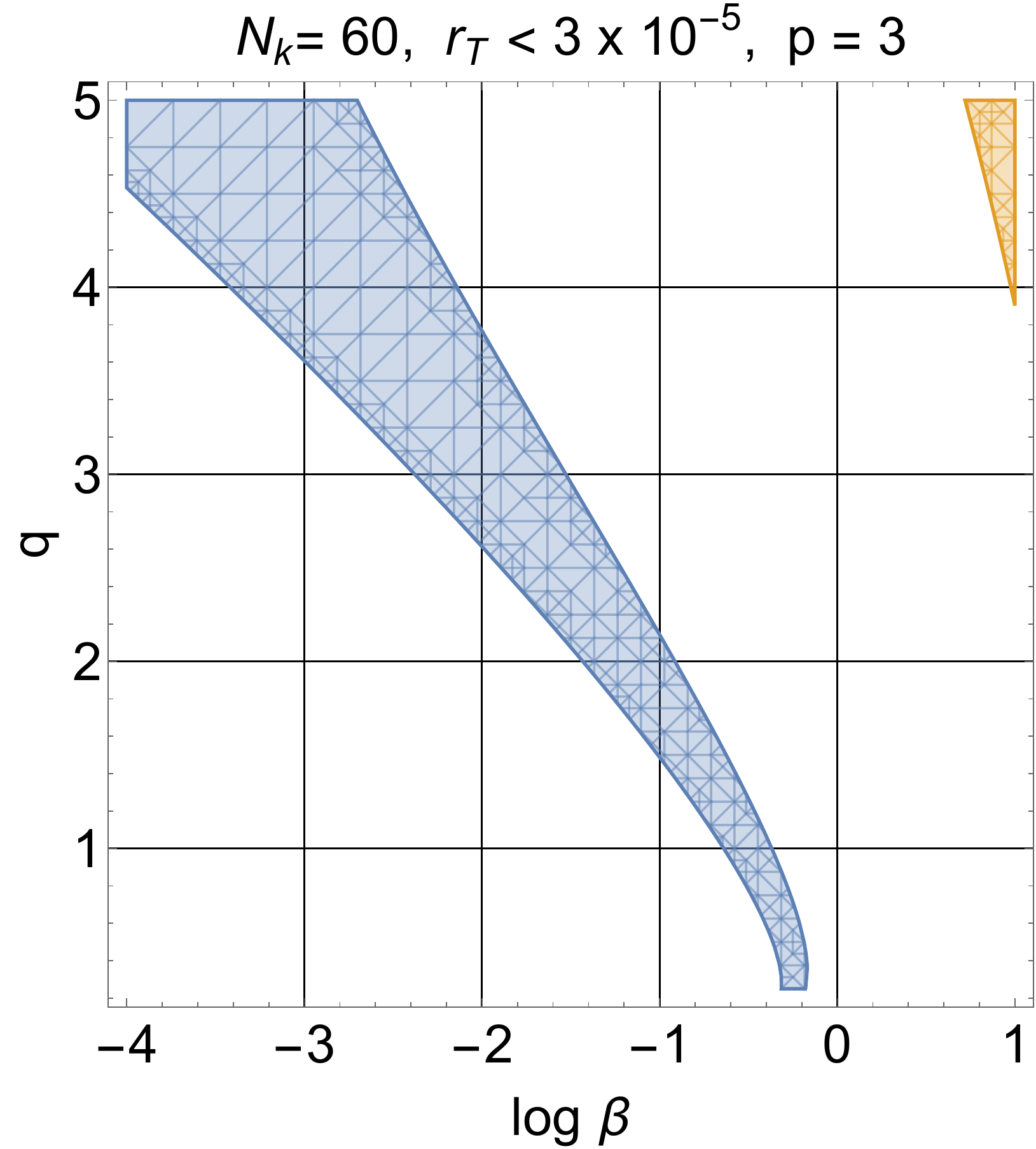}
\caption[a]{The same logic of Fig. \ref{FIGU2} can be illustrated 
for a larger value of the parameter $p$. In this case a reduction of $3$ orders of magnitude 
of $r_{T}$ (from $3\times10^{-2}$ to $3\times 10^{-5}$) implies that the overlap 
between the two regions completely disappears. This means that none of the parameters 
appearing in the right plot would be phenomenologically acceptable in case 
future limits would imply $r_{T} < 3 \times 10^{-5}$; this happens since in the central 
region of the plot the scalar spectral index is correctly reproduced but the tensor to scalar 
ratio is always larger than $3\times 10^{-5}$ and the two regions never overlap.}
\label{FIGU3}      
\end{figure} 
As the value of $p$ increases this progressive reduction of the overlap already pointed out in Fig. \ref{FIGU2} is further exacerbated and this conclusion follows from Fig. \ref{FIGU3} where $p\to 3$. If we compare the right plots of Figs. \ref{FIGU2} and \ref{FIGU3} we see that the distance between the two non-overlapping regions increases. Similarly if we compare the left plots of Figs. \ref{FIGU2} and \ref{FIGU3} we have that the overlap between the allowed regions is comparatively smaller for $p=3$ than for $p=1$. 
A final interesting observation is illustrated in Fig. \ref{FIGU4} where we imposed 
$r_{T}< 3\times 10^{-5}$ for two different values of $p$. By looking 
at Fig. \ref{FIGU3} the allowed regions looked completely absent; it happens, however, that  for comparatively larger values of $q$ 
 the constraints on $r_{T}$ and the requirements on $n_{s}$ are 
 concurrently satisfied\footnote{As already mentioned the results obtained so far, especially for the large $q$-values, 
are only partially accurate. Indeed as we saw in Eq. (\ref{TS25}) 
at the end of inflation the oscillatory regime for $q \gg 1$ implies 
that $0 < \delta < 1$. In this case the total number of $e$-folds 
may be larger than $60$ and this correction may have a relevant 
impact as we shall see in  section \ref{sec4}. }.

The same analysis leading to Figs. \ref{FIGU2}, \ref{FIGU3} and \ref{FIGU4} can be repeated in the case of similar potentials like the one of Eq. (\ref{PP4}). Using again the procedure outlined in appendix \ref{APPB} we can then deduce, for instance, that
\begin{equation}
n_{s}(k) \simeq 1 - \frac{3}{\gamma^2 N_{k}^2} - \frac{2}{N_{k}}, \qquad r_{T}(k) \simeq \frac{8}{\gamma^2\, N_{k}^2},\qquad n^{low}_{T}(k) \simeq -\frac{1}{\gamma^2 \, N_{k}}.
\label{PPS3}
\end{equation}
Equation (\ref{PPS3}) holds for $\Phi_{k}> 1$ and $N_{k}\gg 1$; in this limit, Eq. (\ref{PPS3}) reproduces 
in fact the result obtainable in the case of the potential 
\begin{equation}
v(\Phi) = \bigl(1 - e^{-\beta\Phi})^{2 q}, \qquad \beta > 0, \qquad q >0, 
\label{PPS4}
\end{equation}
where the analog of Eq. (\ref{PPS3}) is obtained by simply replacing $\gamma \to \beta$ (see, in this respect, the recent analysis of Ref. \cite{SS1}). The main difference between the example of Eq. (\ref{PP2}) and the examples of Eqs. (\ref{PP4}) and (\ref{PPS4}) is the $q$ dependence: while in the case of Eqs. (\ref{PP4}) and (\ref{PPS4}) the spectral indices and the tensor-to-scalar-ratio do not depend on $q$, Eqs. (\ref{PPS1})--(\ref{PPS2}) show the opposite. 
\begin{figure}[!ht]
\centering
\includegraphics[height=8cm]{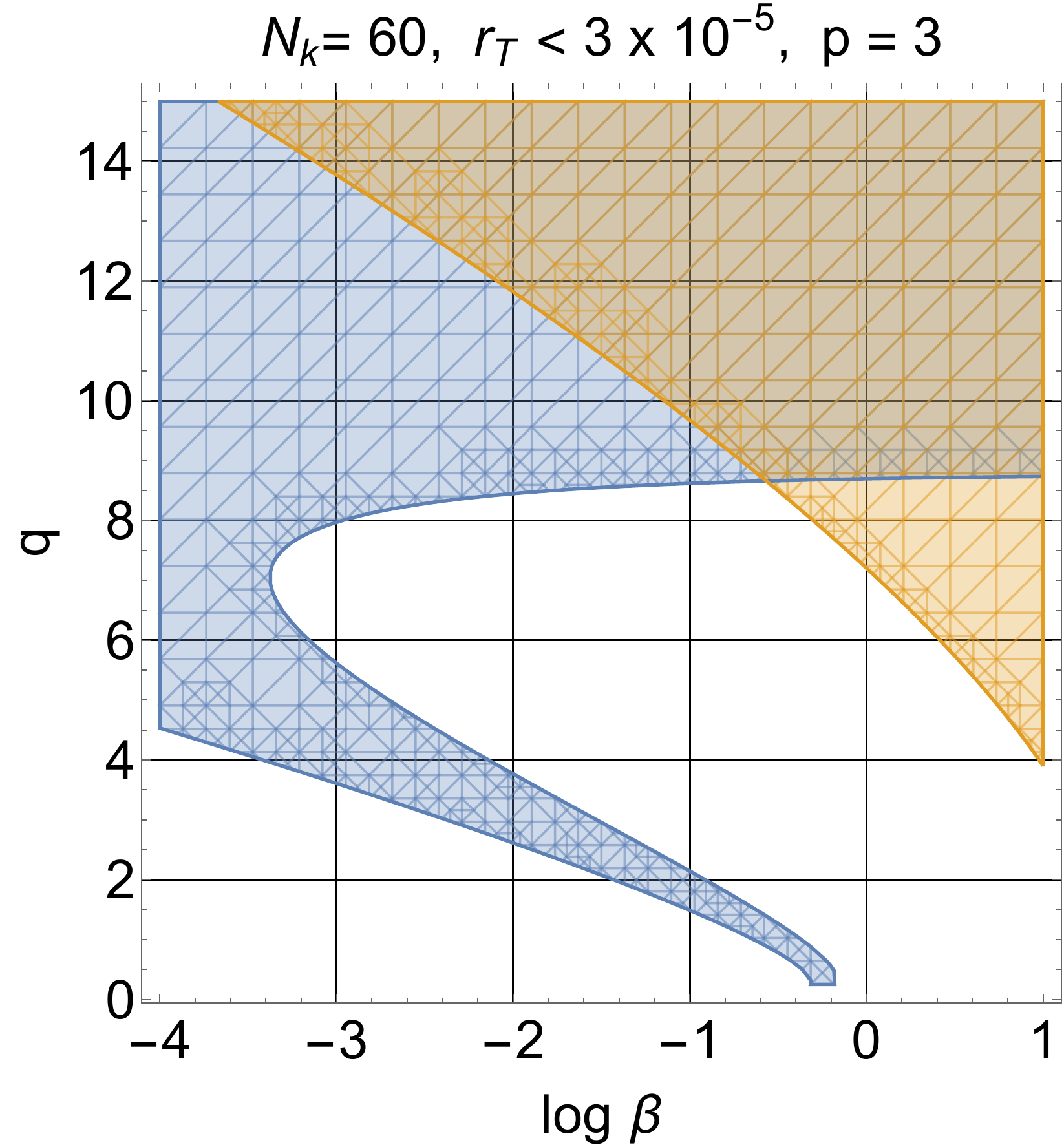}
\includegraphics[height=8cm]{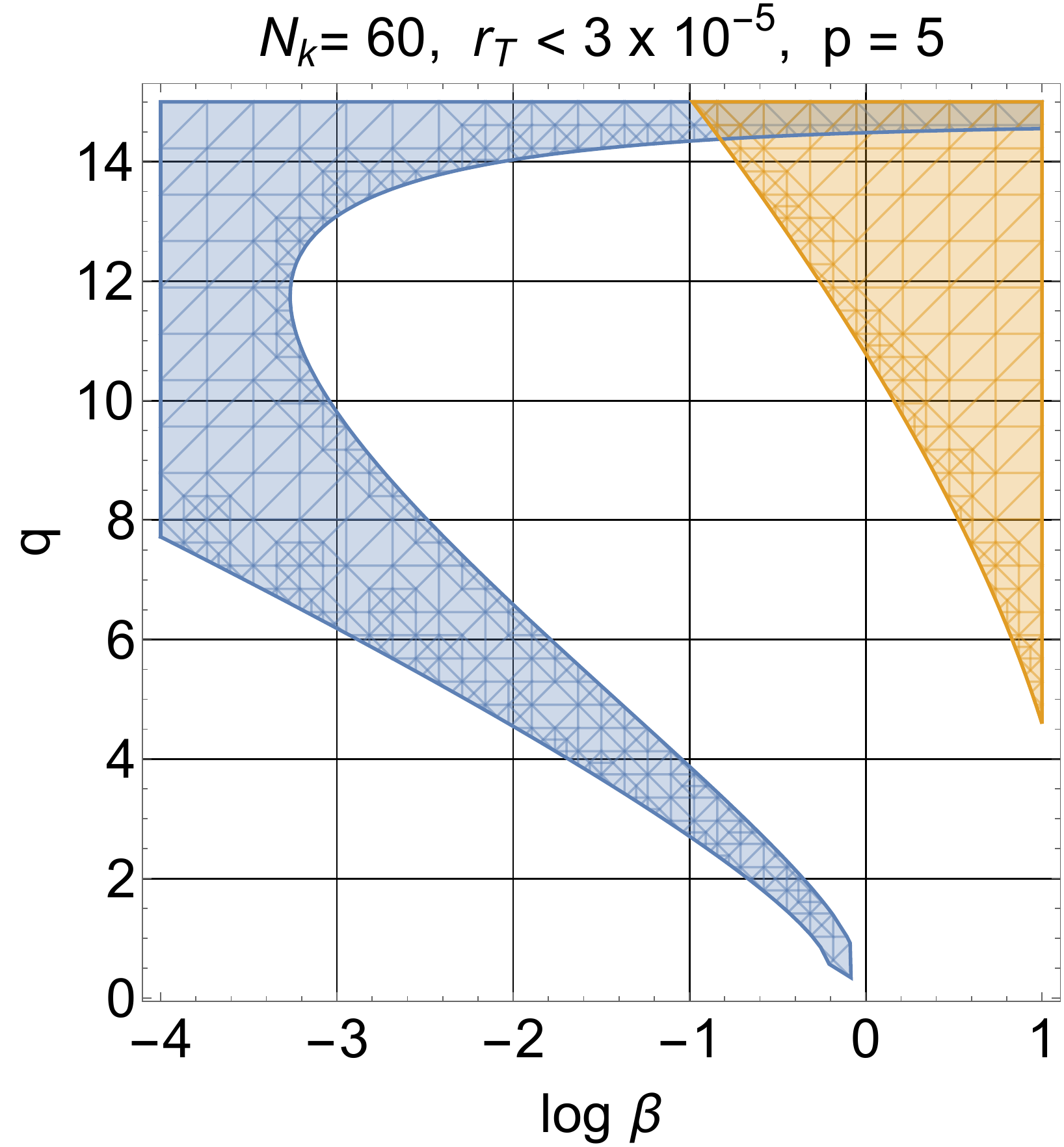}
\caption[a]{In both plots of this figure we imposed $r_{T} < 3 \times 10^{-5}$ for two different values of $p$. The left plot of this figure coincides with the left plot 
of Fig. \ref{FIGU3} but the range of $q$ is larger. From both plots we see 
that the region where the constraints are simultaneously satisfied 
moves towards large $q$-values. We can then recall from Eq. (\ref{TS25}) that for $q\gg1$ 
the inflaton oscillations effectively lead to a phase expanding at a rate that is slower than radiation (i.e. $\delta<1$).}
\label{FIGU4}      
\end{figure} 
The plateau-like potentials seem to imply a certain degree of fine-tuning 
that has been pointed out also at the level of the initial conditions 
of the inflaton. It is true that for generic initial 
conditions of $\varphi$ at the beginning of inflation 
both the kinetic energy and the spatial gradients should be of the 
same order and both larger than the potential $V(\varphi)= {\mathcal O}(M^4)$.
Since the kinetic energy of the inflaton redshifts faster than the spatial gradients,
it can happen that inflation is prevented by the dominance of the spatial inhomogeneities, unless some amount of fine-tuning is invoked \cite{STP1,STP2}.

If $n_{s}(k)$ and $r_{T}(k)$ approximately follow from Eq. (\ref{PPS3}) the tensor-to-scalar-ratio cannot be excessively reduced by keeping the agreement of the scalar spectral index with the observational data. 
It is true that in the case of a long post-inflationary stage 
$N_{k}$ may be larger than ${\mathcal O}(60)$ but besides this possibility (separately analyzed in section \ref{sec4} after the derivation of the 
lower bounds on $r_{T}$), a further reduction of $r_{T}(k)$ may occur either when the consistency relations are broken or for a different class of inflaton potentials. The consistency relations can be broken because of the protoinflationary dynamics \cite{fluid1,fluid2} and 
they can also be modified because of the running associated 
with the scalar spectral index \cite{HT3};  in both cases we could have that $r_{T} < 10^{-4}$. In the context of hilltop potentials $r_{T}$ can be reduced \cite{HT1} and it has even be argued that, in this framework, it is possible to construct models with parametrically small $r_{T}$ \cite{HT2}. These examples are constructed in terms of certain classes of fast-roll potentials where $\eta$ (see Eq. (\ref{PP5})) is actually constant \cite{FR1,FR2,FR3}. All these situations have a counterpart in the 
present framework but their impact will not be explicitly discussed here.

\renewcommand{\theequation}{3.\arabic{equation}}
\setcounter{equation}{0}
\section{The spikes of the spectral energy density}
\label{sec3}
Instead of scanning the range of $r_{T}(k)$ and of the post-inflationary expansion rates of Fig. \ref{FIGU1} it seems more plausible to identify preliminarily all the physical situations where the spectral energy density exceeds the predictions of the concordance scenario for frequencies larger than ${\mathcal O}(10^{-2})$ nHz. From the phenomenological constraints we can argue that the most stringent limits are obtained when $h_{0}^2 \Omega_{gw}(\nu,\tau_{0})$  develops a maximum either in the high-frequency region (i.e. approximately between few MHz and the GHz) or in the audio band. A further spike in the nHz range could only develop if 
 $h_{0}^2 \Omega_{gw}(\nu,\tau_{0})$ sharply increases between ${\mathcal O}(10^{-2})$ nHz and few nHz; this frequency range is however too narrow for an appreciable growth of the spectral energy density\footnote{For the sake of conciseness the nHz region is not be explicitly treated in this section but the corresponding constraints are anyway analyzed at the end of section \ref{sec4}.}.

\subsection{Spikes in the ultra-high-frequency region}
\begin{figure}[!ht]
\centering
\includegraphics[height=8cm]{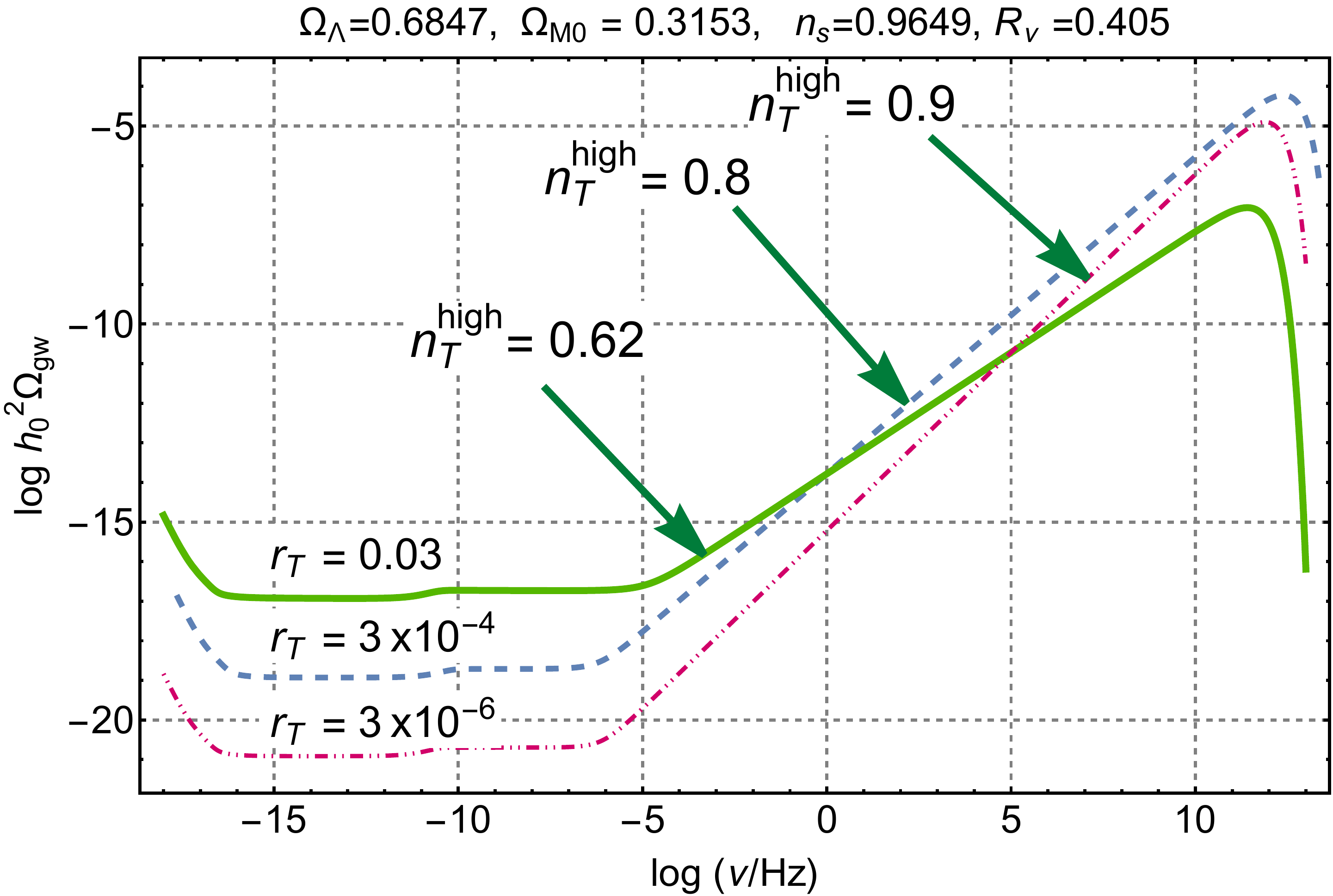}
\caption[a]{For three different choices of the tensor 
to scalar ratio $h_{0}^2 \Omega_{gw}(\nu,\tau_{0})$ is illustrated as a function of the  comoving frequency; common logarithms are employed on both axes. The high-frequency spectral indices have been chosen with the purpose of demonstrating that 
lower values of $r_{T}$ do not necessarily imply a smaller signal at high-frequencies. On the contrary, even though in all the examples we assumed that the consistency relations are enforced, the late-time parameters correspond to the fiducial values of the concordance paradigm. At low-frequencies all the most relevant sources of suppression have been taken into account and, in particular, the free-streaming of neutrinos.}
\label{FIGU5}      
\end{figure}

The presence of a broad spike in the high-frequency domain implies that the profile of the comoving horizon of Fig. \ref{FIGU1} consists of a 
unique post-inflationary stage extending between $H_{1}$ and $H_{r}$: only in this case we could have a single frequency domain 
where the spectral energy density always increases up to the GHz range. In this class of scenarios  
(illustrated in Fig. \ref{FIGU5}) $h_{0}^2 \Omega_{gw}(\nu,\tau_{0})$ comprises 
two separated frequency regions: a quasi-flat plateau (typically arising for $ \nu < \nu_{r}$) and a high-frequency hump approximately corresponding to $\nu = {\mathcal O}(\nu_{max})$. From the considerations developed in Eqs. (\ref{APA9ca})--(\ref{APA9e}) $\nu_{max}$ and 
$\nu_{r}$ can be estimated as follows:
\begin{equation}
\nu_{max} = \xi^{\frac{\delta -1}{2(\delta+1)}} \, \, \overline{\nu}_{max}, \qquad
\nu_{r} = \sqrt{\xi} \, \overline{\nu}_{max},\qquad \xi= H_{r}/H_{1}.
\label{MH0}
\end{equation}
Besides $\nu_{max}$ and $\nu_{r}$ there is also a more conventional third frequency region for $\nu< \nu_{eq}$. In this domain, the corresponding  wavelengths exited the comoving horizon  during the early stages of inflation and reentered after matter-radiation equality;
this bunch of wavelengths is slightly larger than  the shaded stripe illustrated in Fig. \ref{FIGU1}.
The equality frequency $\nu_{eq}$ is given by:
\begin{equation}
\nu_{eq} =  \frac{k_{\mathrm{eq}}}{2 \pi} = 1.597\times 10^{-17} \biggl(\frac{h_{0}^2\,\Omega_{M0}}{0.1411}\biggr) \biggl(\frac{h_{0}^2\,\Omega_{R0}}{4.15 \times 10^{-5}}\biggr)^{-1/2}\,\, \mathrm{Hz},
\label{MH6}
\end{equation}
where $k_{eq} = 0.0732\, h_{0}^2\,\Omega_{M0}\, \mathrm{Mpc}^{-1}$ and, as usual, $\Omega_{M0}$ is the present fraction in dusty matter.

The slopes of $h_{0}^2 \Omega_{gw}(\nu,\tau_{0})$ in the different frequency regions follow from Eq. (\ref{TS16}). For the wavelengths that exited the Hubble radius during inflation and reentered in the radiation stage we have that the structure of the turning point is singular (i.e. $\epsilon(\tau_{re}) \to 2$); this means that, thanks to Eq. (\ref{TS18}):
\begin{equation}
k \simeq a_{ex} \, H_{ex} \simeq - \frac{1}{( 1 - \epsilon_{k}) \tau_{ex}},
\qquad\qquad k\, \tau_{re} \ll 1.
\label{MH0a}
\end{equation}
Because of Eq. (\ref{MH0a}) the term ${\mathcal H}_{re}/k \gg 1$ dominates in Eq. (\ref{TS16}) and  $\Omega_{gw}(\nu,\tau)\propto (\nu/\nu_{r})^{n_{T}^{low}}$  for $\nu< \nu_{r}$.
If the wavelengths 
reenter instead before radiation was dominant (and when the background approximately 
expands as $a(\tau) \simeq (\tau/\tau_{1})^{\delta}$ with $\delta \neq 1$)
for $\nu_{r} < \nu < \nu_{max}$ we have that $\Omega_{gw}(\nu,\tau)\propto (\nu/\nu_{r})^{n_{T}^{high}}$ where $n_{T}^{high}$ is the  high-frequency spectral index. The explicit expressions of $n_{T}^{low}$ and $n_{T}^{high}$ are given by:
\begin{equation}
n_{T}^{low} = - 2 \epsilon_{k} , \qquad n_{T}^{high} = \frac{2 - 4 \epsilon_{k}}{1 - \epsilon_{k}} - 2 \delta,
\label{MH1}
\end{equation}
If the consistency relations are enforced the $n_{T}^{high}$ can also be expressed as:
\begin{equation}
n_{T}^{high}(r_{T}, \delta_{k}) = \frac{32 - 4 r_{T}}{16 - r_{T}} - 2 \delta = 2 ( 1- \delta) + {\mathcal O}(r_{T}), \qquad r_{T} = r_{T}(k).
\label{MH2}
\end{equation}
The high-frequency spectral index of Eq. (\ref{MH2}) depends both on $r_{T}$ and $\delta$; however, as long as $r_{T} \ll {\mathcal O}(10^{-2})$ the corrections induced by the finite value of $r_{T}$ can be safely neglected. For the sake of conciseness we shall be using the stenographic notation $r_{T}(k) = r_{T}$ where it is understood that $k= {\mathcal O}(k_{p})$ and the corresponding scales are the CMB wavelengths.
The spectral energy density induced by a modified post-inflationary history 
bears the mark of the evolution of the comoving horizon of Fig. \ref{FIGU1} \cite{ST1}. In particular, when the post-inflationary expansion rate is 
{\em slower} than radiation $\delta< 1$ the high-frequency spectral index is 
positive and $h_{0}^2 \Omega_{gw}(\nu, \tau_{0})$ is comparatively 
 larger than in the case $ \delta >1$ where the background 
expands faster than radiation\footnote{When $\delta > 1$ the high-frequency spectral index of Eq. (\ref{MH2}) gets negative: in this case there are, in practice, no further constraints besides the low-frequency limits that translate into the upper bound on $r_{T}$.
This means that at high-frequencies $h_{0}^2 \Omega_{gw}(\nu, \tau_{0})$
always decreases as a function of $\nu$ and the maximal frequency of the 
spectrum is much smaller than ${\mathcal O}(100)$ MHz.}. The case $\delta <1$ has been analyzed 
long ago \cite{AA7} (see also \cite{AA8,AA9}) and a particular realisation is provided by a stage dominated by the kinetic term of the inflaton-quintessence field \cite{ST2} (see also \cite{FORD1,spok}). The enhancement of the spectral energy density
occurring when the background expand slower than radiation 
arises in a number of apparently different contexts (see e.g. \cite{ST4,ST5,ST6}) 
that reproduce however the same basic dynamical situation 
of Ref. \cite{AA7,AA8,AA9} where the inflationary phase is followed 
by a stiff stage of expansion. If the post-inflationary evolution is dominated by the inflaton oscillations, the averaged evolution 
of the comoving horizon may mimic the timeline of a stiff epoch \cite{osc,T2}.
Recalling the results of Eqs. (\ref{TS24})--(\ref{TS25}) and (\ref{TS26}) we have 
that the averaged expansion of the background is slower than radiation 
provided $q > 2$ assuming the shape of the potential is the one of Eq. (\ref{TS24}).
It is relevant to mention, in this respect, that the techniques of Ref. \cite{T2} 
can be used to compute the high-frequency slope before and after the averaging 
suggested in Ref. \cite{osc}. It turns out that, in both situations, the high-frequency spectral slope is exactly the same. 

In Fig. \ref{FIGU5} the three curves illustrate the spectral energy density in critical units for different sets of parameters that have been chosen, in short, as follows: 
{\it (i)} for the full curve (labeled by $r_{T} =0.03$) the high-frequency spectral index is $n_{T}^{high} =0.62$ and the related $\xi$ is
$\xi = 10^{-28}$; {\it (ii)} the dashed curve corresponds to $r_{T} = 3 \times 10^{-4}$ while $n_{T}^{high}$ and $\xi$ are, respectively,  $0.8$ and 
$10^{-30}$; {\it(iii)} for the dot-dashed spectrum we have $r_{T} = 3 \times 10^{-6}$, $n_{T}^{high} =0.9$ and always $\xi= 10^{-30}$.
Except for the full line of Fig. \ref{FIGU5}, the remaining two spectra correspond to much lower values of $r_{T}$. However, as we can see, in spite of a reduction of $r_{T}$ the signal in the ultra-high-frequency region remains quite large and it must be constrained by the big-bang nucleosynthesis bound \cite{BBN1,BBN2,BBN3} as well as by the direct limits of wide-band interferometers (see in particular \cite{www3,www4} and \cite{DF} for a review). 

A reduction of $r_{T}$ does not necessarily entail a suppression of the signal in the in MHz and GHz bands. On the contrary, if we look at the dot-dashed curve in Fig. \ref{FIGU5} (corresponding to $r_{T} = 3 \times 10^{-6}$) the maximal value of $h_{0}^{2} \Omega_{gw}(\nu_{max}, \tau_{0})$ in this case is larger than for $r_{T} = 0.03$ (full curve in Fig. \ref{FIGU5}). As we shall see more specifically in section \ref{sec4} the dashed and the dot-dashed curves in Fig. \ref{FIGU5} are actually excluded 
by the nucleosynthesis bound and this shows, once more, that the limits on the high-frequency spike translate indirectly into a constraint on $r_{T}$.  It actually happens that larger values of $r_{T}$ are  less constrained than the lower ones and, from a purely qualitative viewpoint, the examples of Fig. \ref{FIGU5} may even suggest that, when the values of $r_{T}$ are too small, the amplitude of the high-frequency spike is even more restricted. 

As the values of the tensor-to-scalar-ratio get progressively reduced the accurate determination of the low-frequency plateau of Fig. \ref{FIGU5} becomes more essential and they are not only determined by $r_{T}$ but also by the neutrino free-streaming that suppresses $h_{0}^2\,\Omega_{gw}(\nu,\tau_{0})$  for   $\nu < \nu_{bbn}$ \cite{STRNU0,STRNU1,STRNU2,STRNU3} where $\nu_{bbn}$ is the frequency corresponding to the big-bang nucleosynthesis:
\begin{equation}
\nu_{bbn}= 2.3\times 10^{-2} \biggl(\frac{g_{\rho}}{10.75}\biggr)^{1/4} \biggl(\frac{T_{bbn}}{\,\,\mathrm{MeV}}\biggr) \,\,
\biggl(\frac{h_{0}^2 \Omega_{R0}}{4.15 \times 10^{-5}}\biggr)^{1/4}\,\,\mathrm{nHz}.
\label{MH3}
\end{equation}
The spectra of Fig. \ref{FIGU5} correspond to the fiducial parameters the last Planck data release and the simplest possibility has been considered namely the case of three massless neutrinos where $R_{\nu} = \rho_{\nu}/(\rho_{\gamma} + \rho_{\nu}) =0.405$, as indicated on top of each plots \cite{RR1,RR2,RR3}. In Eq. (\ref{MH3}) $g_{\rho}$ is the effective number of relativistic species associated with the energy density\footnote{ Other sources of suppression (taken into account in Fig. \ref{FIGU5} and in the remaining plots) include the late-time dominance of dark energy and the evolution of relativistic species (see e.g. \cite{DF} for a review). }.
  
To make sure that radiation sets in before big-bang nucleosynthesis the condition  $\xi \geq 10^{-38}$ should be imposed (see also the discussion prior to Eq. (\ref{APA2b})); but we can already see from Fig. \ref{FIGU5} that the combined effect of a reduction of $r_{T}$ and of an increase of $n_{T}^{high}$ for a sufficiently small value of $\xi$ (e.g. $\xi = {\mathcal O}(10^{-30})$) may be incompatible with the limits on the relativistic species at the nucleosynthesis time requiring:
\begin{equation}
h_{0}^2 \, \int_{\nu_{bbn}}^{\nu_{max}} \,\Omega_{gw}(\nu,\tau_{0}) d\ln{\nu} < 5.61\times 10^{-6} \biggl(\frac{h_{0}^2 \,\Omega_{\gamma0}}{2.47 \times 10^{-5}}\biggr) \, \Delta N_{\nu},
\label{MH4}
\end{equation}
where $\Omega_{\gamma0}$ is the (present) critical fraction of CMB photons. Equation (\ref{MH4}) sets an indirect constraint  on the extra-relativistic species possibly present at the time of nucleosynthesis. Since Eq. (\ref{MH4}) is also relevant in the context of neutrino physics for historic reasons the limit is  expressed in terms of $\Delta N_{\nu}$ (i.e. the contribution of supplementary neutrino species). The actual bounds on $\Delta N_{\nu}$ range from $\Delta N_{\nu} \leq 0.2$ to $\Delta N_{\nu} \leq 1$ so that the integrated spectral density in Eq. (\ref{MH4}) must range, at most, between  $10^{-6}$ and $10^{-5}$. For the present purposes it  is also useful to mention that the numerical results of Fig. \ref{FIGU5} can be parametrized as:
\begin{equation}
h_{0}^2 \, \Omega_{gw}(\nu, \tau_{0}) = {\mathcal N}_{\rho} \, r_{T} \biggl(\frac{\nu}{\nu_{p}}\biggr)^{n^{low}_{T}} \, \, {\mathcal T}^2_{low}(\nu/\nu_{eq}) \,  {\mathcal T}^2_{high}(\nu/\nu_{r}, \delta), 
\label{MH5}
\end{equation}
where ${\mathcal N}_{\rho} = 4.165 \times 10^{-15}$ for $h_{0}^2\,\Omega_{R0} = 4.15\times 10^{-5}$ and $\nu_{p} = k_{p}/(2\pi) = 3.092\,\,\mathrm{aHz}$.  In Eq. (\ref{MH5}) ${\mathcal T}^2_{low}(\nu/\nu_{eq})$ and ${\mathcal T}^2_{high}(\nu/\nu_{r}, \delta)$ denote, respectively, the low-frequency transfer function and its high-frequency counterpart. Both transfer functions are directly  computed in terms of the spectral energy density \cite{CC1} (see also \cite{T2}), not for the spectral amplitude\footnote{In general the transfer function for the spectral 
energy density does not coincide with the transfer function computed for the 
spectral amplitude \cite{T2};  it is obtained by integrating numerically the mode functions across the radiation-matter transition for each $k$-mode
and by computing $\Omega_{gw}(\nu,\tau)$ for different frequencies. The advantage of the transfer function for the energy density is that while $\Omega_{gw}(\nu,\tau)$ is a mildly oscillating function of $k \tau$, the spectral amplitude exhibits  larger oscillations that need to be averaged, as originally suggested in \cite{T3a,T3b}.}. The expression of ${\mathcal T}_{low}(\nu/\nu_{eq})$ is:
\begin{equation}
{\mathcal T}_{low}(\nu/\nu_{eq}) = \sqrt{1 + c_{2}\biggl(\frac{\nu_{eq}}{\nu}\biggr) + b_{2}\biggl(\frac{\nu_{eq}}{\nu}\biggr)^2},\qquad c_{eq}= 0.5238, \qquad b_{eq}=0.3537.
\label{MH7}
\end{equation}
Unlike ${\mathcal T}_{low}(\nu/\nu_{eq})$, the high-frequency transfer function 
${\mathcal T}_{high}(\nu/\nu_{r}, \delta)$ depends on the value of $\delta$ so that it does not have a general form.  However, as long as $\nu> \nu_{r}$, the high-energy transfer function
can be approximated as ${\mathcal T}_{high}^{2} \to (\nu/\nu_{r})^{n^{high}_{T}}$ so that 
the full spectral energy density at high frequency becomes, in this case,
\begin{equation}
h_{0}^2 \, \Omega_{gw}(\nu, \tau_{0})  = {\mathcal N}_{\rho} \, r_{T} \biggl(\frac{\nu}{\nu_{p}}\biggr)^{n^{low}_{T}} \, \, {\mathcal T}^2_{low}(\nu_{r}/\nu_{eq}) \, \biggl(\frac{\nu}{\nu_{r}}\biggr)^{n^{high}_{T}}, \qquad \qquad \nu_{r} \leq \nu \leq \nu_{max}. 
\label{MH8}
\end{equation}
Equation (\ref{MH8}) rests on the observation that ${\mathcal T}_{low}(\nu_{r}/\nu_{eq}) \to 1$ for $\nu \geq \nu_{r}$;
 in the same limit it is also true that $n^{low}_{T} = - r_{T}/8 \ll 1$ and, in this situation, the prefactor  is practically frequency-independent so that we can write:
\begin{equation}
h_{0}^2 \, \Omega_{gw}(\nu, \tau_{0})  = \overline{{\mathcal N}}_{\rho}(r_{T}, \nu)  \biggl(\frac{\nu}{\nu_{r}}\biggr)^{n^{high}_{T}}, \qquad\qquad \nu > \nu_{r},
\label{MH9}
\end{equation}
where $\overline{{\mathcal N}}_{\rho}(r_{T}, \nu)$ is now defined as
\begin{equation}
 \overline{{\mathcal N}}_{\rho}(r_{T}, \nu) = {\mathcal N}_{\rho} \, r_{T} \biggl(\frac{\nu}{\nu_{p}}\biggr)^{n^{low}_{T}} \, \, {\mathcal T}^2_{low}(\nu_{r}/\nu_{eq}), \qquad n^{low}_{T} = - r_{T}/8 \ll 1.
 \label{MH10}
 \end{equation}
As before we can remark that, as long as $r_{T}< {\mathcal O}(10^{-2})$, 
$\partial \ln{\overline{{\mathcal N}}_{\rho}}/\partial\ln{\nu} \ll 1$ with a residual 
(mild) frequency dependence coming from neutrino free-streaming.
For simplified analytic estimates this dependence can be however ignored, at least in the first approximation.  For instance, along this perspective we can estimate $\overline{{\mathcal N}}_{\rho}$ between ${\mathcal O}(10^{-16})$ and ${\mathcal O}(10^{-17})$. 

\subsection{The spectral energy density at the maximum}
The spectral energy density can be estimated for
$\nu ={\mathcal O}(\nu_{max})$  by observing that the maximal frequency of the corresponds to the production of a single graviton pair (see \cite{boundVV2} and discussion therein). If we introduce the mean number of produced graviton pairs per 
in each mode of the field (denoted hereunder by $\overline{n}(\nu, \tau_{0})$) the spectral energy density in critical units 
becomes:
\begin{equation}
\lim_{\nu \to \nu_{max}} \, \Omega_{gw}(\nu, \tau_{0}) \to \frac{128 \, \pi^2}{3} \, \frac{\nu_{max}^4}{H_{0}^2 \, M_{P}^2} \, \, \overline{n}(\nu_{max}, \tau_{0}).
\label{NM1}
\end{equation}
By definition for $\nu \to \nu_{max}$ we have that $ \overline{n}(\nu_{max}, \tau_{0}) = {\mathcal O}(1)$. Recalling Eq. (\ref{MH0}) and the expressions derived in the appendix \ref{APPA} (see e.g. Eqs. (\ref{APA8})--(\ref{APA8a}) and discussion thereafter),  Eq. (\ref{NM1})  becomes: 
\begin{equation}
 \Omega_{gw}(\nu_{max}, \tau_{0}) = \frac{8}{3 \pi} \, \Omega_{R0} \biggl(\frac{H_{1}}{M_{P}}\biggr)^{4/(\delta+1)} \, \biggr(\frac{H_{r}}{M_{P}}\biggr)^{2(\delta -1)/(\delta+1)}.
\label{RRR1}
\end{equation}
Equation (\ref{RRR1}) holds in the case of a single post-inflationary phase
but it can be generalized to include all the relevant physical cases 
corresponding to the timeline of Fig. \ref{FIGU1}. The result of Eq. (\ref{RRR1}) slightly overestimates the actual amplitude of the spectral energy density since it does not take into account the late-time sources of suppression related, for instance, with the neutrino free streaming. By now recalling the considerations of appendix \ref{APPB} (and, in particular, Eq. (\ref{APB6})) we can trade $H_{1}$ for $H_{k}$ and 
use that $(H_{k}/M_{P}) \simeq \sqrt{ \pi\, \epsilon_{k} \, {\mathcal A}_{{\mathcal R}}}$. If we now impose the consistency relations we see 
that the maximum of the spectral energy density also depends on $r_{T}$. 

 \subsection{Spikes in the audio band}
 As discussed in the previous subsections, 
 when a single post-inflationary stage precedes the radiation epoch $h_{0}^2\,\Omega_{gw}(\nu,\tau_{0})$ consists of three separate branches. However the spectral energy density may include 
multiple frequency domains \cite{ST1} and, in this situation, a maximum may develop in the audio band.
Besides the standard aHz region (i.e. $\nu_{p} < \nu < \nu_{eq}$) and part of the intermediate branch (for $ \nu_{eq} < \nu < \nu_{r}$), the slopes in the two supplementary ranges (i.e. $\nu_{r} < \nu < \nu_{2}$ and  $\nu_{2} < \nu < \nu_{max}$) depend on the values of the expansion rates in that region (i.e. $\delta_{1}$ and  $\delta_{2}$). The different typical frequencies 
of this case are discussed in appendix \ref{APPA} (see, in particular, Eqs. (\ref{APA9ca})--(\ref{APA9cb})) and here the focus will be on the aspects that are directly relevant for the derivation of a bound on $r_{T}$.
 \begin{figure}[!ht]
\centering
\includegraphics[height=8cm]{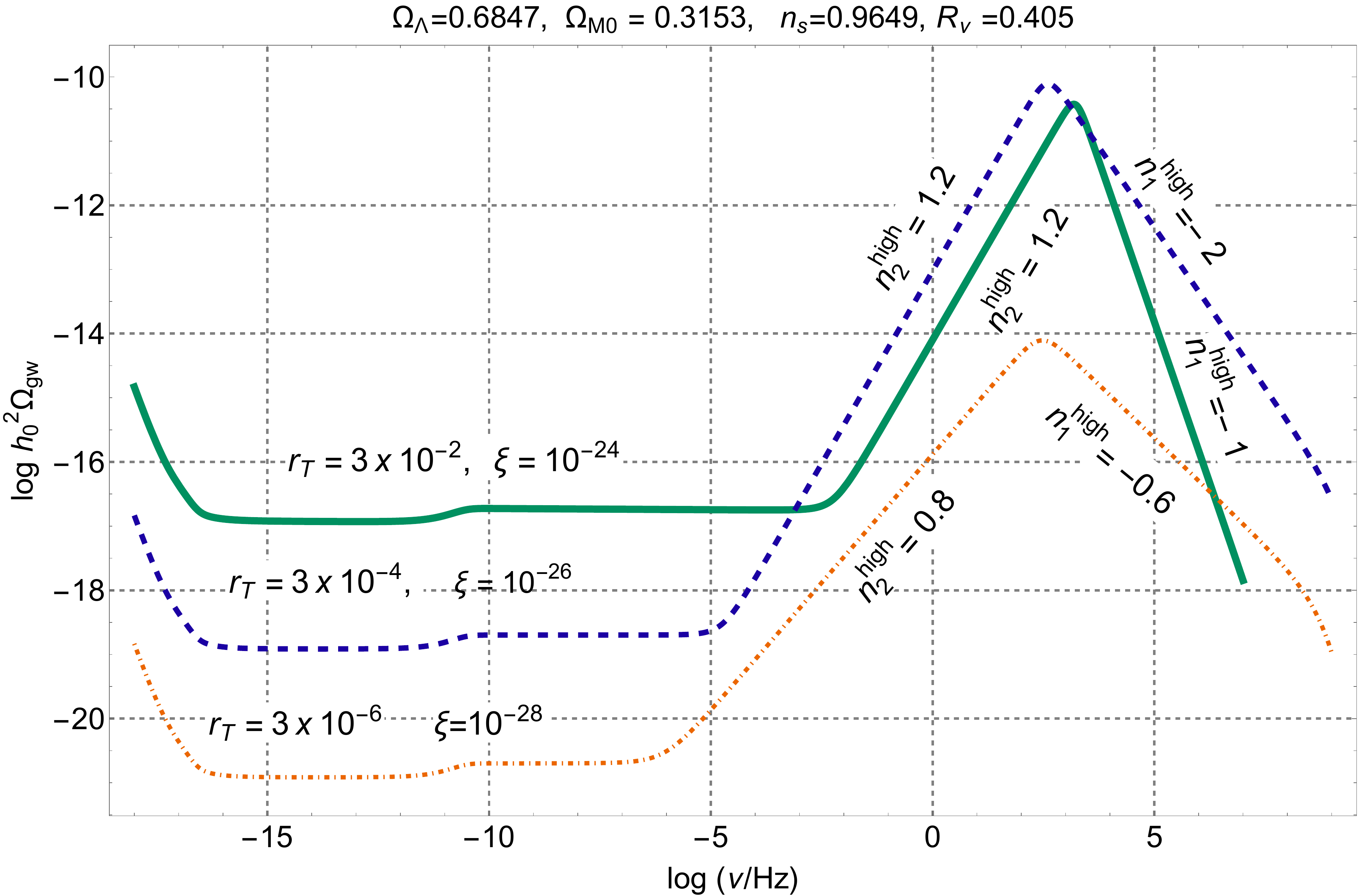}
\caption[a]{The spectral energy density is illustrated with the same notations 
of Fig. \ref{FIGU5}. The values of $r_{T}$
selected in this plot purposely match the ones of Fig. \ref{FIGU5} but, in the present situation, the spike of $h_{0}^2 \, \Omega_{gw}(\nu,\tau_{0})$ falls in the audio band. The various parameters have been selected by requiring that $\nu_{2}$ (i.e. the frequency of the spike) is such that $\nu_{2} = {\mathcal O}(\nu_{audio})$. This is one of the most constraining cases 
since the direct bounds of wide-band detectors fall into the audio band \cite{www1,www2,www3,www4} (see also the discussion of section \ref{sec4}). Note that the maximum corresponds to frequencies $\nu = {\mathcal O}(\nu_{2})$ and not to $\nu_{max}$. Typical frequencies $\nu = {\mathcal O}(\nu_{max})$ are barely visible 
in rightmost region of the plot (see, in particular, the final part of the dot-dashed curve).}
\label{FIGU6}      
\end{figure}

Three different examples have been illustrated 
in Fig. \ref{FIGU6} and the selected values of $r_{T}$ reproduce the ones appearing in Fig. \ref{FIGU5}. With a unified notation the spectral slopes (denoted in Fig. \ref{FIGU6} by $n^{high}_{1}$ and $n^{high}_{2}$) are:
\begin{equation}
n^{high}_{i} = \frac{32 - 4 r_{T}}{16 - r_{T}} - 2 \delta_{i}, \qquad \qquad r_{T} \ll 1, \qquad\qquad i = 1,\,\,2.
\label{AU1}
\end{equation}
The spectral energy density of Fig. \ref{FIGU6} 
follows from the general evolution of the comoving horizon 
of Fig. \ref{FIGU1} for $n=3$ (see also Eqs. (\ref{APA9ca})--(\ref{APA9cb})) and this means that the post-inflationary evolution 
(prior to radiation dominance) consists of two successive stages 
where the background first expands faster than radiation (i.e. $\delta_{1}>1$) and then slows down (i.e. $\delta_{2} < 1$). According 
to the general arguments of Ref. \cite{AA7} we have from Eq. (\ref{AU1}) 
that the spectral energy density decreases for $\nu> \nu_{2}$ (i.e. 
$n_{1}^{high} <0$) while it increases at lower frequencies 
(i.e. $n_{2}^{high} >0$ for $\nu< \nu_{2}$).  If $r_{T}\ll 0.03$ \cite{RR1,RR2,RR3} we have, in practice, that Eq. (\ref{AU1}) 
reduces to:
\begin{equation}
n^{high}_{i} = 2 ( 1 - \delta_{k}^{(i)}) + {\mathcal O}(r_{T}), \qquad\qquad i = 1,\,\,2.
\label{AU2}
\end{equation}
The result of Eq. (\ref{AU2}) follows directly from the expression of 
$\Omega_{gw}(\nu,\tau_{0})$ given in Eq. (\ref{TS16}) with the 
caveat that the comoving wavelengths reentering 
between $a_{1}$ and $a_{2}$ in Fig. \ref{FIGU1} (and controlling the frequencies $\nu> \nu_{2}$) obey $k \tau_{re} = {\mathcal O}(1)$ and not $k\tau_{re}\ll 1$ as in the 
case of the wavelengths doing their second crossing during the radiation stage of expansion.
In this case the spectral slopes are given by Eq. (\ref{AU2}) and, in particular,  $n_{1}^{high} = 2( 1 -\delta_{1})<0$ for the wavelengths reentering before $a_{2}$.  Conversely for the wavelengths reentering between $a_{2}$ and $a_{r}$ we have $n_{2}^{high} = 2( 1 -\delta_{2})>0$. If the 
timeline is reversed (and $\delta_{1}<1$ while $\delta_{2} >2$) 
instead of a spike $h_{0}^2 \Omega_{gw}(\nu,\tau_{0})$ exhibits a 
trough but this timeline would be comparatively less constrained than the 
one of Fig. \ref{FIGU6}.  

All in all the two high-frequency 
branches of the spectral energy density can be parametrized as:
\begin{eqnarray}
h_{0}^2\,\Omega(\nu,\tau_{0}) &=& \overline{{\mathcal N}}_{\rho}(r_{T},\nu) \biggl(\frac{\nu}{\nu_{r}}\biggr)^{n^{high}_{2}}, \qquad\qquad \nu_{r} < \nu < \nu_{2}, 
\label{AU3a}\\
h_{0}^2\,\Omega(\nu,\tau_{0}) &=& \overline{{\mathcal N}}_{\rho}(r_{T},\nu) \biggl(\frac{\nu_{2}}{\nu_{r}}\biggr)^{n_{2}^{high}}\biggl(\frac{\nu}{\nu_{2}}\biggr)^{-|n^{high}_{1}|}, \qquad\qquad \nu_{2} < \nu < \nu_{max},
\label{AU3b}
\end{eqnarray}
where we are implicitly assuming that $n^{high}_{1}<0$ and $n^{high}_{2}>0$ and $\overline{{\mathcal N}}_{\rho}(r_{T},\nu)$ has been already defined in Eq. (\ref{MH10}). The spectral energy density given of Eqs. (\ref{AU3a})--(\ref{AU3b}) exhibits a maximum for $\nu= {\mathcal O}(\nu_{2}$) but when $\delta_{1} \to 1$ the maximum is replaced by a plateau since $h_{0}^2\,\Omega_{gw}(\nu, \tau_{0})$ flattens out (i.e. $n^{high}_{1} \to 0$  for $\nu > \nu_{2}$) \cite{ST1}.  We then illustrated the situations that are phenomenologically more constraining; on this basis it is now possible to derive the lower bounds on $r_{T}$. 

\renewcommand{\theequation}{4.\arabic{equation}}
\setcounter{equation}{0}
\section{Lower bounds on $r_{T}$}
\label{sec4}
We established that even if $r_{T}$ controls the overall normalization of the spectral energy density, the tensor-to-scalar-ratio also enters the typical frequencies of the spectrum through $\overline{\nu}_{max}$ and $\xi$ (see Eqs. (\ref{APA8})--(\ref{APA8a}) and discussions thereafter); for this reason the position and the amplitudes of the spikes of $h_{0}^2 \Omega_{gw}(\nu,\tau_{0})$ are affected both by the post-inflationary evolution and by $r_{T}$. If the background expands faster than radiation the maximal frequency may well shift from the MHz region to the audio band; conversely if the expansion rate is slower than radiation $\nu_{max}$ may easily reach the GHz region. To stress this point we note that, if the consistency relations are enforced,  Eq. (\ref{RRR1}) can be written as\footnote{The numerical factor  ${\mathcal C}(\delta) =(1/6)(16/\pi)^{(\delta-1)/(\delta+1)} $  appearing in Eq. (\ref{RRR2}) does not alter the scaling properties of $\Omega_{gw}(\nu,\tau_{0})$ for different values of  $r_{T}$.}
\begin{equation}
h_{0}^2 \Omega_{gw}(\nu_{max}, \tau_{0}) = {\mathcal C}(\delta)  h_{0}^2 \Omega_{R0} (r_{T}\,\, {\mathcal A}_{{\mathcal R}})^{2/(\delta+1)} \, \biggr(\frac{H_{r}}{M_{P}}\biggr)^{2(\delta -1)/(\delta+1)} \propto r_{T}^{2/(\delta+1)}.
\label{RRR2}
\end{equation}
In the second relation of Eq. (\ref{RRR2}) all the factors have been neglected except for the overall dependence on $r_{T}$ and, by looking at this result, it could be argued that any suppression of $r_{T}$ also reduces the high-frequency spike of the spectral energy density (see also Fig. \ref{FIGU5} and discussions therein). However a reduction of $r_{T}$ does not necessarily entail a suppression of the maximum of the spectral energy density since $H_{r}$, $\delta$ and $r_{T}$ come from different 
physical considerations and are therefore independently assigned.

Let us first examine the case where, prior to radiation, the post-inflationary 
evolution consists of a single stage expanding at a rate that is slower than radiation (i.e. $\delta <1$). This case is illustrated in Fig. \ref{FIGU5} and $h_{0}^2 \,\Omega_{gw}(\nu,\tau_{0})$ sharply
increases below $\nu_{max}$ but Eq. (\ref{RRR2}) also suggests that a suppression of few orders of magnitudes in $r_{T}$ may be compensated by an increase of the overall normalization that contains $(H_{r}/M_{P})$. More specifically, when $H_{r}/M_{P} \ll 1$ and $\delta <1$ the term $(H_{r}/M_{P})^{2(\delta -1)/(\delta+1)}$ appearing in Eq. (\ref{RRR2}) overwhelms the reduction provided by $r_{T}^{2/(\delta+1)}$ unless $r_{T}$ is really too small. If $\overline{\nu}_{max} = {\mathcal O}(270)$ MeV for a pivotal value of the tensor-to-scalar-ratio, a reduction of $r_{T}$ by $4$ orders of magnitude only shifts $\overline{\nu}_{max} \to {\mathcal O}(27)$ MHz. This is therefore 
a minor effect if compared with the modifications of the post-inflationary evolution . Finally $r_{T}$ also impacts the high-frequency slopes of the spectral energy density but this effect is negligible: we may consider, in this respect Eqs. (\ref{MH2}) and (\ref{AU1}) where the corrections induced on the spectral slopes are insignificant as long as $r_{T} < {\mathcal O}(10^{-2})$.
 
Even though, taken singularly, the different effects are easily estimated, the overall result of concurrent modifications of $r_{T}$ and of the other parameters is not obvious and this is why we now present a numerical determination of $r_{T}^{min}$, i.e. the minimal value of $r_{T}$ compatible with the potential presence of a high-frequency spike. By definition for $r_{T}> r_{T}^{min}$ the spectral energy density is larger than in the concordance scenario for the same value of $r_{T}$. This means that in the audio band and in the high-frequency domain the signal could be potentially detected by future instruments. Conversely for $r_{T} \leq r_{T}^{min}$ the relic gravitons are invisible both in the aHz domain and also at higher frequencies. 

A first necessary requirement for the estimate of $r_{T}^{min}$ is that the frequency corresponding to radiation dominance must always exceed $\nu_{bbn}$ already introduced in Eq. (\ref{MH3}): 
\begin{equation}
\nu_{r} = \,\sqrt{\xi} \,\,\overline{\nu}_{max} > \,\nu_{bbn}, \qquad \xi = H_{r}/H_{1}.
\label{BB1}
\end{equation}
In the case of a single post-inflationary phase $\xi = H_{r}/H_{1}$ whereas in the presence of multiple post-inflationary phases $\xi$ is given by the product of the individual $\xi_{i}$ (see, in this respect, Eqs. (\ref{APA9d})--(\ref{APA9e}) and the discussion thereafter). For a maximum in the audio band (see Fig. \ref{FIGU6}) we have $\xi= \xi_{1} \, \xi_{2}$ where $\xi_{1} = H_{2}/H_{1}$ and $\xi_{2} = H_{r}/H_{2}$. 
Equation (\ref{BB1}) applies both for of a single maximum 
and in the case of multiple spikes but in the two situations 
the overall expression of $\xi$ is different. Using together Eqs. (\ref{APA8a})
and (\ref{BB1}) we can also deduce the following bound on $H_{r}/M_{P}$:
\begin{equation}
\sqrt{H_{r}/M_{P}} \geq \frac{k_{bbn}}{(2 \, \Omega_{R\,0})^{1/4}} \bigl(H_{0}\, M_{P}\bigr)^{-1/2} \, {\mathcal C}(g_{s}, g_{\rho}),
\label{BB4}
\end{equation}
where, by definition, $2\pi\, k_{bbn}= \nu_{bbn}$. If the explicit values of the various terms are inserted into Eq. (\ref{BB4}) we may obtain that $\sqrt{H_{r}/M_{P}} \geq 1.95 \times 10^{-22}/{\mathcal C}(g_{s}, g_{\rho})$ for $h_{0}^2 \Omega_{R\,0} = 4.15 \times 10^{-5}$. The interesting point to stress, in this respect, is that while $\overline{\nu}_{max}$ and $\xi$ individually depend upon $r_{T}$, the bound of Eq. (\ref{BB1}) does not depend on the tensor-to-scalar-ratio. Indeed we have that while $\xi$ scales as $r_{T}^{-1/4}$, $\overline{\nu}_{max}$ is proportional to $r_{T}^{1/4}$.

A further general requirement determining  $r_{T}^{min}$ follows from the current limits (summarized in Tab. \ref{TABLE1}) on the presence of relic graviton backgrounds in the audio band \cite{www1,www2,www3,www4}.
\begin{table}[!ht]
\centering
\caption{Recent limits on the relic gravitons obtained by wide-band interferometers.}
\vskip 0.4 cm
\begin{tabular}{||c|c|c||}
\hline
\rule{0pt}{4ex}  $\sigma$ & frequency range if $\nu_{ref}$ [Hz] & Bound \\
\hline
\hline
$0$ &  $20-81.9$ & $\overline{\Omega}_{0} < 6 \times 10^{-8}$ Ref. \cite{www3}\\
$2/3$ & $20-95.2$ & $\overline{\Omega}_{2/3} < 4.8 \times 10^{-8}$ Ref. \cite{www3}\\
$3$ & $20-301$ & $\overline{\Omega}_{3} < 7.9 \times 10^{-9}$ Ref. \cite{www3}\\
$0$ & $20-76.6$ & $ \overline{\Omega}_{0} <    5.8\times10^{-9}$  Ref. \cite{www4}\\
$2/3$ & $20-90.6$ & $  \overline{\Omega}_{2/3} < 3.4\times 10^{-9}$  Ref. \cite{www4}\\
$3$ & $20-291.6$ & $\overline{\Omega}_{3} < 3.9\times 10^{-10}$ Ref. \cite{www4}\\
\hline
\hline
\end{tabular}
\label{TABLE1}
\end{table}
The parametrization of the spectral energy density adopted by 
Refs. \cite{www3,www4} is in fact given by 
$\Omega_{gw}(\nu,\tau_{0}) = \overline{\Omega}_{\sigma} \, (\nu/\nu_{ref})^{\sigma}$ so that, for instance, $\overline{\Omega}_{0}$ denotes the amplitude 
of the spectral energy density at $\nu_{ref}$ for a scale-invariant slope.
The three specific cases constrained in Refs. \cite{www3,www4} are given in Tab. \ref{TABLE1} however it is possible to find an interpolating formula valid for a handful of spectral indices (see e.g. \cite{boundVV2}). By following here this 
approach we can broadly adopt the bounds of Tab. \ref{TABLE1} and require 
\begin{equation}
10^{-12} \leq h_{0}^2 \, \Omega_{gw}(\nu_{LVK}, \tau_{0}) < 10^{-10}, \qquad\qquad \nu_{LVK} = {\mathcal O}(100)\,\, \mathrm{Hz},
\label{BB3}
\end{equation}
where $\nu_{LVK}$ denotes the Ligo-Virgo-Kagra frequency which can be estimated in terms of $\nu_{ref}$. The most sensitive region for the detection 
of relic gravitons in the audio band is, grossly speaking, below $0.1$ kHz since, in this band, the overlap reduction function has its first zero \cite{DF}.
Equation (\ref{BB3}) requires, in practice, that the bounds coming from wide-band interferometers 
are satisfied while, in the same frequency range, $h_{0}^2 \, \Omega_{gw}(\nu,\tau_{0})$ is larger than $10^{-12}$. We cannot foresee when 
the corresponding sensitivity will be reached by wide-band detectors 
but the requirement of Eq. (\ref{BB3}) follows from some of the optimistic claims suggested by the observational 
collaborations\footnote{The alternative would be that relic gravitons backgrounds 
will not be accessible in the audio band; while this possibility cannot be excluded, in what follows we shall entertain a less pessimistic attitude which is motivated by the steady increase of the sensitivity to relic gravitons in the last $20$ years. We must actually recall that in $2004$ wide-band detectors gave limits implying $h_{0}^2 \Omega_{gw}(\nu, \tau_{0}) < {\mathcal O}(1)$ \cite{www1} while today the same limits improved by roughly $10$ orders of magnitude \cite{www3,www4}. } \cite{www4}. 
By following a similar attitude, we shall enforce  the big-bang 
nucleosynthesis bound for all the ranges of the spectrum (and, in particular, at high 
frequencies) but also assume that that $h_{0}^2 \, \Omega_{gw}(\nu, \tau_{0}) > {\mathcal O}(10^{-8})$ for typical frequencies $\nu = {\mathcal O}(\nu_{max})$. This means, on the one hand, that the bound of Eq. (\ref{MH4}) is enforced while, at the same time, it is not excluded that the signal from the relic gravitons could be interesting for a direct detection with microwave cavities \cite{boundVV2} in a 
range that encompasses the MHz and the GHz regions.  In this regime 
coupled microwave cavities with superconducting walls
 \cite{CAV1,CAV2,CAV3}, waveguides \cite{WG1,WG2} or even small interferometers \cite{SI1,SI2,SI3} could be used for direct detection even if the current sensitivities should not be overestimated without reason (see, in this respect, \cite{boundVV2}).
 
\subsection{Spike in the ultra-high-frequency region}
\begin{figure}[!ht]
\centering
\includegraphics[height=8cm]{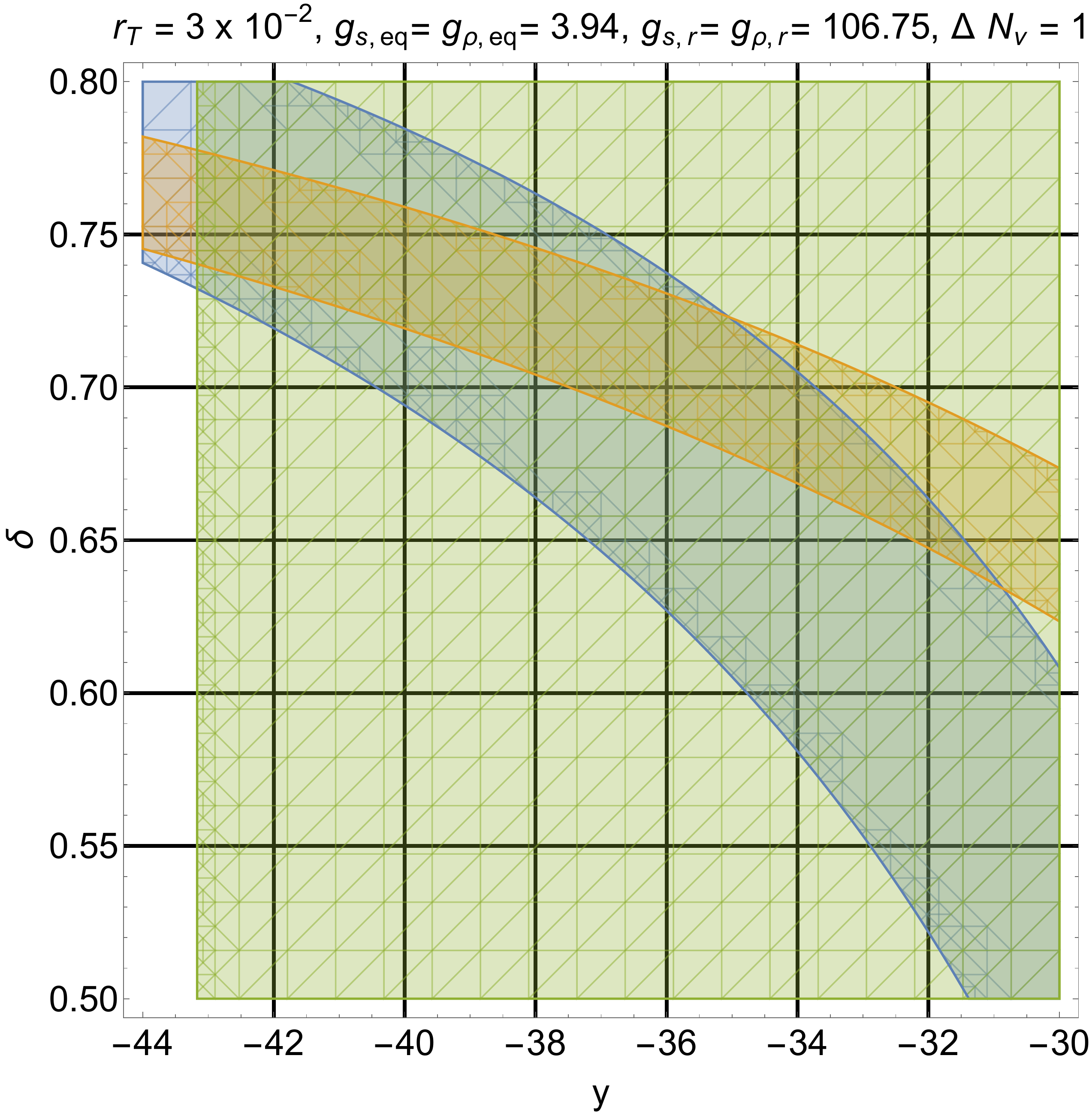}
\includegraphics[height=8cm]{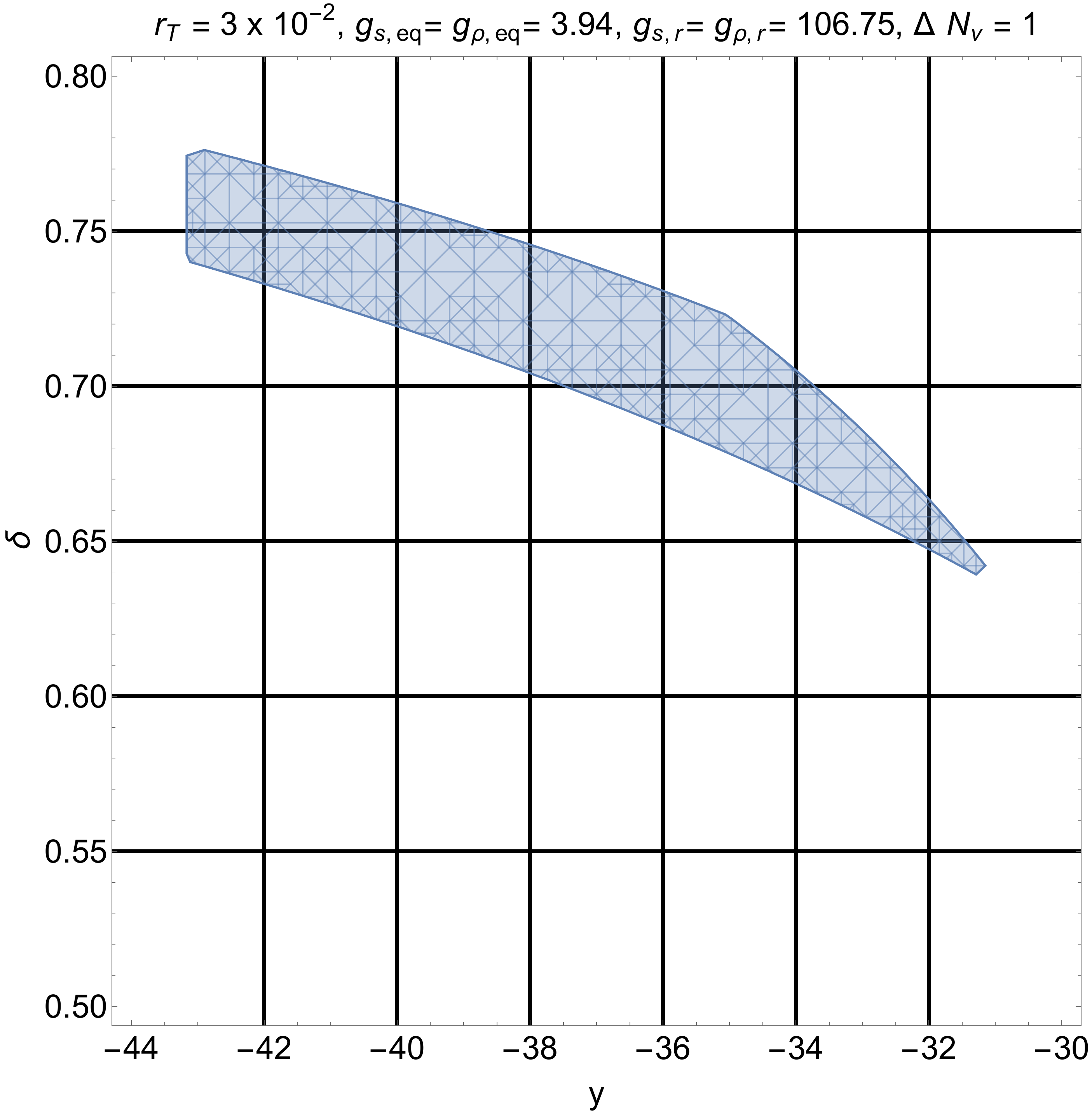}
\caption[a]{Since $\xi$ contains a dependence on $r_{T}$, 
it is preferable to plot (on the horizontal axis) the common logarithm 
of $H_{r}/M_{P}$ (i.e. $y = \log{(H_{r}/M_{P})}$) which does 
not contain $r_{T}$. We assumed here that $r_{T} \to 0.03$; in the plot at the {\em left} 
all the different limits (see Eqs. (\ref{BB1}), (\ref{BB3}) and (\ref{BB5})) have been superimposed  in the plane $(y,\,\delta)$. In the plot at the right  the intersection of the three regions (defined, respectively, by Eqs. (\ref{BB1}), (\ref{BB3}) and (\ref{BB5})) is directly illustrated.  The values of the late-time parameters considered in this and in the following figure lead approximately to ${\mathcal C}(g_{s}, g_{\rho}) = 0.75$; these terms are included even if they are not essential for the quantitative assessment of the various bounds.}
\label{FIGU7}     
\end{figure}
The interplay between the high-frequency determinations of the spectral energy density and the low-frequency limits on $r_{T}$ can be analytically deduced by considering the MHz and the GHz bands where 
the frequencies $\nu ={\mathcal O}(\nu_{max})$ are approximately located. 
In this case the most constraining class of bounds follows by requiring that $\nu \to \nu_{max}$ 
in Eq. (\ref{BB3}) and by examining a (single) post-inflationary 
phase parametrized by $\delta$ and $\xi$. Equation (\ref{MH4})
should then be satisfied together with the condition that $h_{0}^2 \Omega_{gw}(\nu_{max}, \tau_{0}) > 10^{-8}$;
putting everything together we therefore obtain the approximate condition
\begin{equation}
10^{-8} \leq h_{0}^2 \Omega_{gw}(\nu_{max}, \tau_{0}) < 5.61\times 10^{-6} \,\Delta N_{\nu}\, n_{T}^{high}.
\label{BB5}
\end{equation}
In Eq. (\ref{BB5}) the high-frequency spectral index appears if the parametrize the high-frequency behaviour with a single slope 
$n_{T}^{high}$; this is actually the most constraining instance even if the spikes may also arise at lower frequencies where, however, the corresponding amplitude is necessarily smaller since the frequency range for a potential growth of $h_{0}^2 \Omega_{gw}(\nu,\tau_{0})$ is narrower. With these specifications we actually have from Eq. (\ref{MH4}) that\footnote{The upper limit of Eq. (\ref{BB5}) follows from Eq. (\ref{BB5a}) in the case $n_{T}^{high}>0$ and $\nu_{r} > \nu_{bbn}$ (which is the same condition already discussed in Eq. (\ref{BB1})).}: 
\begin{equation}
h_{0}^2 \int_{\nu_{bbn}}^{\nu_{max}} \Omega_{gw}(\nu, \tau_{0}) \, d \ln{\nu} = \frac{h_{0}^2 \, \Omega_{gw}(\nu_{max},\tau_{0})}{n_{T}^{high}}\biggl[ 1 - \biggl(\frac{\nu_{bbn}}{\nu_{r}}\biggr)^{n_{T}^{high}}\biggr] <5.61\times 10^{-6} \,\Delta \,N_{\nu}.
\label{BB5a}
\end{equation}
 In Fig. \ref{FIGU7} we illustrate the allowed region of the parameter space in the plane defined by $\delta$ and by $y = \log{H_{r}/M_{P}}$.  Since  $\xi= H_{r}/H_{1}$ implicitly contains a dependence upon $r_{T}$,  the various constraints are illustrated in the $(y,\,\delta)$ plane where $y$ denotes the common logarithm of $(H_{r}/M_{P})$.
In the left plot of Fig. \ref{FIGU7} all the different constraints are superimposed while the overlapping region is directly illustrated in the plot at the right.  

The shaded area in the right 
plot of Fig. \ref{FIGU7} corresponds then to $r_{T} = 3\times 10^{-2}$ which is of the order 
of one of the most stringent constraints on $r_{T}$ available at the moment\footnote{In Fig. \ref{FIGU7} 
we also fixed $g_{s,\, eq} = g_{\rho, \, eq} = 3.94$ and $g_{s,\, r} = g_{\rho, \, r} = 106.75$
implying that ${\mathcal C}(g_{s}, g_{\rho}) =0.75$. Different values 
of the late-time parameters have a comparatively small effect on the shape of the allowed region.} \cite{RR1,RR2,RR3}.
The logic is now to decrease progressively the value of $r_{T}$ and, in doing so, the 
area of the overlap of Fig. \ref{FIGU7} is expected to shrink.
\begin{figure}[!ht]
\centering
\includegraphics[height=8cm]{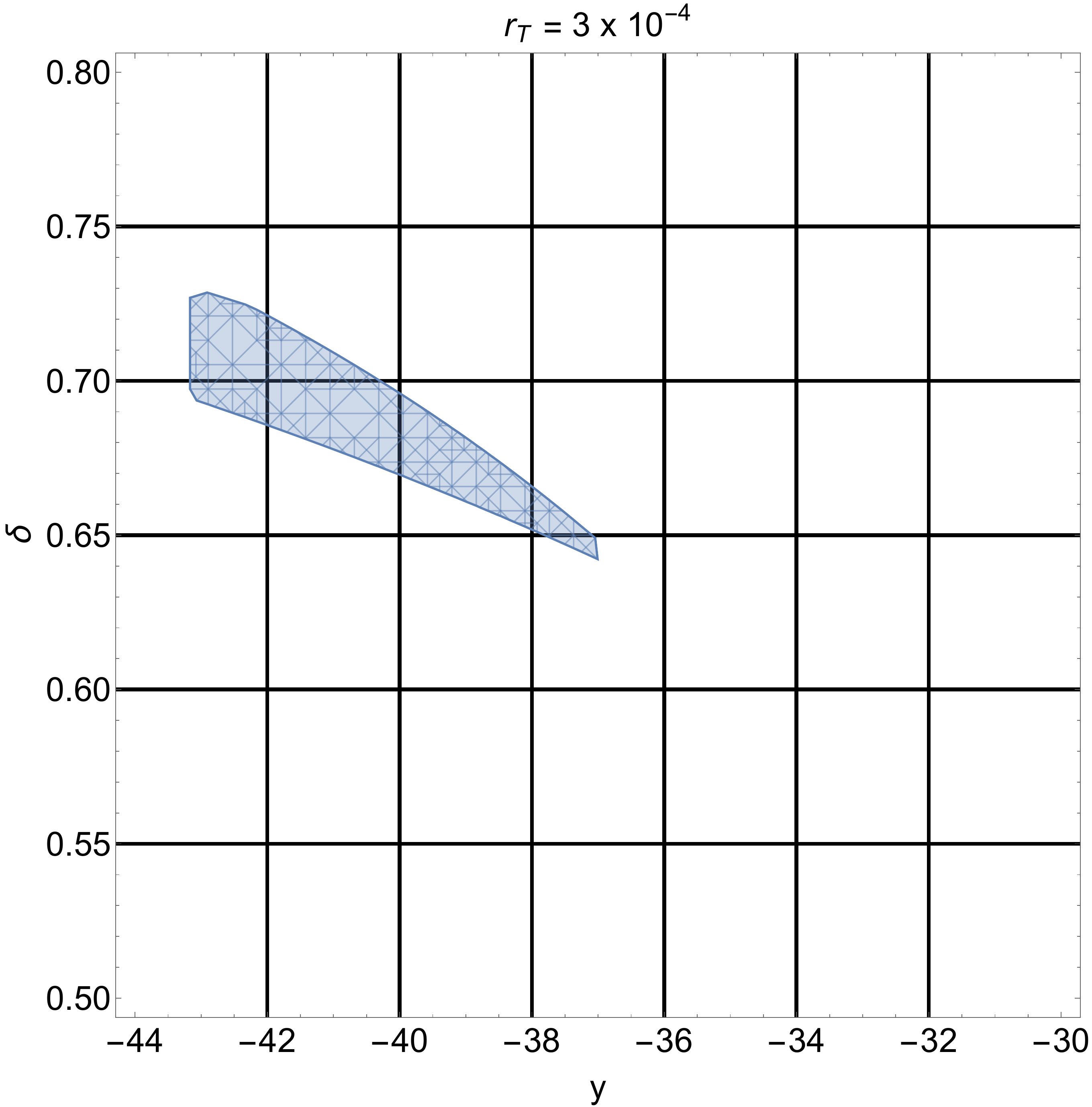}
\includegraphics[height=8cm]{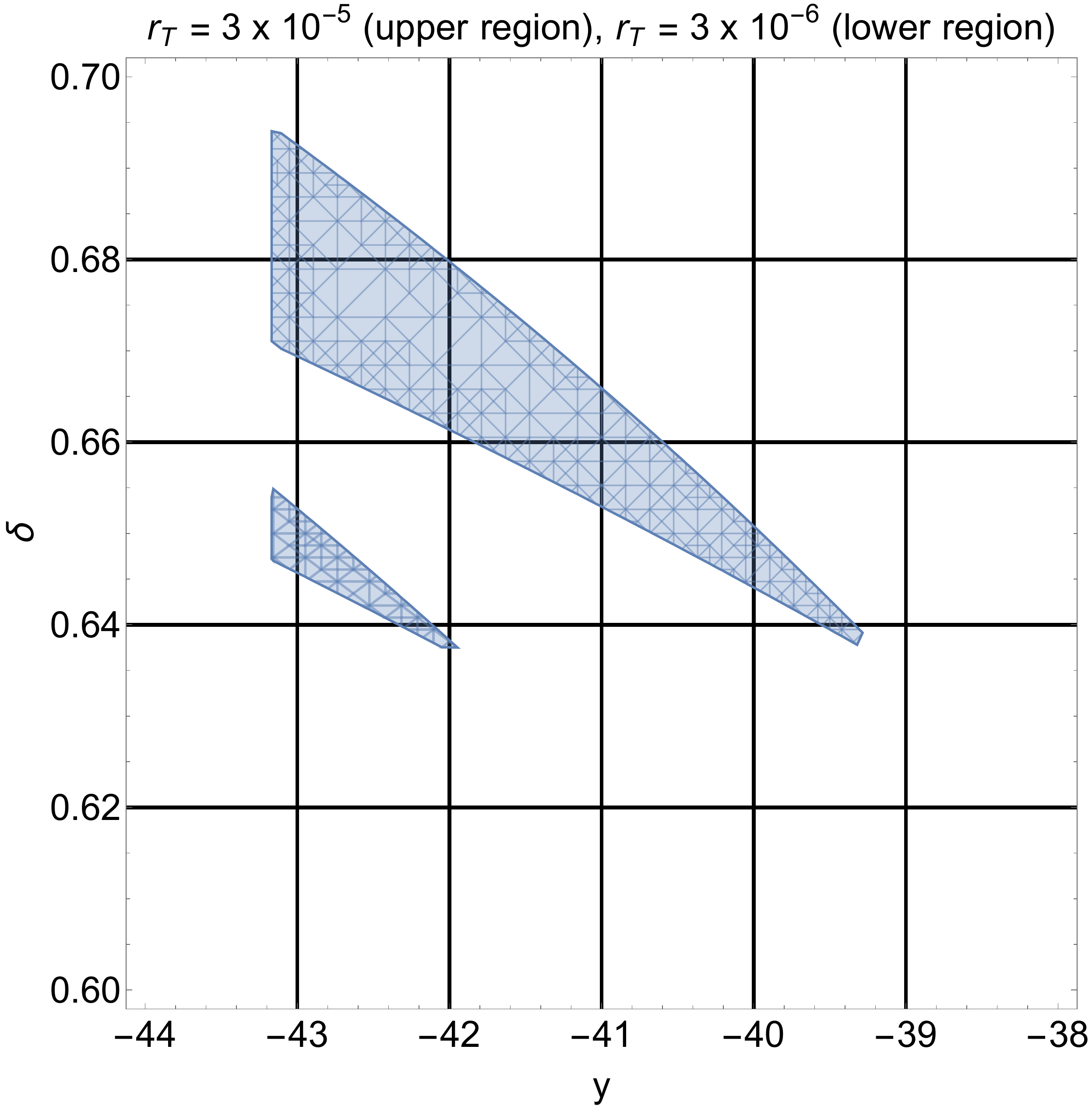}
\caption[a]{The value of $r_{T}$ is gradually reduced while all the remaining late-time  parameters are kept fixed. In the left plot the shaded region corresponds to $r_{T} = 3\times 10^{-4}$. In 
the right plot there are two different regions: in the upper spot $r_{T} = 3 \times 10^{-5}$ while in the lower spot $r_{T}= 3\times 10^{-6}$. If the value of $r_{T}$ is further reduced to $3 \times 10^{-7}$ 
the intersection of the different constraints illustrated in Fig. \ref{FIGU7} completely disappears and this occurrence implies that $r_{T}^{min} = {\mathcal O}(10^{-6})$. To characterize better the different regions the two plots have different ranges of $\delta$: this is while the upper region of the right plot seems superficially larger than the area of the left plot.  }
\label{FIGU8}     
\end{figure}
If the allowed region originally present in Fig. \ref{FIGU7} completely disappears, the value of $r_{T}$ will be by definition smaller than $r_{T}^{min}$. The gradual decrease is illustrated in Fig. \ref{FIGU8} where the values of $r_{T}$ are reduced first to $ 3 \times 10^{-4}$ (plot at the left) and then to $3 \times 10^{-5}$ (upper part of the right plot). The lower region of the right plot corresponds to $r_{T} = 3\times 10^{-6}$. If the value of $r_{T}$ is further reduced to $r_{T} = 3 \times 10^{-7}$ the lower spot completely disappears from the figure. We therefore conclude that the tensor-to-scalar-ratio should exceed $r_{T}^{min}= {\mathcal O}(10^{-6})$ if the spectral energy density of the relic gravitons exhibit a potential signal in high-frequency band. 

\newpage
\subsection{Spike in the audio band}
The analysis of the high-frequency region can be repeated at lower frequencies and, in particular, in the audio band. By looking at Eqs. (\ref{AU3a})--(\ref{AU3b})
the most constraining situation is the one where the frequency of the signal falls exactly in the audio band (i.e. $\nu_{2} = {\mathcal O}(\nu_{audio})$). 
If, prior to radiation dominance, the expansion rate 
is first faster than radiation and then slows down the spectral energy 
density exhibits a local maximum in $\nu_{2}$ and this is 
the situation already illustrated in Fig. \ref{FIGU6}. The high-frequency slopes of $h_{0}^2 \, \Omega_{gw}(\nu,\tau_{0})$
are positive for $\nu < \nu_{2}$ (i.e. $n_{2}^{high} >0$) and negative in the complementary frequency range (i.e. $n_{1}^{high}<0$ for $\nu>\nu_{2}$). If the timeline 
of the comoving horizon is inverted (and the background expands 
first slower than radiation and then faster) in $\nu_{2}$ there will be a 
minimum with a  spectral energy density even smaller than in the 
case of the concordance scenario where $h_{0}^2 \Omega_{gw}(\nu,\tau_{0}) = {\mathcal O}(10^{-17})$. 
\begin{figure}[!ht]
\centering
\includegraphics[height=8cm]{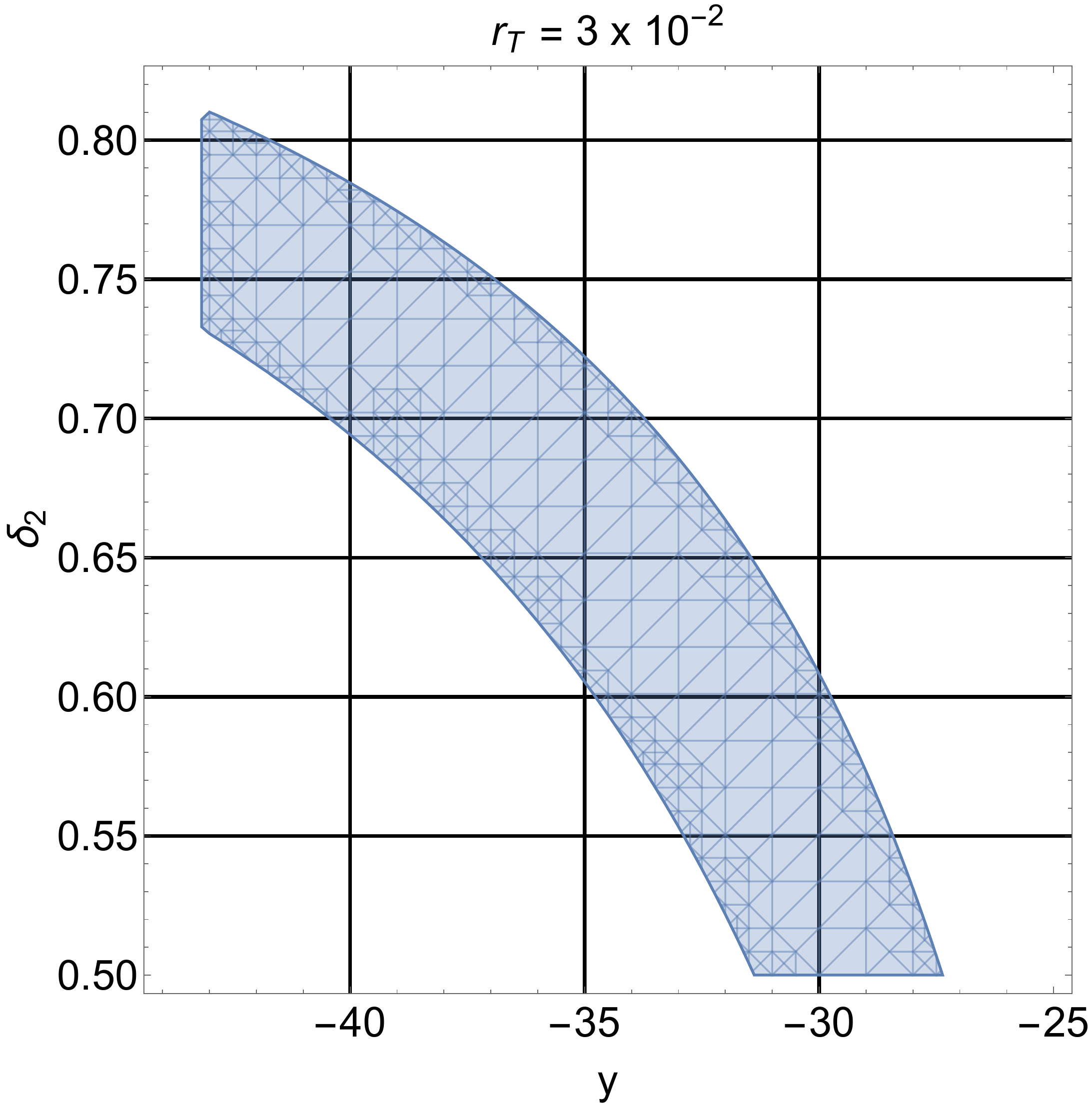}
\includegraphics[height=8cm]{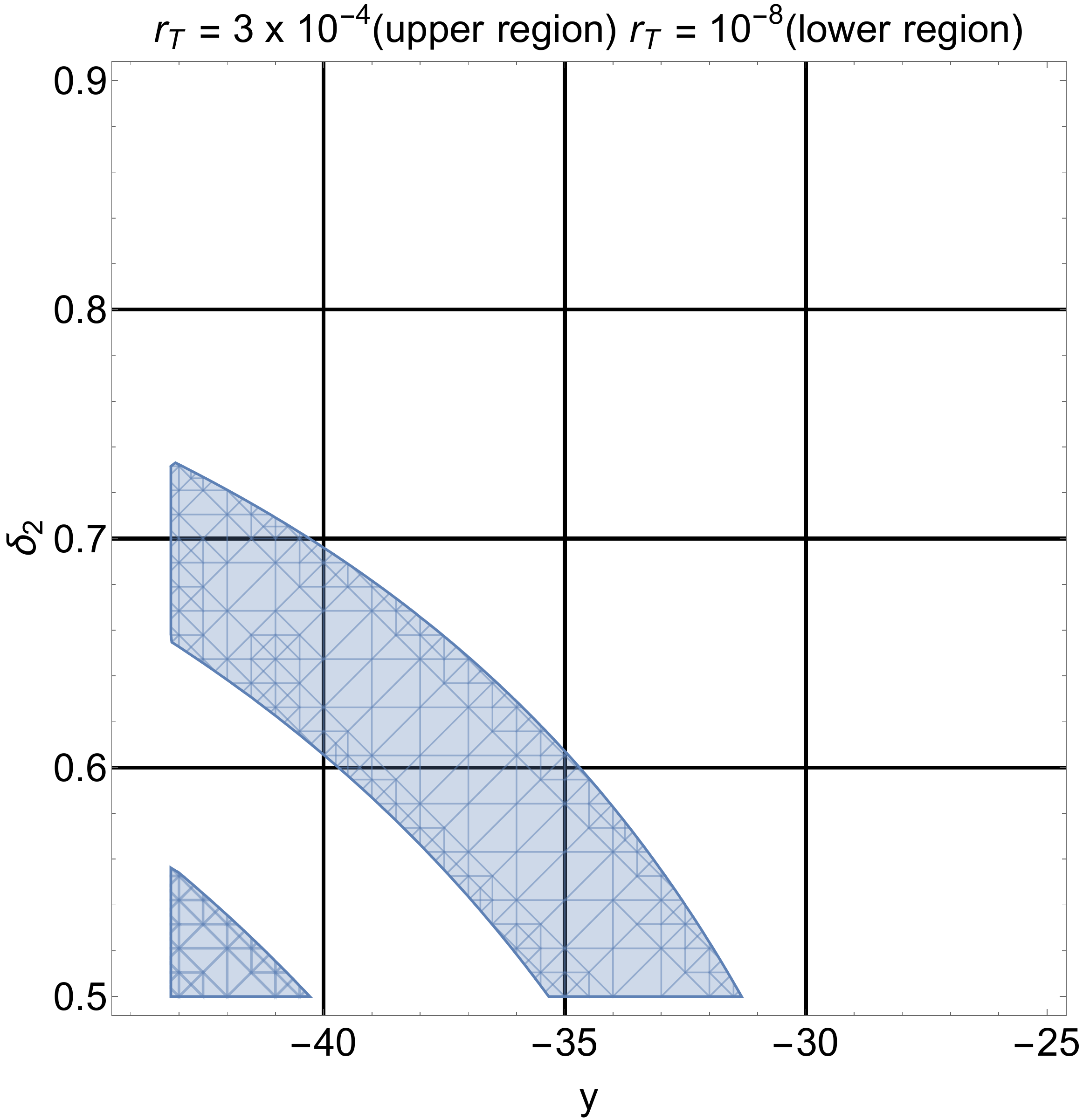}
\caption[a]{We analyze the case of spike in the audio band. In the two plots  the shaded regions 
correspond, respectively, to $r_{T} = 3\times 10^{-2}$ (plot at the left) and to 
$r_{T} = 3 \times 10^{-4}$ (plot at the right). The lower region in the plot at the right is obtained 
for $r_{T}= 10^{-8}$.}
\label{FIGU9}     
\end{figure}
The constraints at higher frequencies can be neglected 
since the signal is automatically small for  $\nu = {\mathcal O}(\nu_{max})$ and 
the bounds stemming from the expansion rate at the nucleosynthesis time are automatically satisfied. 
The  results of the analysis are summarized in Fig. \ref{FIGU9} 
where the late-time parameters have been selected as in Figs. \ref{FIGU7} and \ref{FIGU8}.
In the left plot of Fig. \ref{FIGU9} we have chosen $r_{T} = 3\times 10^{-2}$ while in the 
right plot the two shaded spots correspond, respectively, to $r_{T} = 3\times 10^{-4}$ (upper region) and 
to $r_{T} = 10^{-8}$ (lower region). If $r_{T}$ is further reduced below $10^{-8}$ 
the different requirements are not concurrently satisfied and we must therefore 
conclude that, in this case, $r_{T}^{min} = {\mathcal O}(10^{-7})$. 

There is the possibility, in principle, that the spike of the spectral energy density might occur 
in nHz region. In this instance the current measurements of the PTA could be relevant to set a 
limit on $r_{T}$. The PTA reported in fact a tentative signal in the nHz band: 
\begin{equation}
10^{-8.86}  \,\,b_{0}^2\,\, <  h_{0}^2\,\Omega_{gw}(\nu)< \,\,b_{0}^2 \,\,10^{-9.88} , \qquad 3\,\,\mathrm{nHz} \, < \nu< \,100 \,\, \mathrm{nHz},
\label{BBP1}
\end{equation}
where $b_{0}$ depends on the different experiments. The Parkes Pulsar Timing Array (PPTA) collaboration \cite{PPTA} suggests $b_{0}= 2.2$; 
the  International Pulsar Timing Array (IPTA) gives $b_{0}= 2.8$ \cite{IPTA} while the EPTA (European Pulsar Timing Array) \cite{EPTA} would measure $b_{0} = 2.95$.  These results are compatible with the NANOgrav 12.5 yrs data \cite{NANO} implying $b_{0} =1.92$. From the average of the four measurements presented so far we obtain $\overline{b}_{0} = 2.467$ which implies\footnote{If  $b_{0} \to 1$, Eq. (\ref{BBP1}) would suggest that the energy density in the nHz domain is comparatively smaller than the Ligo-Virgo-Kagra constraint for a flat spectrum \cite{www4}; however, according to current determinations, $b_{0} > 1$. }
\begin{equation} 
10^{-9.09} \biggl(\frac{\overline{b}_{0}}{2.467}\biggr)^2 \leq h_{0}^2 \, \Omega_{gw}(\nu) \leq 10^{-8.07} \biggl(\frac{\overline{b}_{0}}{2.467}\biggr)^2.
\label{BBP2}
\end{equation}
The constraints
of Eqs. (\ref{BBP1})--(\ref{BBP2}) are only marginally relevant in the present case. Indeed, recalling Eq. (\ref{BB1}) 
we have that, at most, $\nu_{r} = {\mathcal O}(10^{-2})$ nHz. To be relevant for Eq. (\ref{BBP1})  the spectral energy density 
should be sufficiently large for $\nu = {\mathcal O}(100)$ nHz. Taking into account that, at most, $h_{0}^2\Omega_{gw}(\nu_{r}, \tau_{0}) \leq {\mathcal O}(10^{-17})$ it follows that a spike for $\nu_{spike} = {\mathcal O}(100)$ nHz can only be $h_{0}^2\Omega_{gw}(\nu_{spike}, \tau_{0}) \leq {\mathcal O}(10^{-13})$ since, at most, the slope of the spectral energy density between $\nu_{r}$ and $\nu_{spike}$ 
can be linear\footnote{This means that a spike on the nHz band is always smaller than the figures suggested by Eqs. (\ref{BBP1})--(\ref{BBP2}). The nature of the PTA  observations is still under debate and it cannot be excluded that the potential signal is not related with the relic gravitons.}. Before concluding this discussion it is appropriate 
to stress that the values of the various $\delta_{i}$ have been always assumed to be 
larger than $1/2$. The reason for this choice is that this is what happens in all the relevant physical situations. In the case of a stiff fluid we have that $1/2 \leq \delta < 1$;
smaller values of $\delta$ correspond to sound speeds of the plasma larger than the 
speed of light. In the case of the scalar potentials discussed in Eq. (\ref{TS25}) 
$\delta \to 1/2$ for $q\gg 1$. 

\subsection{Constraints, specific potentials and compensations}
The results obtained so far assume the enforcement of the consistency relations and, in the opposite case, it is always possible to reduce the spectral energy density in the aHz region by keeping a comparatively larger signal in the higher frequency range. The bounds obtained here suggest 
\begin{eqnarray}
r_{T} \,> \,r_{T}^{min} &=& {\mathcal O}(10^{-7}), \qquad \nu={\mathcal O}(\nu_{audio}),
\nonumber\\
r_{T}\,> \,r_{T}^{min} &=& {\mathcal O}(10^{-6}), \qquad \nu={\mathcal O}(\nu_{max}).
\label{AAA1}
\end{eqnarray}
The reference to the frequency range simply reminds that, in the two cases, the most constraining situation correspond to the presence of a spike around either $\nu_{audio}$ or  $\nu_{max}$. The results of Eq. (\ref{AAA1}) are now illustrated  by few examples based 
on the different classes of potentials discussed in section \ref{sec2} and in appendix \ref{APPB}. In this respect the first point to bear in mind is that the number of $e$-folds depends both on $r_{T}$ and on the post-inflationary evolution but the relative impact of the two contributions is different. 
We recall that $N_{k}$ actually measures the number 
of $e$-folds since the crossing of the CMB scale and if the post-inflationary 
evolution is slower than radiation $N_{k}$ is larger than $60$; similarly if $r_{T}$ is smaller also $N_{k}$ diminishes.  Let us examine, for the sake of concreteness, the expression of $N_{k}$ given in Eq. (\ref{APA7}) which we rewrite by enforcing the consistency relations\footnote{If we compute $N_{k}$ from Eq. (\ref{APA7}) and use the consistency relations together with the explicit value of ${\mathcal C}(g_{s}, g_{\rho})$ we would obtain $59.58$ and this is why we wrote ${\mathcal O}(60)$ in Eq. (\ref{AAA2}); as we shall see these differences are immaterial 
for this discussion.} 
\begin{equation}
N_{k} = {\mathcal O}(60) + \frac{1}{4} \ln{\biggl(\frac{r_{T}}{0.03}\biggr)} + \frac{1}{2}\sum_{i=1}^{n-1} \, \biggl(\frac{\delta_{i} -1}{\delta_{i} + 1}\biggr) \, \ln{\xi_{i}}.
\label{AAA2}
\end{equation}
\begin{figure}[!ht]
\centering
\includegraphics[height=8cm]{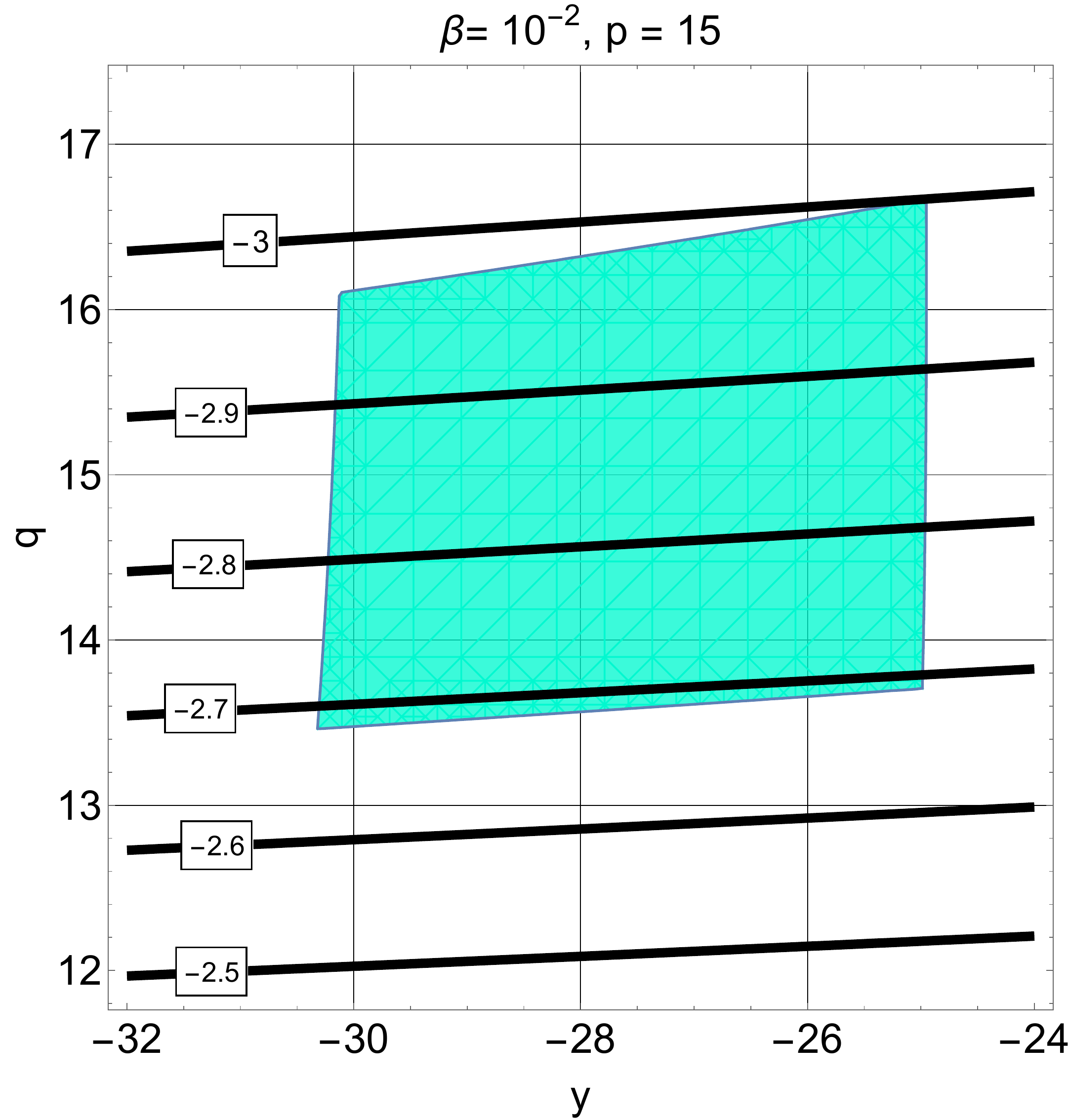}
\includegraphics[height=8cm]{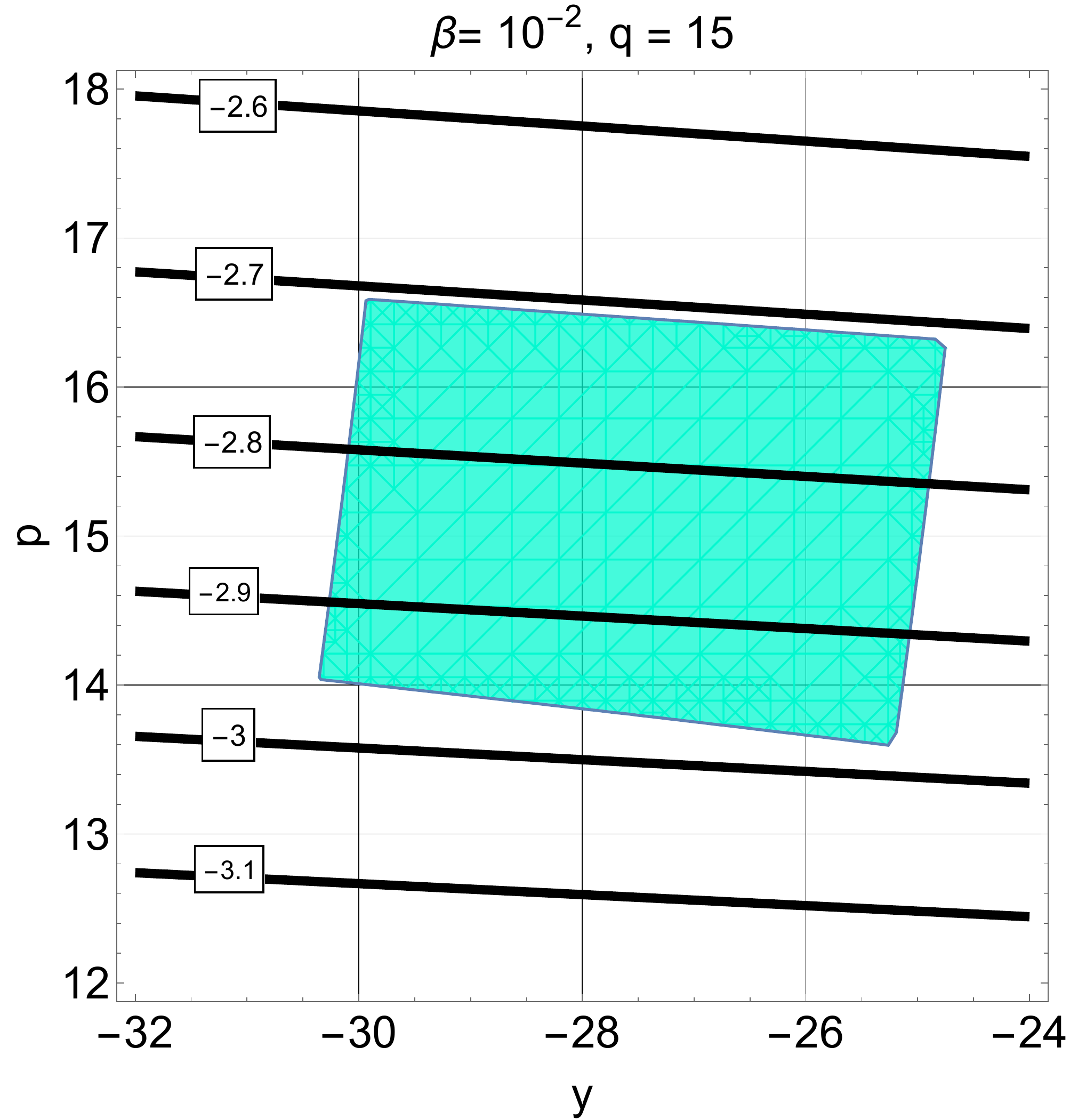}
\caption[a]{We evaluate the allowed values of $n_{s}(N_{k})$ and $r_{T}(N_{k})$. As in the previous plots $y$ denotes the common logarithm of $H_{r}/M_{P}$ (i.e. $y= \log{H_{r}/M_{P}}$). In the plot at the left the value of $p$ is fixed while in the plot at the right $q$ is fixed. In the shaded regions $n_{s}(N_{k})$ and $r_{T}(N_{k})$ fall within the 
constraints of Eq. (\ref{PP1}) and simultaneously  lead to a large signal for frequencies 
${\mathcal O}(\nu_{max})$. In both plots we can verify that $r_{T}(N_{k}) > r_{T}^{min}$ implying the validity 
of the general bound deduced in Eq. (\ref{AAA1}).}
\label{FIGU10}     
\end{figure}

For the reasons explained in Eqs. (\ref{APB4}) and (\ref{APB6}) we used that $H_{k}/H_{1} = {\mathcal O}(1)$ since this 
way of writing $N_{k}$ is more suitable. As already stressed in different frameworks \cite{AA7,AA8,liddle} the existence of post-inflationary 
phases with sound speed stiffer than radiation entails an {\em increase} both of $N_{k}$ and of $N_{max}$ (see Eqs. (\ref{APA3a})--(\ref{APA5}) and discussion therein). Within the notations 
of Eq. (\ref{AAA2}) we have indeed that when some of the $\delta_{i} <1$ (and the background 
expands slower than radiation) $N_{k} \gg 60$ depending on the various $\xi_{i}$ which are always, by definition, smaller (or even much smaller) than $1$.  In the past various upper bounds on the potential increase of $N_{k}$ have been set; within these attempts we can conceivably assume that $N_{k}$ and $N_{max}$ may increase by a factor\footnote{This estimate follows from Eq. (\ref{AAA2}) by considering the smallest 
value of $\delta$ compatible with standard sources (i.e. $\delta \to 1/2$) and by taking the minimal value of $\xi$ compatible with the $\Lambda$CDM 
paradigm; we get in this case $\Delta N_{k}= (1/6) \ln{\xi_{min}} = {\mathcal O}(15)$ for $\xi_{min} = {\mathcal O}(10^{-38})$. } ${\mathcal O}(15)$.  When $N_{k}$ is larger 
than in the standard case the tensor to scalar 
ratio and the scalar spectral index may be comparatively 
more suppressed and their values must then be 
confronted again with Eqs. (\ref{PP1})--(\ref{PP2}). 
\begin{figure}[!ht]
\centering
\includegraphics[height=8cm]{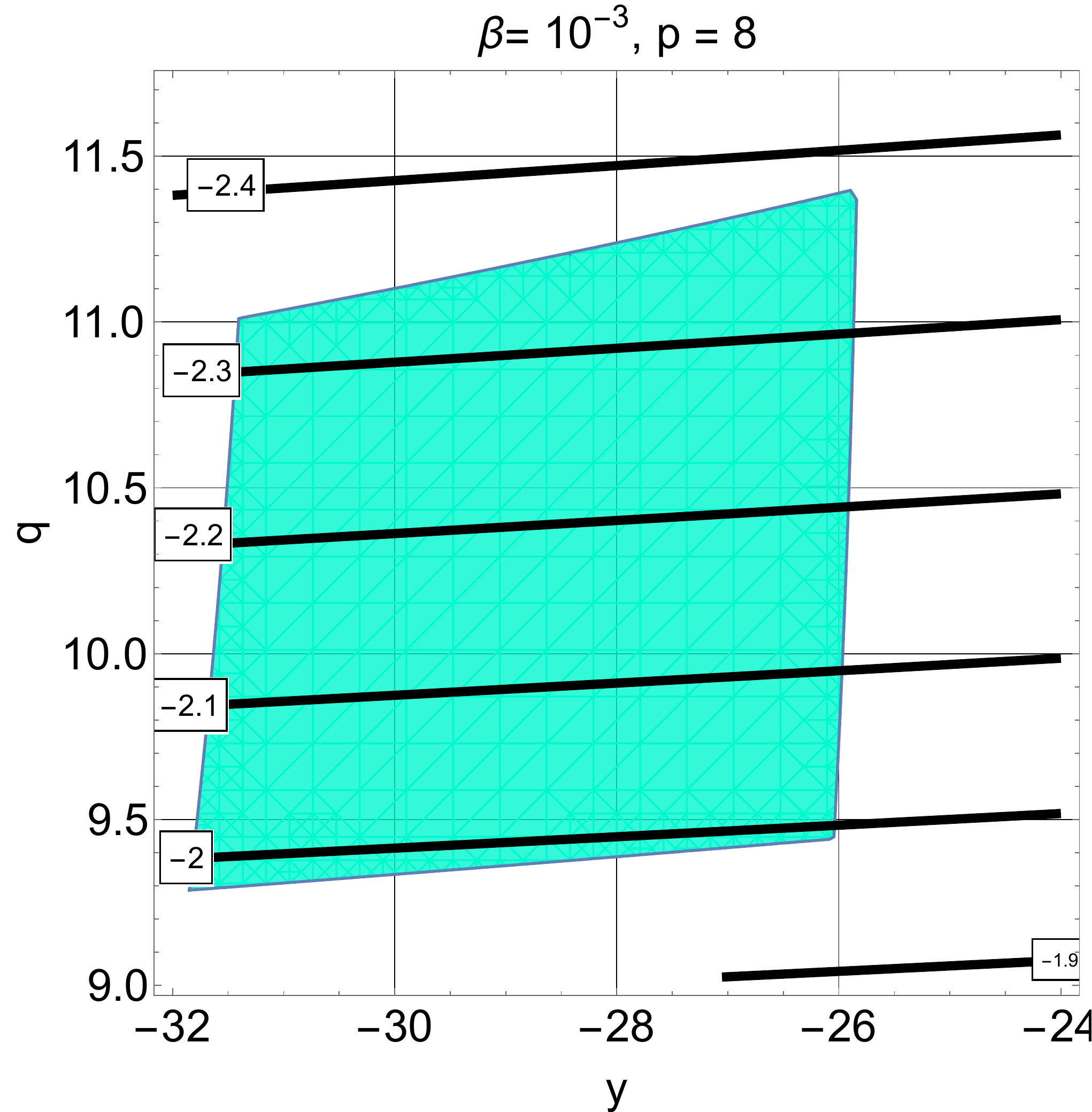}
\includegraphics[height=8cm]{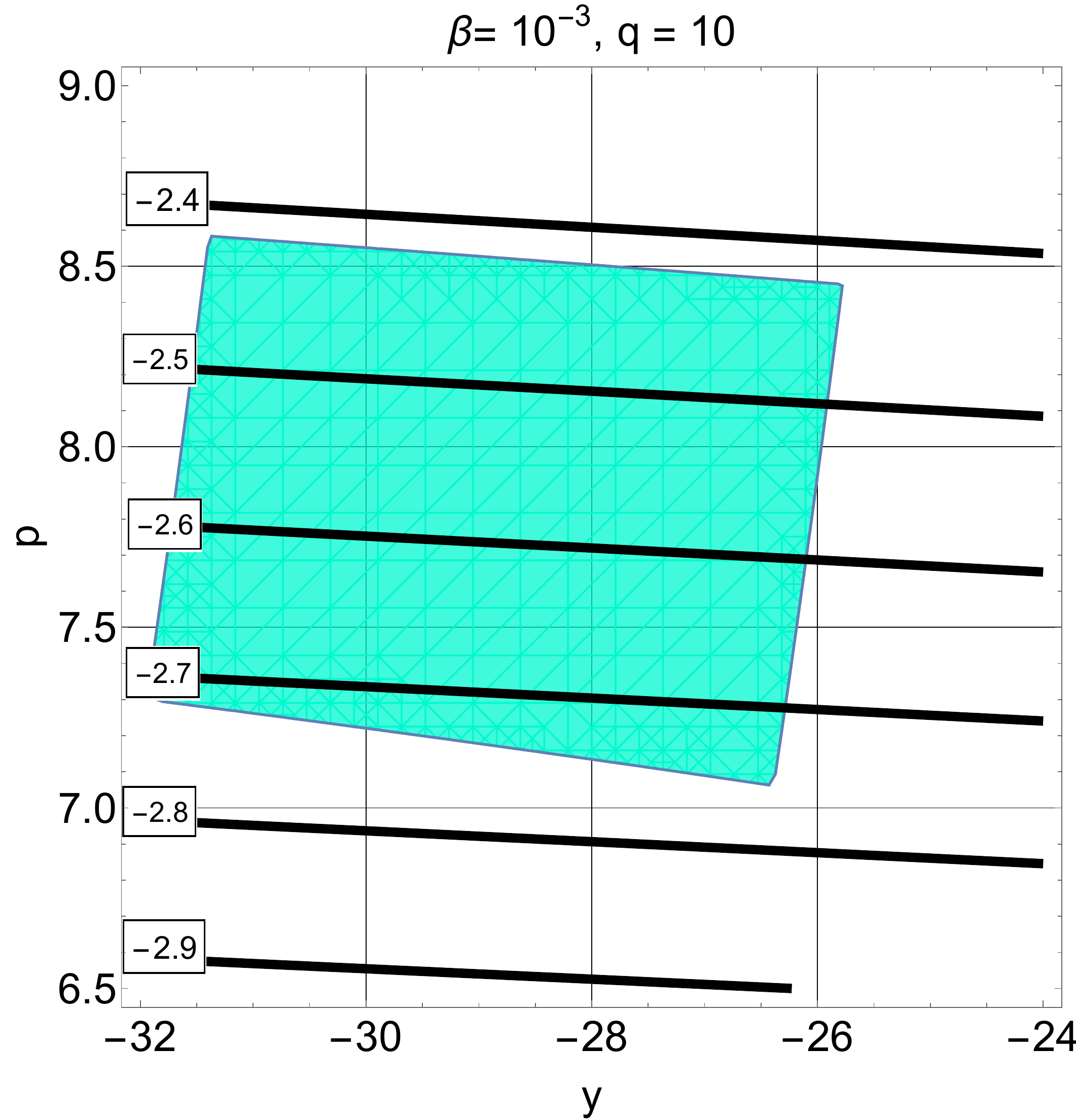}
\caption[a]{The same notations of Fig. \ref{FIGU10} have been followed for a different choice of the parameters. In both cases $r_{T}(N_{k}) > r_{T}^{min}$. We see that, also in this case, the values of $r_{T}$ in the allowed region are larger than $r_{T}^{min}$. }
\label{FIGU11}     
\end{figure}

Indeed  as we specifically discussed in the previous 
examples (see e.g. (\ref{PPS1})--(\ref{PPS2})) the scale dependence 
of $r_{T}$ and $n_{s}$ is mediated by $N_{k}$, i.e. 
$r_{T}(k) =r_{T}(N_{k})$ and $n_{s}(k) =n_{s}(N_{k})$. It becomes therefore a quantitative issue if the amount of reduction of $r_{T}(N_{k})$ for $N_{k} \gg {\mathcal O}(60)$ is compatible with $n_{s}(N_{k})$ and with the (upper) bounds on $r_{T}$ of Eqs. (\ref{PP1})--(\ref{PP2}). 
Based on the bounds of Eq. (\ref{AAA1}) and on the results of the previous sections we then expect $r_{T}(N_{k}) > r_{T}^{min} ={\mathcal O}(10^{-7})$ 
(or even ${\mathcal O}(10^{-6})$) provided the signal of high-frequency gravitons is still sufficiently large. In this respect the first test is reported in 
Figs. \ref{FIGU10}  and \ref{FIGU11} where we examine the plateau-like potential of Eq. (\ref{PP3}) in the plane defined by $y= \log{H_{r}/M_{P}}$ 
and by $q$. While for $\Phi\gg 1$ the inflationary phase fixes $n_{s}(N_{k})$ and $r_{T}(N_{k})$ according to Eqs. (\ref{PPS1})--(\ref{PPS2}),
for $\Phi < 1$ there is an oscillating stage whose length is here taken as a 
free parameter. During this stage the expansion rate $\delta$ follows
from Eqs. (\ref{TS23})--(\ref{TS24}).  In Figs. \ref{FIGU10} and \ref{FIGU11} the shaded areas correspond to the region where the scalar spectral index and the $r_{T}$ obey the constraints 
of Eq. (\ref{PP1}) and all the other high-frequency limits. We also require 
that $10^{-8} < h_{0}^2 \, \Omega_{gw}(\nu,\tau_{0}) < 10^{-6}$ for $\nu = {\mathcal O}(\nu_{max})$ (with the caveat that also $\nu_{max}$ depends on $\delta$).

In the left plot of Fig. \ref{FIGU10} we fixed $p$ while in the right plot 
$q$ is fixed. The same strategy has been adopted in Fig. \ref{FIGU11}
for a different set of parameters. The various curves appearing 
in Figs. \ref{FIGU10} and \ref{FIGU11} correspond to the 
common logarithms of $r_{T}(N_{k})$. We see, as expected, 
that these values are all larger than $r_{T}^{min}$.
It is relevant to stress that  Figs. \ref{FIGU10} and \ref{FIGU11}
have been obtained by assuming the dynamical determination 
of $N_{k}$ as suggested by Eq. (\ref{AAA2}). On the contrary 
in the case of  Figs. \ref{FIGU3} and \ref{FIGU4} $N_{k} = 60$ 
as if the post-inflationary evolution was absent. Both illustrative strategies have their own virtues but we think that the former is more consistent than 
the latter especially in the case where a long oscillating stage with $q> 2$ 
implies an expansion rate that is slower than radiation.

So far we explored the regime of relatively large $r_{T}$ and we now 
consider the opposite limit by requiring that the high-frequency signal is (systematically) small while all the other low-frequency constraints are satisfied. As already mentioned in section \ref{sec2} a potential 
candidate for rather small $r_{T}$ is represented by the hilltop 
models in their different versions \cite{HT1,HT2,FR1,FR2,FR3}.
Here we are considering a slightly different perspective 
by focussing on the potentials of Eqs. (\ref{PP3}) and (\ref{PP4}) 
where, however, $N_{k}$ includes the dependence 
of a post-inflationary stage expanding slower than radiation 
as suggested by Eq. (\ref{AAA2}).

\begin{figure}[!ht]
\centering
\includegraphics[height=8cm]{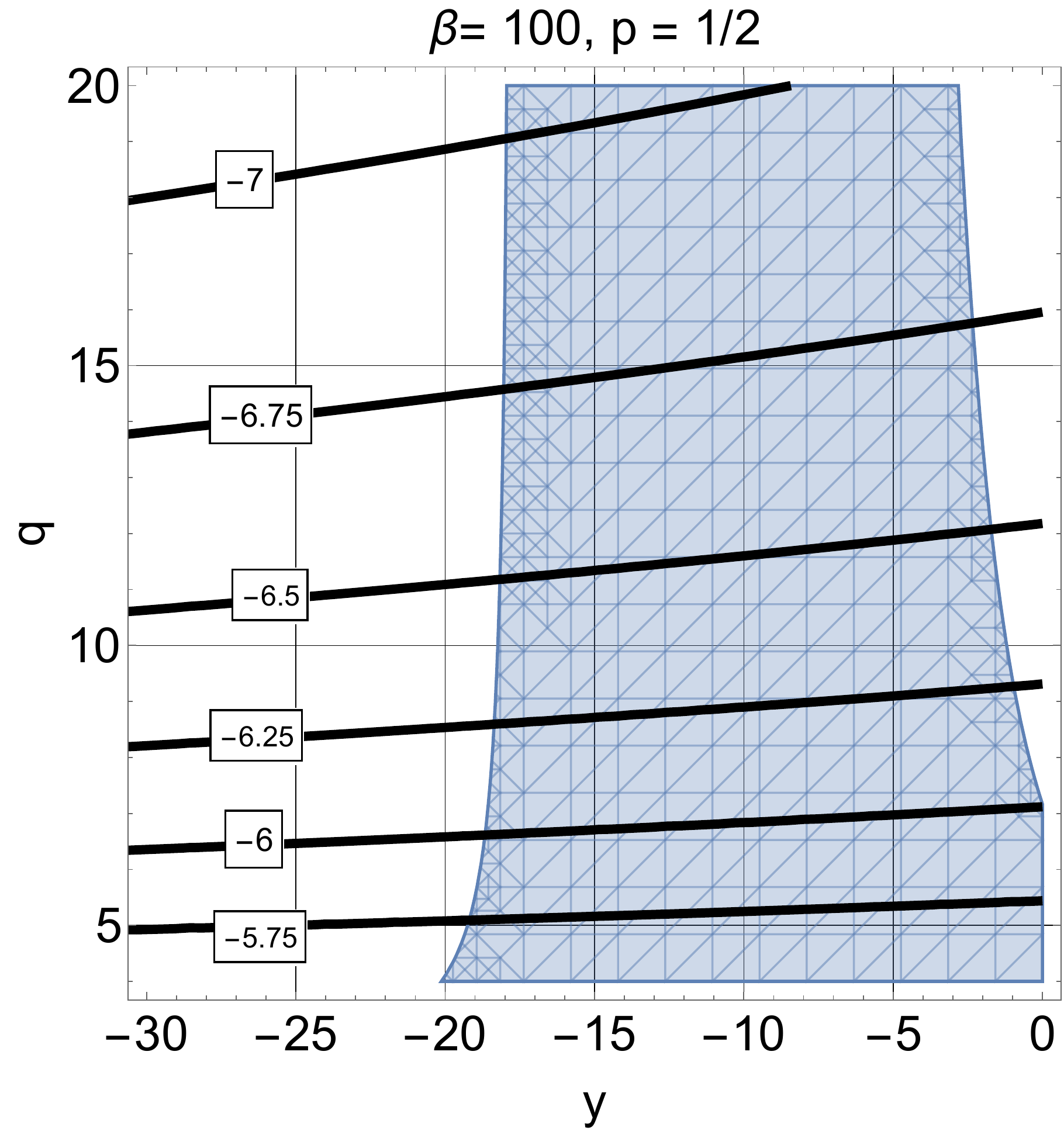}
\includegraphics[height=8cm]{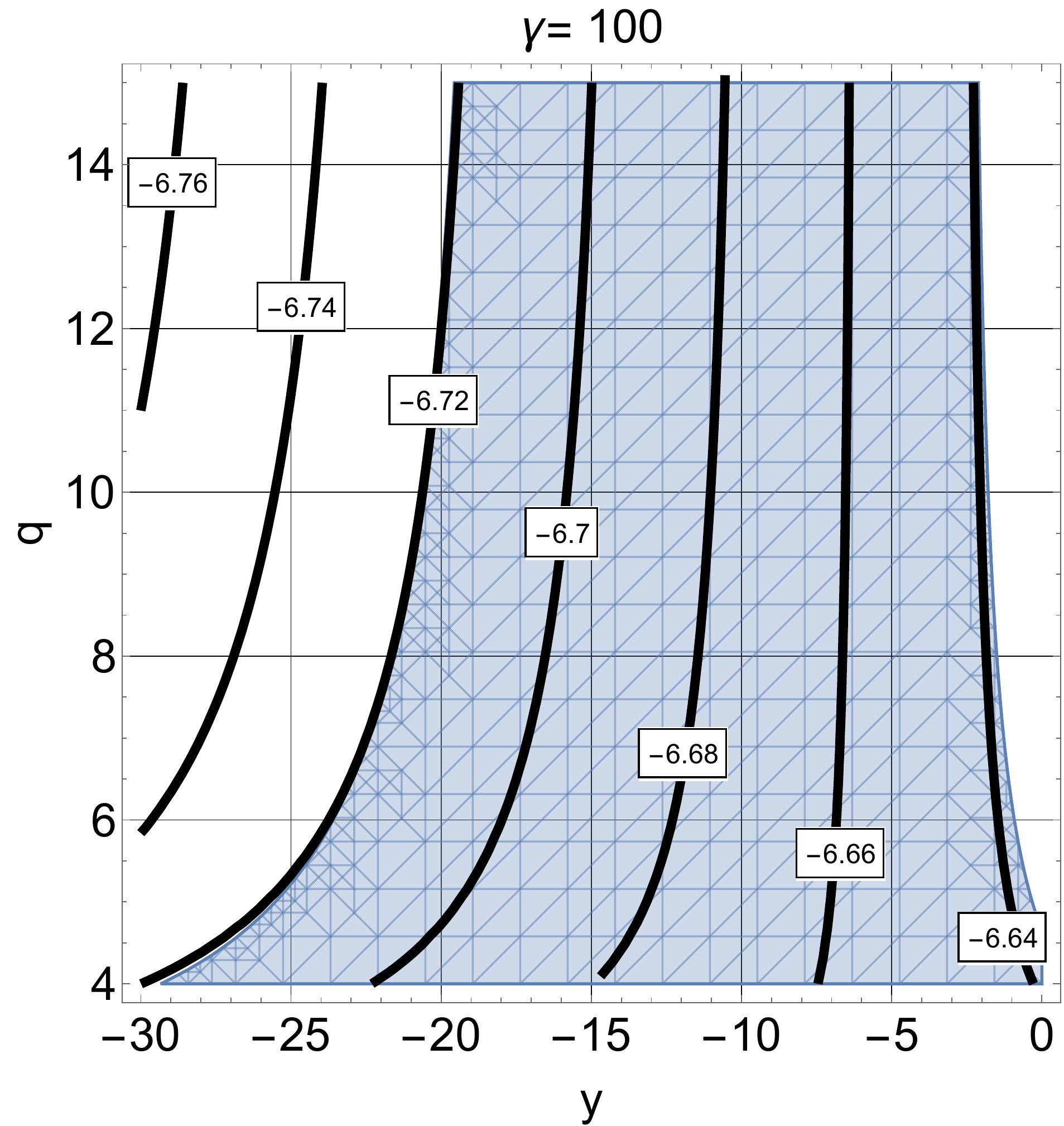}
\caption[a]{We illustrate the situation where the high-frequency 
signal is small while the low-frequency bounds of Eq. (\ref{PP1}) 
are concurrently satisfied. In the plot at the left we always consider 
the potential of Eq. (\ref{PP3}) while at the right the potential is given by Eq. (\ref{PP4}). In both plots the shaded area corresponds to $10^{-25} < h_{0}^2 \, \Omega_{gw}(\nu_{max}, \tau_{0}) < 10^{-15}$ and the various labels denote the common logarithm of $r_{T}$. We see, as expected, that $r_{T} < {\mathcal O}(r^{min}_{T})$.  }
\label{FIGU12}     
\end{figure}
This analysis is illustrated in Fig. \ref{FIGU12} where the low-frequency 
constraints of Eq. (\ref{PP1}) are supplemented by the requirement 
$10^{-25} < h_{0}^2 \, \Omega_{gw}(\nu_{max}, \tau_{0}) < 10^{-15}$.
According to the general arguments given above we should have that 
$r_{T}(N_{k}) = {\mathcal O}(10^{-6})$ (or smaller). According to Fig. 
\ref{FIGU12} this is what happens. More precisely in the left plot of Fig. \ref{FIGU12} we consider the potential of Eq. (\ref{PP3}) but for a 
range of parameters different from the ones discussed before.
The shaded area now defines the region of the parameter space where 
the spectral energy density is smaller than $10^{-15}$ (but larger than $10^{-25}$). This is approximately the standard signal of the concordance 
paradigm. In the plot at the right the notations are the same but the 
potential is the one of Eq. (\ref{PP4}).

There is finally an extreme situation where a drastic decrease of $r_{T}$ is compensated by the contribution of the post-inflationary stage of expansion. Indeed, by looking at Eq. (\ref{AAA2}) we might 
argue that the decrease in the number of $e$-folds caused 
by a reduction of $r_{T}$ could compensated by a 
corresponding increase of the term ${\mathcal D}(\delta_{i}, \xi_{i})$
so that, overall, $N_{k} = {\mathcal O}(60)$. This cancellation 
implies that the post-inflationary stage must expand slower than radiation 
and, in this case,
\begin{equation} 
r_{T} \simeq {\mathcal D}^2(\delta_{i}, \, \xi_{i}) = \prod_{i=1}^{n-1} \, \xi^{- 2 (\delta_{i} -1)/(\delta_{i} +1)} \ll {\mathcal O}(0.03).
\label{FFF1}
\end{equation}
For a single post-inflationary stage we would have that $r_{T} \simeq \xi^{- 2 (\delta-1)/(\delta +1)}$ with $0<\delta <1$. From Eqs. (\ref{MH5}) and (\ref{MH9}) we therefore have that the high-frequency value of the spectral energy density is approximately given by $h_{0}^2 \Omega_{gw}(\nu_{max}, \tau_{0}) = {\mathcal O}(10^{-18})$, slightly smaller than in the 
case of the concordance scenario. This means that a very small 
$r_{T}$ (e.g. much smaller than ${\mathcal O}(10^{-6})$) can be compensated by a long post-inflationary stage expanding slower 
than radiation. In this case, however, the spectral energy density 
is just shifted below the signal of the concordance paradigm.

\newpage 
\renewcommand{\theequation}{5.\arabic{equation}}
\setcounter{equation}{0}
\section{Concluding remarks}
\label{sec5}
The limits on the relic graviton backgrounds in different domains of comoving frequencies have been combined with the purpose of narrowing the range of variation of the tensor-to-scalar-ratio. In the concordance paradigm the low-frequency tail of the aHz region is maximized by considering the largest $r_{T}$ compatible with the available observational data. In this investigation we explored the opposite possibility suggesting that the low-frequency gravitons could remain practically invisible in the aHz region while their spectral energy density exceeds the predictions of the concordance paradigm at higher frequencies. As a result, depending on the frequency domain of the spike the lower bound on the  tensor-to-scalar-ratio ranges between ${\mathcal O}(10^{-6})$ and ${\mathcal O}(10^{-7})$.

The obtained results do not depend on the shape of the inflationary potential since the relic gravitons couple predominantly to the space-time curvature and not to the matter sources. We nonetheless enforced the consistency relations and corroborated the obtained bound with the detailed discussion of various classes of inflationary potentials admitting a modified post-inflationary evolution and
eventually leading to a high-frequency spike. In the present framework the number of $e$-folds corresponding to the exit of the CMB wavelengths can be larger than ${\mathcal O}(60)$ if the post-inflationary expansion rate is slower than radiation. If the consistency relations are instead violated (for instance because of the properties on the initial state or for some other reason), $r_{T}$ can be further reduced without affecting the high-frequency signal and the lower
bounds derived here do not apply. The general logic remains however intact since the reduction of the tensor-to-scalar-ratio would not preclude spectral energy densities exceeding the ones of the concordance scenario at high-frequencies. 

Since the concurrent determinations stemming from the audio band and from the aHz region involve the post-inflationary evolution, in the present approach any potential upper limit at high-frequency  simultaneously constrain the expansion rate prior to radiation dominance and the tensor-to-scalar-ratio.  This perspective implicitly encourages a synergy between experiments  scrutinizing different branches of the graviton spectrum. The fruitful dialogue between the instruments sensitive to small and to intermediate frequencies could be extended, in principle, also to conceptually different kinds of detectors in the MHz domain as repeatedly suggested in the past.

\section*{Acknowledgements} 
I wish to thank T. Basaglia, A. Gentil-Beccot, S. Reyes, S. Rohr and J. Vigen of the CERN Scientific Information Service for their valuable help. Various discussions with the late E. Picasso on high-frequency gravitons are also acknowledged.

\newpage
\begin{appendix}

\renewcommand{\theequation}{A.\arabic{equation}}
\setcounter{equation}{0}
\section{The number of $e$-folds and the maximal frequency }
\label{APPA}
\subsection{Number of $e$-folds and crossing of the pivot scales}
In what follows and in the bulk of the paper\footnote{As already mentioned in the introduction $\ln{x}$ denotes 
the natural logarithm of a generic variable $x$; $\log{x}$ denotes instead the common logarithm
of the same quantity.} $N_{k} = \ln{(a_{1}/a_{k})}$
denotes the number of $e$-folds corresponding to the crossing of a given 
wavelength and $a_{1}$ conventionally sets the end of the inflationary stage of expansion. 
Since the inflationary epoch may be followed by a sequence of different
phases all characterized by their own expansion rates $\delta_{i}$ (i.e.  $a_{i}(\tau) \propto \tau^{\delta_{i}}$
in the conformal time parametrization) the values of $N_{k}$ are implicitly determined from:
\begin{equation}
\frac{k}{a_{k}\, H_{k}} = e^{N_{k}} \, \biggl(\frac{H_{1}}{H_{k}}\biggr) \frac{k}{a_{1}\, H_{1}},
\label{APA1}
\end{equation} 
and the general evolution of the comoving horizon is illustrated in Fig. \ref{FIGU1}.
When the given wavelength crosses the Hubble radius in Fig. \ref{FIGU1} (i.e. $k \simeq a_{k}\, H_{k}$) 
Eq. (\ref{APA1}) fixes the value of $N_{k}$ not only in terms of $H_{1}$ (the expansion rate at the end of inflation) but also as a function of the subsequent expansion history. We can appreciate this statement by 
expressing the last term at the right-hand side of Eq. (\ref{APA1}) as:
\begin{equation}
 \frac{k}{a_{1}\, H_{1}}= \frac{k}{a_{0}\, H_{0}}\, \biggl(\frac{a_{0}\, H_{0}}{a_{eq} \, H_{eq}}\biggr) \,  \biggl(\frac{a_{eq}\, H_{eq}}{a_{r} \, H_{r}}\biggr) \,  \biggl(\frac{a_{r}\, H_{r}}{a_{1} \, H_{1}}\biggr),
\label{APA2}
\end{equation}
where $H_{r}$ is the expansion rate at the moment of radiation dominance while $H_{eq}$ and $H_{0}$ denote, respectively, 
the expansion rates at equality and at the present time; we remind that, in the notations of Fig. \ref{FIGU1}, $H_{n} = H_{r}$ and $a_{n} = a_{r}$, as already discussed in the bulk of the paper. From Eqs. (\ref{APA1})--(\ref{APA2}) we can also deduce
\begin{equation}
\frac{k}{a_{k} \, H_{k}} = \biggl(\frac{k}{a_{0} \, H_{0}} \biggr) \frac{\sqrt{H_{0}/M_{P}}}{( 2 \,\pi \, \Omega_{R\,0}\, \epsilon_{k} \, {\mathcal A}_{{\mathcal R}})^{1/4}} \frac{ \sqrt{H_{1}/H_{k}}}{{\mathcal C}(g_{s}, g_{\rho})} \, \, {\mathcal D}(\delta_{i}, \xi_{i}) \,\, e^{N_{k}}.
\label{APA2a}
\end{equation}
In Eq. (\ref{APA2a}) ${\mathcal C}(g_{s}, g_{\rho})$ and  ${\mathcal D}(\delta_{i}, \xi_{i})$ depend, respectively, on the radiative evolution after $a_{r}$ and on the post-inflationary history between the end of inflation and the moment of radiation dominance:
\begin{equation}
{\mathcal C}(g_{s}, \, g_{\rho}) = \bigl(g_{s,\, eq}/g_{s,\, r}\bigr)^{1/3} \, \bigl(g_{\rho,\, r}/g_{\rho,\,eq}\bigr)^{1/4},
\qquad\qquad {\mathcal D}(\delta_{i}, \, \xi_{i}) = \prod_{i=1}^{n-1}\, 
\xi_{i}^{- \frac{(\delta_{i} -1)}{2\, (\delta_{i} +1)}}.
\label{APA4}
\end{equation}
The first expression of Eq. (\ref{APA4}) follows by considering the radiation-dominated evolution between $a_{r}$ and $a_{eq}$; in this case we exactly obtain that $(H_{r}/H_{eq})^{1/2}= (a_{eq}/a_{r}) {\mathcal C}(g_{s}, g_{\rho})$ since in a stage of local thermal equilibrium, the entropy density is conserved and the total energy density depends on $g_{\rho}$ (i.e. the number of relativistic degrees of freedom in the plasma) while $g_{s}$ denotes the effective number of relativistic degrees of freedom appearing in the entropy density. In the standard situation where $g_{s,\, r}= g_{\rho,\, r} = 106.75$ and $g_{s,\, eq}= g_{\rho,\, eq} = 3.94$ we have that ${\mathcal C}(g_{s}, \, g_{\rho})= 0.75$. The contribution of ${\mathcal C}(g_{s}, \, g_{\rho})$ to Eq. (\ref{APA2a}) is numerically not essential for the determination of $N_{k}$ and the term ${\mathcal D}(\delta_{i}, \xi_{i})$ has a more prominent effect. Indeed, as already mentioned, in Eq. (\ref{APA4}) $\delta_{i}$ estimates the expansion rate in each of the post-inflationary stages and $\xi_{i} = H_{i+1}/H_{i}< 1$ measures their relative duration. In the limit of a single post-inflationary phase all the $\delta_{i}$ collapse to a single $\delta$ and ${\mathcal D}(\delta, \xi) = \xi^{- (\delta -1)/[2 (\delta+1)]}$ where $ \xi = H_{r}/H_{1} < 1$.  

The constraints from big-bang nucleosynthesis suggest an absolute lower limit on $\xi$ (i.e. $H_{r} \geq 10^{-44} \, M_{P}$) since the plasma must be dominated by radiation as soon as the formation of light nuclei starts and, for the same reason, the product of all the $\xi_{i}$ equals $H_{r}/H_{1}$, i.e. by definition $\xi_{1}\, \xi_{2} \,.\,.\,. \,\xi_{n-1} \,\xi_{n} =\xi = H_{r}/H_{1}$. The actual value of $\xi=H_{r}/H_{1}$ ultimately depends on $H_{1}$ whose explicit estimate follows by remarking that\footnote{The result of Eq. (\ref{APA2b}) has been implicitly used in Eq. (\ref{APA2a}) and it is 
further analyzed in the subsequent appendix \ref{APPB}.}
\begin{equation}
\biggl(\frac{H_{k}}{M_{P}}\biggr) = \sqrt{\pi\, \epsilon_{k}\, {\mathcal A}_{{\mathcal R}}},
\qquad\qquad \biggl(\frac{H_{k}}{H_{1}}\biggr) = {\mathcal O}(1).
\label{APA2b}
\end{equation}
Besides the number of $e$-folds associated with the crossing of a given scale $k$ there is also a second relevant notion, namely the {\em maximal number of $e$-folds presently accessible to large-scale observations} ($N_{max}$ in what follows). This quantity is determined by fitting the (redshifted) inflationary event horizon inside the current Hubble patch, i.e. $H_{b}^{-1} (a_{0}/a_{i}) \simeq H_{0}^{-1}$ where $H_{b}$ denotes the expansion rate in the initial stages of inflation:
\begin{equation}
e^{N_{max}} = (2\, \Omega_{R0})^{1/4} \, \sqrt{\frac{H_{b}}{H_{0}}}\, \frac{{\mathcal C}(g_{s}, g_{\rho})}{{\mathcal D}(\delta_{i}, \xi_{i})}, \qquad\qquad H_{b} = {\mathcal O}(H_{1}).
\label{APA3}
\end{equation}
In Eq. (\ref{APA3}) $H_{b}$ can be estimated from $H_{1}$ owing from the small variation 
of the Hubble rate during inflation and the same observation can be actually made in the case of $H_{k}$, as we shall 
discuss in appendix \ref{APPB}. As in the case of $N_{k}$, in Eq. (\ref{APA3}) we assumed that the post-inflationary expansion rate is partitioned in $n$ subsequent epochs expanding at different rates but the last stage always coincides with radiation so that, by definition, $H_{n} = H_{r}$. The values of $N_{k}$ and $N_{max}$ are related as:
\begin{equation}
N_{k} = N_{max} - \ln{\biggl(\frac{k}{a_{0}\, H_{0}}\biggr)} -\ln{\frac{H_{k}}{H_{1}}}.
\label{APA3a}
\end{equation}
As we shall see more specifically in appendix \ref{APPB} $H_{k} = {\mathcal O}(H_{1})$ 
for all the comoving wavelengths relevant in the case of CMB physics so that, 
in practice, $N_{max}$ coincides with $N_{k}$ when $k = {\mathcal O}(a_{0}\, H_{0})$.
The explicit values of $N_{k}$ and $N_{max}$ enter the estimates the inflationary observables for a specific choice of the potential; for instance the value of $N_{max}$ is:
\begin{eqnarray}
N_{max}  &=& 61.55 - \ln{\biggl(\frac{h_{0}}{0.7}\biggr)} + \frac{1}{4} \ln{\biggl(\frac{\epsilon}{0.001}\biggr)} +\frac{1}{4} \ln{\biggl(\frac{{\mathcal A}_{{\mathcal R}}}{2.41\times 10^{-9}}\biggr)} + \ln{{\mathcal C}(g_{s}, \, g_{\rho})}
\nonumber\\
&+&   \frac{1}{4} \ln{\biggl(\frac{h_{0}^2 \, \Omega_{R0}}{4.15\times 10^{-5}}\biggr)}
+ \frac{1}{2}\sum_{i}^{n-1} \, \biggl(\frac{\delta_{i} -1}{\delta_{i} + 1}\biggr) \, \ln{\xi_{i}}.
\label{APA5}
\end{eqnarray}
Equation (\ref{APA5}) depends on the slow-roll parameter $\epsilon$ which we have taken to be 
constant and scale-independent. However if the consistency relations are enforced we can trade\footnote{In this case, if $r_{T} = {\mathcal O}(0.06)$ 
we would have that $N_{max}  = 61.88$ always assuming the typical values of the other parameters; if the explicit value of ${\mathcal C}(g_{s}, g_{\rho})$ is taken into account 
the value of $N_{max}$ is insignificantly reduced to $N_{max}=\,61.61$.} 
$\epsilon$ for $r_{T}$ since $r_{T} \simeq 16 \epsilon$. 
Since the values of $N_{k}$ and $N_{max}$ are related we have from Eqs. (\ref{APA2})--(\ref{APA3}) that 
\begin{equation}
e^{N_{k}} = ( 2 \, \Omega_{R\,0})^{1/4} \sqrt{\frac{H_{k}}{H_{1}}} \sqrt{\frac{H_{k}}{H_{0}}}\, \frac{{\mathcal C}(g_{s}, g_{\rho})}{{\mathcal D}(\delta_{i}, \xi_{i})}
\biggl(\frac{k}{a_{0} \, H_{0}}\biggr)^{-1}.
\label{APA6}
\end{equation}
We now recall (see also appendix \ref{APPB}) that $H_{k}/M_{P} = \sqrt{\pi \, \epsilon_{k} \, {\mathcal A}_{{\mathcal R}}}$; Eq. (\ref{APA6}) can then be  written, after taking the natural logarithm of both sides,
\begin{eqnarray}
N_{k} &=& 59.4  + \frac{1}{4} \ln{\biggl(\frac{\epsilon_{k}}{0.001}\biggr)} +\frac{1}{4} \ln{\biggl(\frac{{\mathcal A}_{{\mathcal R}}}{2.41\times 10^{-9}}\biggr)} + \ln{{\mathcal C}(g_{s}, \, g_{\rho})}  - \ln{\biggl(\frac{k}{0.002\,\,\mathrm{Mpc}^{-1}}\biggr)}
\nonumber\\
&+&   \frac{1}{4} \ln{\biggl(\frac{h_{0}^2 \, \Omega_{R0}}{4.15\times 10^{-5}}\biggr)}
+ \frac{1}{2}\sum_{i=1}^{n-1} \, \biggl(\frac{\delta_{i} -1}{\delta_{i} + 1}\biggr) \, \ln{\xi_{i}} - \frac{1}{2} \ln{\biggl(\frac{H_{1}}{H_{k}}\biggr)}.
\label{APA7}
\end{eqnarray}
If we set $\delta_{i} =1$ into Eqs. (\ref{APA5})--(\ref{APA7}) we obtain the 
standard result  implying that $ N_{max} = {\mathcal O}(60)$; in this case the whole 
post-inflationary stage collapses to a single radiation-dominated phase extending down 
to the $H_{r}$. If  $\delta_{i}<1$ we have instead $N_{max} > 60$ and this happens since 
all the $\xi_{i}$ are, by definition, all smaller than $1$; for the same reason $N_{max} < 60$ when $\delta_{i}>1$. For a single phase expanding slower than radiation $N_{max}$ can be as large as $75$ and in all the intermediate situations (where there are different phases expanding either faster or slower than radiation) $N_{max}$ depends on the relative duration of the various epochs and on their expansion rates.  What is true for $N_{max}$ is also true for $N_{k}$ by virtue of Eq. (\ref{APA3a}). In the case of a specific potential admitting a post-inflationary stage not dominated by radiation the correct value of $N_{k}$ employed to evaluate the inflationary observables must then follow from Eq. (\ref{APA7}).

\subsection{The typical frequencies of the spectrum}
As in the case of $N_{k}$ and $N_{max}$, all the
typical frequencies depend on $r_{T}$ and on the post-inflationary expansion rate (see also the 
cartoon of Fig. \ref{FIGU1}). Starting from $\nu_{max}$ we have:
\begin{equation}
\nu_{max} =  \prod_{i\,=\,1}^{n-1} \, \xi_{i}^{\frac{\delta_{i} -1}{2 (\delta_{i}+1})}\,\, \overline{\nu}_{max} = {\mathcal D}^{-1}(\delta_{i},\xi_{i})\, \overline{\nu}_{max}.
\label{APA8}
\end{equation}
Note that when all the $\delta_{i} \to 1$ the value of $\nu_{max}$
coincides with $\overline{\nu}_{max}$ whose explicit value is:
\begin{equation}
\overline{\nu}_{max} = \frac{M_{P}}{2 \pi} \,\bigl(2\, \Omega_{R0}\bigr)^{1/4} \, \sqrt{\frac{H_{0}}{M_{P}}}\, \sqrt{\frac{H_{1}}{M_{P}}}\, \, {\mathcal C}(g_{s}, g_{\rho}).
\label{APA8a}
\end{equation}
Both $\nu_{max}$ and $\overline{\nu}_{max}$ are computed from the 
smallest wavelength that crosses the Hubble radius of \ref{FIGU1} and immediately 
reenters; this is why Eq. (\ref{APA8a}) depends upon $H_{1}$. From a quantum mechanical 
viewpoint the maximal frequency corresponds to the production of a single graviton pair.
In view of a direct estimate of $\overline{\nu}_{max}$ we recall again that $H_{1} = {\mathcal O}(H_{k})$ (see Eq. (\ref{APA2b}) and discussion therein):
\begin{equation}
\overline{\nu}_{max} = 195.38 \, {\mathcal C}(g_{s}, g_{\rho}) \, \biggl(\frac{{\mathcal A}_{{\mathcal R}}}{2.41\times 10^{-9}}\biggr)^{1/4}\,\,
\biggl(\frac{\epsilon_{k}}{0.001}\biggr)^{1/4} \,\, \biggl(\frac{h_{0}^2 \, \Omega_{R\,0}}{4.15\times 10^{-5}}\biggr)^{1/4} \,\,\mathrm{MHz}.
\label{AP9}
\end{equation}
In case the consistency relations are enforced, we can always trade $\epsilon_{k}$ for $r_{T}$ so that for a typical value $r_{T}= {\mathcal O}(0.06)$ the value of $\overline{\nu}_{max}$ becomes\footnote{As already mentioned in connection with $N_{k}$ and $N_{max}$, for typical values of the relativistic degrees of freedom, ${\mathcal C}(g_{s}, \, g_{\rho})= \,{\mathcal O}(0.75)$. More precisely for $g_{s,\, eq} = g_{\rho,\, eq} = 3.94$ and $g_{s,\, r} = g_{\rho,\, r} = 106.75$ we have, from Eq. (\ref{AP9}), that $\overline{\nu}_{max} = 148.41 \,\, \mathrm{MHz}$ 
while $\overline{\nu}_{max} = 206.53\,\, \mathrm{MHz}$ from Eq. (\ref{APA9b}). }: 
\begin{equation}
\overline{\nu}_{max} = 271.88 \, {\mathcal C}(g_{s}, g_{\rho}) \, \biggl(\frac{{\mathcal A}_{{\mathcal R}}}{2.41\times 10^{-9}}\biggr)^{1/4}\,\,
\biggl(\frac{r_{T}}{0.06}\biggr)^{1/4} \,\, \biggl(\frac{h_{0}^2 \, \Omega_{R\,0}}{4.15\times 10^{-5}}\biggr)^{1/4} \,\,\mathrm{MHz}.
\label{APA9b}
\end{equation}
The frequency $\nu_{max}$ be complemented by  the other frequencies of the spectrum and since
$\overline{\nu}_{max}$ depends on $r_{T}$, also all the other frequencies are sensitive 
to the specific value of the tensor-to-scalar-ratio. Let us then suppose that, before radiation dominance,
the post-inflationary epoch consists of thee separate phases; this means that the final spectrum 
is characterized by three typical frequencies: $\nu_{1} = \nu_{max}$, $\nu_{2}$ and $\nu_{3} = \nu_{r}$.
The expression of $\nu_{max}$ follows from Eq. (\ref{APA8}) in the case $n=3$
\begin{equation}
\nu_{max} = \nu_{1} = \prod_{i\,=\,1}^{2} \, \xi_{i}^{\frac{\delta_{i} -1}{2 (\delta_{i}+1})}\,\, \overline{\nu}_{max}.
\label{APA9ca}
\end{equation}
Similarly we can easily compute $\nu_{2}$ and $\nu_{r}$:
\begin{eqnarray}
\nu_{2} &=& \sqrt{\xi_{1}} \,\, \xi_{2}^{(\delta_{2} -1)/[2 (\delta_{2}+1)]} \, \,\overline{\nu}_{max},
\nonumber\\
\nu_{r} &=&\nu_{3} = \sqrt{\xi_{1}}\, \sqrt{\xi_{2}}\, \overline{\nu}_{max} = \sqrt{\xi} \,\overline{\nu}_{max},
\label{APA9cb}
\end{eqnarray}
where, by definition, $\xi= \, \xi_{1} \, \xi_{2}$. In the case of $n$ intermediate phases taking place prior to $a_{r}$ the generic intermediate frequency $\nu_{m}$ can be expressed as 
\begin{eqnarray}
\nu_{m} &=& \sqrt{\xi_{1}} \, .\,.\,.\, \sqrt{\xi_{m-1}}\,  \prod_{m\,=\,1}^{n-1} \xi_{i}^{\frac{\delta_{i} -1}{2 (\delta_{i}+1})}\,\, \overline{\nu}_{max},
\label{APA9d}\\
\nu_{r} &=& \nu_{n} = \sqrt{\xi_{1}}\,\sqrt{\xi_{2}} \, .\,.\,.\, \sqrt{\xi_{n-2}}\,\sqrt{\xi_{n-1}} \,\overline{\nu}_{max}.
\label{APA9e}
\end{eqnarray}
Recalling the remarks presented before Eq. (\ref{APA2b}), since 
the different phases must not last below $H_{r}$,  the product of all the $\xi_{i}$ equals $H_{r}/H_{1}$, i.e. by definition $\xi_{1}\, \xi_{2} \,.\,.\,. \xi_{n-1} \,\xi_{n} =\xi = H_{r}/H_{1}$.
Therefore, in case the consistency relations are enforced, Eqs. (\ref{APA9ca})--(\ref{APA9cb}) and (\ref{APA9d})--(\ref{APA9e}) show that both the maximal and the intermediate frequencies of the spectrum depend on $r_{T}$ through $\xi$. 

\renewcommand{\theequation}{B.\arabic{equation}}
\setcounter{equation}{0}
\section{The expansion rate at horizon crossing}
\label{APPB}
The value of the expansion rate when the wavelengths associated with the pivot 
wavenumber $k_{p}$ cross the comoving Hubble radius determines both the number of $e$-folds and the typical frequencies of the spectrum. In the present investigation the wavelengths ${\mathcal O}(2\pi/k_{p})$ are referred to as the CMB wavelengths. Since at horizon crossing $(H_{k}/H_{1}) = {\mathcal O}(1)$ it is natural to set $H_{k} = {\mathcal O}(H_{1})$ in the expressions of $N_{k}$ 
and $N_{max}$ [see Eqs. (\ref{APA5})--(\ref{APA7}) and discussion therein]. If $H_{k}/H_{1} = {\mathcal O}(1)$ it also follows that $V_{1}^{1/4}/H_{k} \gg 1$ where $V_{1}$ denotes throughout this appendix the value of the inflaton potential at $H_{1}$. If the estimates are presented in terms of $V_{1}^{1/4}/H_{k}$, the values 
of $N_{k}$ and $N_{max}$ are systematically larger than in the case where $H_{k}$ 
is measured in units of $H_{1}$. In this appendix we also take the opportunity of introducing the relevant notations that are employed throughout the main discussion; in particular, at the end of this appendix  the scale-dependence of the slow-roll parameters is explicitly analyzed since the related results are relevant for various examples discussed in the text.

\subsection{The r\^ole of $H_{k}$ and $H_{1}$}
It is well known that the power spectrum of curvature inhomogeneities 
during inflation can be expressed  in two complementary ways:
\begin{equation}
P_{{\mathcal R}}(k,\tau) = \frac{ |k \tau|^2}{\pi\, \epsilon(\tau)} \frac{H^2(\tau)}{M_{P}^2} \equiv  \frac{ |k \tau|^2}{\pi\, \epsilon[\varphi(\tau)]} \frac{H[\varphi(\tau)]}{M_{P}^2}.
\label{APB1}
\end{equation}
The difference between the first and second equality of Eq. (\ref{APB1}) is that the Hubble rate and the slow-roll parameter $\epsilon$ are regarded either as functions of the conformal time coordinate or rather as fuctionals of the inflaton field $\varphi$. 
When the given wavelength crosses the Hubble radius (i.e. $k\tau = {\mathcal O}(1)$) the previous expression can be written as 
\begin{equation}
P_{{\mathcal R}}(k,\,1/k) = \frac{1}{\pi \epsilon_{k}} \frac{H_{k}^2}{M_{P}^2},\qquad H_{k} = H(1/k), \qquad \epsilon_{k} = \epsilon(1/k),
\label{APB2}
\end{equation}
where $H_{k}$ denotes expansion rate at horizon crossing. We now consider the crossing of the scales relevant for CMB physics and, for these scales, the power spectrum of curvature inhomogeneities is customarily expressed as $P_{{\mathcal R}}(k,\,1/k)= {\mathcal A}_{{\mathcal R}} \, (k/k_{p})^{n_{s}-1}$ where $k_{p}$ is the (conventional) pivot scale while $n_{s}$ is the (scalar) spectral index. If $k$ is comparable 
with $k_{p}$ we therefore have that 
\begin{equation}
\frac{H_{k}}{M_{P}} \simeq \sqrt{\pi \, \epsilon_{k} \, {\mathcal A}_{{\mathcal R}} } = \frac{\sqrt{\pi \, r_{T} \, {\mathcal A}_{{\mathcal R}}}}{4}.
\label{APB3}
\end{equation}
The second expression of Eq. (\ref{APB3}) follows from the consistency relations; we also note that, by definition, in Eq. (\ref{APB3}) $r_{T}= r_{T}(k_{p})$. 
If we assume that $r_{T} \leq 0.03$ and ${\mathcal A}_{{\mathcal R}} = {\mathcal O}(10^{-9})$ it 
is clear that $H_{k}/M_{P} \ll 1$. However it turns out that $H_{k} = {\mathcal O}(H_{1})$ 
and from the physical viewpoint it is easy to see that $H_{k} = {\mathcal O}(H_{1})$ since the expansion rate decreases very little during inflation. We want however to be more specific and eventually compare $H_{1}/H_{k}$ with $V_{1}^{1/4}/H_{k}$. 
From $a_{k} H_{k} = k$ we can actually obtain the following chain of equalities:
\begin{equation}
H_{k} = \frac{H_{1}}{1 - \epsilon_{k}} \biggl|\frac{k}{a_{1} \, H_{1}} \biggr|^{ - \frac{\epsilon_{k}}{1 - \epsilon_{k}}} = H_{1}  \biggl|\frac{k}{a_{1} \, H_{1}} \biggr|^{ - \epsilon_{k}}\biggl[1 + {\mathcal O}(\epsilon_{k})\biggr].
\label{APB4}
\end{equation}
But for typical wavenumbers $k= {\mathcal O}(k_{p})$ it turns out that $k\,\ll \, |a_{1} \, H_{1}|$; more specifically we can estimate the value of $k_{p}/(a_{1}\, H_{1})$ and obtain:
\begin{equation}
\frac{k_{p}}{a_{1}\, H_{1}} =\frac{10^{-25.85}}{{\mathcal D}(\delta_{i}, \xi_{i})} \,\biggl(\frac{k_{p}}{0.002\,\, \mathrm{Mpc}^{-1}}\biggr) \biggl(\frac{r_{T}}{0.03}\biggr)^{-1/4} \biggl(\frac{{\mathcal A}_{{\mathcal R}}}{2.41\times 10^{-9}}\biggr)^{-1/4} \, \biggl(\frac{h_{0}^2 \, \Omega_{R0}}{4.15\times 10^{-5}}\biggr)^{-1/4}.
\label{APB5}
\end{equation}
For a post-inflationary history dominated by radiation all the $\delta_{i}$ go to $1$ and $k_{p} = {\mathcal O}(10^{-26}) \, a_{1} \, H_{1}$ and when the expansion rate is slower than radiation the ratio $k_{p}/(a_{1} \, H_{1})$ is even smaller. 
Therefore, if we insert Eq. (\ref{APB5}) into Eq. (\ref{APB4}), we can conclude, as previously anticipated in Eq. (\ref{APA2b}) that 
\begin{equation}
\frac{H_{k}}{M_{P}} \simeq \sqrt{ \pi\, \epsilon_{k} \, {\mathcal A}_{{\mathcal R}}}, \qquad \qquad H_{k} \simeq H_{1}.
\label{APB6}
\end{equation}
Even if, according to Eq. (\ref{APB6}), $H_{k}$ and $H_{1}$ are of the same order,  $V_{1}^{1}$ and $H_{k}$ are rather 
different. To appreciate this statement we recall that, at the and of inflation, $\epsilon \to 1$ which means $H^2 = - \dot{H}$; but this condition can also be translated as $V = \dot{\varphi}^2$ and this implies that $H_{1}^2 \, M_{P}^2 = 4 \pi \, V_{1}$. We can therefore obtain an estimate of $V_{1}^{1/4}$ and the result is: 
\begin{equation}
\frac{V_{1}^{1/4}}{H_{k}} = \frac{\sqrt{H_{1}/H_{k}}}{\sqrt{2 \pi} ( \epsilon_{k} {\mathcal A}_{{\mathcal R}})^{1/4}}  \gg 1.
\label{APB7}
\end{equation} 
If we use  Eq. (\ref{APB7}) and express $N_{k}$ in terms of $V_{1}^{1/4}/H_{k}$ we get 
\begin{eqnarray}
N_{k} &=& 64.902   + \ln{\biggl[\frac{{\mathcal C}(g_{s}, \, g_{\rho})}{0.7596}\biggr]}  - \ln{\biggl(\frac{k}{0.002\,\,\mathrm{Mpc}^{-1}}\biggr)}
\nonumber\\
&+&   \frac{1}{4} \ln{\biggl(\frac{h_{0}^2 \, \Omega_{R0}}{4.15\times 10^{-5}}\biggr)}
+ \frac{1}{2}\sum_{i}^{n-1} \, \biggl(\frac{\delta_{i} -1}{\delta_{i} + 1}\biggr) \, \ln{\xi_{i}} - \ln{\biggl(\frac{V_{1}^{1/4}}{H_{k}}\biggr)}.
\label{APB8}
\end{eqnarray}
If we compare Eqs. (\ref{APA7}) and (\ref{APB8}) we see that, because of Eq. (\ref{APB7}) 
the value of $N_{k}$ gets larger than in the case where the number of $e$-folds is expressed 
as a function of $H_{k}/H_{1}$; since $H_{k}/H_{1} = {\mathcal O}(1)$, Eq. (\ref{APA7}) is more suitable 
for an explicit estimate of $N_{k}$.

\subsection{Specific potentials and slow-roll algebra}
\label{APPB2}
When the consistency relations are enforced the tensor to scalar 
ratio cannot be equally small for all the classes of inflationary potentials and while the 
monomials are clearly excluded, the plateau-like and the hill-top 
potentials may lead to $r_{T}$ that are comparatively smaller. Since different classes of 
potentials have been mentioned in the main discussion, their associated
properties will now be swiftly recalled.  In the case of Eq. (\ref{PP3}) the explicit expressions of the slow-roll parameters 
follow from Eq. (\ref{PP5}) so $\epsilon(\Phi)$ and $\overline{\eta}(\Phi)$ are given by:
\begin{equation}
\epsilon(\Phi) = \frac{2 \, q^2 }{\Phi^2 ( 1 + \beta^2 \Phi^{\frac{4 \,q}{p}})^2}, \qquad\qquad
\overline{\eta}(\Phi) = \frac{2 q\,[2 \, p\, q- p - \beta^2 (p + 4 q) \Phi^{\frac{4 q}{p}}]}{ p \Phi^2 ( 1 +\beta^2 \, \Phi^{\frac{4 q}{p}})^2}.
\label{PP6}
\end{equation}
In this case, according to Eq. (\ref{PP6}), the tensor-to-scalar ratio and the scalar spectral index are given by:
\begin{equation}
r_{T}(\Phi)= \frac{ 32 \, q^2}{\Phi^2 (1 + \beta^2 \Phi^{\frac{4 q}{p}})^2}, \qquad\qquad n_{s}(\Phi) = 1 - \frac{4\, p \, q( 1+ q) + 4 q ( q + 4 p) \beta^2 \Phi^{\frac{4 q}{p}}}{p \Phi^2 ( 1 + \beta^2 \Phi^{\frac{4 q}{p}})^2}.
\label{PP7}
\end{equation}
To compare the physical features of the various potentials when the pivot scales cross the 
comoving Hubble radius it is practical to estimate directly $\epsilon_{k}$ and $\overline{\eta}_{k}$ 
as a function of the number of $e$-fold $N_{k}$ for $k = {\mathcal O}(k_{p})$. For this purpose, as a 
a general observation, we should compute the total number of $e$-folds elapsed since the crossing of the bunch 
of the CMB wavelengths; since this procedure is consistently followed for all the explicit potentials discussed here,
it is useful to discuss it more explicitly in the concrete case of Eq. (\ref{PP3}).
The number of $e$-folds $N_{k}$ is then given by:
\begin{equation}
N_{k} = \int_{\Phi_{f}}^{\Phi_{k}} \biggl(\frac{v}{\partial_{\Phi} v }\biggr) \, d \Phi = 
\int_{\Phi_{f}}^{\Phi_{k}} \frac{\Phi}{ 2 q } \biggl( 1 + \beta^2 \Phi^{\frac{4 q}{p}}\biggr) \, d \Phi,
\label{PP8}
\end{equation}
where $\Phi_{k}$ denotes the value of the field when the scale $k$ crosses the comoving 
Hubble radius while  $\Phi_{f} \to 1$ coincides with the 
end of inflation. Even if, as we saw, different approaches can be envisaged we are here taking the standard practice and require that
\begin{equation}
\epsilon(\Phi_{f}) \to 1 \,\,\,\Rightarrow\,\,\, H^2 = - \dot{H} \,\,\,\Rightarrow \,\,\, V = \dot{\varphi}^2.
\label{PP9}
\end{equation}
For instance if we evaluate $\epsilon(\Phi_{f})$ from Eq. (\ref{PP6}) and require $\epsilon(\Phi_{f}) \to 1$ 
we obtain the condition 
\begin{equation}
\Phi_{f}^2 \bigl( 1 + \beta^2 \Phi_{f}^{4 q/p}\bigr)^2 = 2 q^2.
\label{PP9a}
\end{equation}
We now have two complementary possibilities. If $\beta^2 < 1$ (as we always assumed 
in the explicit evaluations) then $\Phi_{f} \simeq 1/(\sqrt{2}\, q) = {\mathcal O}(1)$. In the opposite case 
(i.e. $\beta>1$) we get $\Phi_{f} \simeq (\sqrt{2}\, q\, \beta^2)^{q/(4 p +q)}$ which is 
again of order $1$. All in all from Eq. (\ref{PP8}) the number of $e$-folds is ultimately given by: 
\begin{equation}
N_{k} = \frac{\Phi_{k}^2 -1}{4 q} + \frac{p \, \beta^2 \, \bigl( \Phi_{k}^{2 + \frac{4 q}{p}} -1\bigr)}{4 q ( p + 2 q)},
\label{PP10}
\end{equation}
where we simply assumed $\Phi_{f} \to 1$.
Since the field value at $\Phi_{k}$ is defined at the time of the crossing during inflation we can 
take the limit $\Phi_{k} \gg 1$ in Eq. (\ref{PP10}) and eventually determine the connection between 
$\Phi_{k}$ and $N_{k}$:
\begin{equation}
N_{k} = \frac{p \, \beta^2}{4 q\, ( p + 2 q)} \, \, \Phi_{k}^{2 + \frac{4 q}{p}}
\qquad \Rightarrow \qquad \Phi_{k} = \biggl[ \frac{4 q (p + 2 q) \, \, N_{k}}{p \, \beta^2}\biggr]^{ \frac{p}{2 (p + 2 q)}}.
\label{PP11}
\end{equation}
Thanks to Eq. (\ref{PP11})  Eqs. (\ref{PP6})--(\ref{PP7}) can be directly expressed in terms of $\Phi_{k}> 1$
\begin{equation}
\epsilon_{k} = \frac{2 q^2}{\beta^4 \, \Phi_{k}^{8 q/p +2}}, \qquad \qquad \overline{\eta}_{k} = - 
\frac{2 q (p + 4 q)}{p \,\beta^2 \, \Phi_{k}^{8 q/p+2}}.
\label{PP12}
\end{equation}
Finally using Eq. (\ref{PP11}) into eq. (\ref{PP12}) we have: 
\begin{equation}
\epsilon_{k} = \frac{ 2 q^2 \, \beta^{- \frac{2 p}{p + 2 q}}}{[ 4 q ( p + 2 q) N_{k}/p]^{\frac{p + 4 q}{p + 2 q}}}, \qquad \qquad \overline{\eta}_{k} = - \frac{ p + 4 q}{2 ( p + 2 q) \, N_{k}}.
\label{PP13}
\end{equation}
The same strategy is used for the other potentials discussed in sections \ref{sec2} and \ref{sec4}.
\end{appendix}

\end{document}